\documentclass[fleqn,usenatbib,twocolumn]{mnras}
\usepackage[utf8]{inputenc}
\usepackage{geometry}
\usepackage{stackengine}
\usepackage{newtxtext,newtxmath}
\usepackage[T1]{fontenc}
\usepackage[normalem]{ulem}
\usepackage{tikz}
\usepackage{algorithm}
\usepackage{algorithmicx}
\usepackage{algpseudocode}
\usepackage[outline]{contour}

\usepackage{chemfig}

\renewcommand{\algorithmiccomment}[1]{\textit{#1}} 
\algrenewcommand{\algorithmiccomment}[1]{#1}

\DeclareRobustCommand{\VAN}[3]{#2}
\let\VANthebibliography\thebibliography
\def\thebibliography{\DeclareRobustCommand{\VAN}[3]{##3}\VANthebibliography}
\usetikzlibrary{calc,fadings,decorations.markings,shapes.geometric}
\usepackage{graphicx}	
\usepackage{amsmath}	
\usepackage{float}
\usepackage{appendix}
\usepackage{xcolor}
\usepackage{multirow}
\usepackage{caption}
\definecolor{G01}{rgb}{0.0000,0.4470,0.7410}
\definecolor{G02}{rgb}{0.9290,0.6940,0.1250}
\definecolor{G07}{rgb}{0.4660,0.6740,0.1880}
\definecolor{G08}{rgb}{0.6350,0.0780,0.1840}
\definecolor{G28}{rgb}{0.8500,0.3250,0.0980}
\definecolor{G29}{rgb}{0.4940,0.1840,0.5560}
\definecolor{cH}{rgb}{1, 1, 0}
\definecolor{cH2}{rgb}{1, 0, 0}
\definecolor{cO}{rgb}{1, 0, 1}
\definecolor{cHO}{rgb}{0, 1, 1}
\definecolor{cH2O}{rgb}{0, 0, 1}
\definecolor{cO2}{rgb}{0, 1, 0}
\definecolor{Hp}{rgb}{0.6784,0.8471,0.9020}
\definecolor{H2p}{rgb}{0.2746,0.5098,0.7059}
\definecolor{H3p}{rgb}{0,0,0.5020}
\definecolor{Op}{rgb}{1.0000,0.7843,0}
\definecolor{HOp}{rgb}{1.0000,0.5882,0}
\definecolor{H2Op}{rgb}{1.0000,0.3922,0}
\definecolor{H3Op}{rgb}{1.0000,0,0}
\definecolor{O2p}{rgb}{0.5647,0.9333,0.5647}
\definecolor{O2Hp}{rgb}{0,0.5020,0}
\usepackage{graphicx}
\graphicspath{{./figures/}}

\title[Ion-neutral chemistry at Ganymede]{Ion-neutral chemistry at icy moons: the case of Ganymede}

\author[A. Beth et al.]{
A. Beth,$^{1}$\thanks{E-mail: arnaud.beth@gmail.com}
M. Galand,$^{1}$
X. Jia,$^{2}$
F. Leblanc$^{3}$
\\
$^{1}$Imperial College London, Department of Physics, London, SW7 2AZ, UK\\
$^{2}$University of Michigan, Department of Climate and Space Sciences and Engineering, Ann Arbor, MI 48109-2143, USA\\
$^{3}$LATMOS/IPSL, Sorbonne Université, UVSQ, CNRS, Paris, France
}

\date{Accepted XXX. Received YYY; in original form ZZZ}

\pubyear{\the\year{}}
\graphicspath{{figures/}}

\begin{document}
\label{firstpage}
\pagerange{\pageref{firstpage}--\pageref{lastpage}}
\maketitle

\begin{abstract}
Icy moons orbiting giant planets are often described as airless bodies though they host an exosphere where collisions between neutral species are scarce. In the case of Ganymede, the neutral composition is dominated by $\mathrm{H_2O}$, $\mathrm{H_2}$, and $\mathrm{O_2}$. Past observations by Galileo showed that Ganymede hosts an ionosphere and those by Juno revealed the presence of $\mathrm{H_3^+}$, an ion species only stemming from ion-neutral collisions. $\mathrm{H_3^+}$ detection suggests that ions and neutrals might still collide and be the source of new ion species on icy moons. We examine Ganymede's ability to host a more diverse ionosphere in terms of ion composition than previously thought and predict its variety. We upgraded our test-particle code of Ganymede’s ionosphere, formerly collisionless, to include ion-neutral collisions in a probabilistic manner. The updated code is applied to three Galileo flybys of Ganymede that were investigated in the absence of chemistry. Both sets of simulations have been compared and the effect of ion-neutral chemistry has been assessed. We show that in the case of an exosphere predominantly composed of $\mathrm{H_2O}$, $\mathrm{H_2}$, and $\mathrm{O_2}$, the ionosphere is made not only of their associated cations but also of $\mathrm{H_3^+}$, $\mathrm{H_3O^+}$, and $\mathrm{O_2H^+}$. Simulations reveal that, depending on the location, the contribution of $\mathrm{H_3^+}$ and $\mathrm{H_3O^+}$ to the ion composition may be significant. Strong dayside/nightside and Jovian/anti-Jovian asymmetries in the ion composition are identified. Our findings are key to interpreting Juno and future JUICE ion mass spectrometer datasets.
\end{abstract}




\section{Introduction}

Ganymede is the largest moon in the Solar System. Amongst its uncommon characteristics, such as bearing an intrinsic magnetic field \citep{Kivelson1996} and a potential subsurface ocean \citep{Kivelson2002,Saur2015}, Ganymede is surrounded by a thin, almost collisionless layer of gas so-called exosphere. This layer of gas is produced through several main processes: sublimation of water ice, sputtering by the harsh Jovian magnetospheric environment \citep{Plainaki2015,Pontoni2021} and ionospheric ions \citep{Carnielli2020b}, and radiolysis \citep{Johnson1990,Marconi2007,Shematovich2016} leading to an exosphere dominated by H$_2$O, O$_2$, and H$_2$, in part supported by remote-sensing observations \citep[e.g.][]{Barth1997,Hall1998,Roth2021,Roth2023}. In addition, there was recent evidence for a local CO$_2$ gas patch over the leading side of the North polar cap \citep{Bockelee2024}.  

The transition from collisional to collisionless environments is defined as the exobase where the scale height of the different neutral species is of the order of the mean free path. In the case of Ganymede, the exobase remains close to the surface except near the subsolar point where sublimation drastically increases the number density of H$_2$O, reducing the mean free path of neutral species \citep{Marconi2007,Leblanc2017,Leblanc2023}. However, Ganymede's atmosphere is often regarded as collisionless, hence models of its atmosphere neglect collisions if possible
\citep[e.g.][]{Vorburger2022,Vorburger2024,Leblanc2023}, allieviating computational resources. 

By ionising this exosphere, through photo- and electron-impact ionisation, Ganymede hosts an ionosphere as well, that has been probed in-situ by several flybys of the Galileo spacecraft \citep{Eviatar2001,Beth2025} and recently by the Juno spacecraft \citep{Kurth2022}. As the spacecraft approached the moon, the plasma density probed in situ via the PWS instrument \citep{Gurnett1992} increased, reaching 200\,cm$^{-3}$ for the G02 flyby at closest approach. Although Galileo did not offer us the ability to infer the ion composition during these flybys, preliminary collisionless modelling that combines DSMC simulations of the exosphere \citep{Leblanc2017,Leblanc2023} and MHD simulations of Ganymede's electromagnetic environments \citep{Jia2008,Jia2009} showed evidence for an ionosphere mainly made of H$_2^+$ and O$_2^+$ ions \citep{Carnielli2019} that agree with the in-situ total plasma number density for some flybys \citep{Beth2025}. There is good agreement in terms of ion energy distribution and shape within the Alfvén wings between simulations and in-situ observations by PLS \citep{Frank1992},  but worse in terms of absolute values, in part due to the lack of field-aligned electric fields in the modelling. In contrast to Galileo, thanks to its time-of-flight ion mass spectrometer, the Juno spacecraft provided further insight with the additional capability to separate masses. Not only were H$_2^+$ and O$_2^+$ indeed unambiguously identified, supporting early findings of \citet{Carnielli2019} and \citet{Beth2025}, but also H$_3^+$ and water-group ions around 16\,u\,q$^{-1}$, a priori dominated by O$^+$, were also detected  \citep{Valek2022,Allegrini2022}.

The serendipitous detection of H$_3^+$ by Juno casts doubt on the ``collisionlessness'' of Ganymede's exo-ionosphere. Indeed, H$_3^+$ cannot be produced through the ionisation of a neutral species. It is primarily found in the interstellar medium and in the thermosphere of gas giants: H$_3^+$ is produced through ion-neutral chemistry, mainly H$_2$+H$_2^+$. This implies that collisions between ion and neutral species may still happen at Ganymede. \citet{Waite2024} attempted to model H$_3^+$, assuming photo-chemical equilibrium (neglecting transport but including dissociative recombination), imposing the electron density profile from \citet{Buccino2022}, and applying a limited chemical network that does not account for all possible ion and neutral species.

Based on the neutral composition of Ganymede's exosphere and their ionised counterpart, three new ions species are produced through ion-neutral collisions only: H$_3^+$ (present around gas giants and in the interstellar medium), H$_3$O$^+$ (found at planets and comets), and O$_2$H$^+$ (not yet detected). In this paper, we address the ability of Ganymede, an icy moon, to host these additional ions, hence, a more diverse ionosphere than previously thought. By updating their  test-particle model from \citet{Beth2025} to include ion-neutral collisions, we show that Ganymede's ionosphere for different configurations, corresponding to those of Galileo's flybys, is not only made of H$_2^+$, O$_2^+$, and H$_2$O$^+$ (along with H$^+$, O$^+$, and HO$^+$ to a lesser extent) but of H$_3^+$ and H$_3$O$^+$ (O$_2$H$^+$ found to be negligible).

We present the update made for the test-particle model to account for ion-neutral collisions in Section~\ref{section2}. The results are presented in Section~\ref{section3}, then discussed in Section~\ref{section4}, followed by the conclusions in Section~\ref{section5}.

\section{Method}\label{section2}

\subsection{Model description}

The present modelling work builds upon \citet{Beth2025}, in which the test particle model is described in detail, including inputs and approach. We summarise here the main steps. First, we use the simulated neutral exospheric number densities of the different neutral species (H, H$_2$, O, HO, H$_2$O, O$_2$, ignoring CO$_2$ and its fragments that are not adequately modelled yet) as a function of the altitude from the surface to 5~$R_G$ ($R_G$: Ganymede's radius), of the different neutral species in the environment of Ganymede from a DSMC model \citep{Leblanc2023}. Second, we generate ions as macroparticles at a rate based on the ionisation frequency, which includes both photo- and electron impact ionisation \citep[values are those from][]{Carnielli2019}. Once produced, macroparticles are transported through the magnetosphere via electric and magnetic fields. These are extracted from MHD simulations at Ganymede performed for the different Galileo flybys \citep{Jia2008,Jia2009}. Based on the time spent in the cells of our simulation grid, one can derive the moments, such as number densities and mean velocities of the different ion species within Ganymede's magnetosphere.

The gyrofrequency of the ion primarily governs the timestep in the simulation $\Delta t$. We have fixed $\Delta t [\text{s}]= 10^{24} m_i[\text{kg}]\approx 1.7\times 10^{-3} m_i[\text{u}]$ where $m_i$ is the ion mass. The original limitation was to resolve at least 1/20 of the gyromotion for good accuracy with the Boris pusher \citep{Boris1970}. At Ganymede, the strongest magnetic field is $\sim$1440\,nT near the poles such that:
\begin{align}
    \Delta t<\min\left\{\dfrac{1}{20}\dfrac{2\pi}{\omega_{ci}}\right\}&=\min\left\{\dfrac{1}{10}\dfrac{m_i\pi}{q_i B}\right\}\nonumber\\
    &=\dfrac{1}{10}\dfrac{m_i\pi}{q_i\max\{B\}}=1.36\times10^{24} m_i\label{dt}
\end{align}
$\Delta t$ must also be constrained by the grid resolution (i.e. fulfilling the Courant-Friedrich-Lewy (CFL) condition). As our grid resolution $\Delta x$ is $\sim 66$\,km and $\Delta t[s]\sim 0.0017 m_i[\text{u}]$, the heaviest ion (O$_2^+$) would need to be faster than $10^3$~km\,s$^{-1}$ to break the CFL condition. That is almost an order of magnitude higher than the speed of the undisturbed Jovian plasma and three orders higher than that of the plasma flow near Ganymede's surface \citep{Beth2025}.

We assume that these ions have no feedback on the field: The MHD simulations are not self-consistent and are purely based on the interaction of the Jovian plasma with Ganymede's magnetic field. \citet{Beth2025} showed that the modelled number densities are still consistent with in-situ Galileo PWS observations for some Galileo flybys (e.g. G01 and G07), and the trends in the ion energy spectra seen with PLS are captured, while discrepancies persist. Their simulations were performed with the collisionless version of the test-particle model. 

\subsection{Implementation of collisions}\label{section22}

Compared with the model of \citet{Beth2025}, we have added ion-neutral collisions that macroparticles can undergo at each timestep while transported through the ionosphere. Initially, the trajectory of the macroparticle is simulated by applying the Boris scheme \citep{Boris1970}. We interpolate the electric and magnetic fields at the location of the macroparticle and use them as input to update the particle's position and velocity from $t$ to $t+\Delta t$. That is repeated until the ion crashes onto Ganymede's surface or leaves the simulation box (see Appendix~\ref{Appendixreaction}). In this updated version of the test-particle model, to account for ion-neutral collisions during the period $[t;t+\Delta t]$, we estimate the probability of colliding with a neutral species $p_\text{collision}$ given by:
\begin{eqnarray}
    p_\text{collision}&=&1-\exp\left(-\sum_n \nu^\text{tot}_{i,n} \Delta t\right)\\
     \nu^\text{tot}_{i,n}&=&n_n <\!\sigma^\text{tot}_{i,n}(\varv_\text{rel.})\varv_\text{rel.}\!>\label{Eq3}
\end{eqnarray}
where $n_n$ is the number density of the neutral species $n$, {$\sigma^\text{tot}_{i,n}$} stands for the total collision cross section between the ion species $i$ and the neutral species $n$, $\nu_{i,n}$ is the collision frequency between both species and ${<\!\sigma^\text{tot}_{i,n}(\varv_\text{rel.}){\varv_\text{rel.}}\!>}$ is defined as
\begin{equation}
<\!\sigma^\text{tot}_{i,n}(\varv_\text{rel.})\varv_\text{rel.}\!>=\int_{\mathbb{R}^3}\,\sigma^\text{tot}_{i,n}(\varv_\text{rel.})\varv_\text{rel.}f_{n}(\vec{V})\,\mathrm{d}^3\vec{V}
\label{Eq5}
\end{equation}
where $f_n$ stands for the velocity distribution function of the targeted neutral and the relative speed $\varv_\text{rel.}=\Vert\vec{\varv}_i-\vec{V}\Vert$ where $\vec{\varv}_i$ and $\vec{V}$ stand for the ion and neutral velocities respectively. This represents the weighted/averaged rate at which an ion collides with neutrals following a velocity distribution function $f_n$. We have ignored the angular dependency of the cross-section for the sake of simplicity. In practice, $f_n$ is assumed to be a normalised Maxwell-Boltzmann velocity distribution defined by a mean bulk velocity $\vec{V}_n$ and a temperature $T_n$. Still Eq.~\ref{Eq5} remains impractical to solve. $<\!\sigma^\text{tot}_{i,n}(\varv_\text{rel.})\varv_\text{rel.}\!>$ can be approximated in two different ways. The first approach is to drop the integral and assume that all neutrals move at $\vec{V}_n$ such that 
\begin{equation}
<\!\sigma^\text{tot}_{i,n}(\varv_\text{rel.})\varv_\text{rel.}\!>\approx\sigma^\text{tot}_{i,n}(\Vert\vec{\varv}_i-\vec{V}_n\Vert)\Vert\vec{\varv}_i-\vec{V}_n\Vert,
\end{equation}
valid as long as $\varv_\text{rel.}\approx\Vert\vec{\varv}_i-\vec{V}_n\Vert\gg \varv_\text{th,n}$, where $\varv_\text{th,n}=\sqrt{2k_BT_n/m_n}$ is the thermal speed of the neutral species $n$. In the limit $\Vert\vec{\varv}_i-\vec{V}_0\Vert\lesssim \varv_\text{th,n}$, a more refined model would be needed to take into account the velocity dispersion of the neutrals. For instance, in the hard sphere approximation, one can evaluate \citep[see e.g.][]{Fahr1967}
\begin{align}
<\!\sigma^\text{tot}_{i,n}(\varv_\text{rel.})\varv_\text{rel.}\!>\approx&\, \sigma^\text{tot}_{i,n}(\Vert\vec{\varv}_i-\vec{V}_n\Vert)\cdot\nonumber\\
&\underbrace{\varv_\text{th,n}\left[\dfrac{\exp(-x^2)}{\sqrt{\pi}}+x\left(1+\dfrac{1}{2x^2}\right)\text{erf}(x)\right]}_{\int_{\mathbb{R}^3}\,\Vert\vec{\varv}_i-\vec{V}\Vert f_{n}(\vec{V})\,\mathrm{d}^3\vec{V}}\label{Eq6}
\end{align}
where $x=\Vert\vec{\varv}_i-\vec{V}_n\Vert/\varv_\text{th,n}$ (consistent with Eq.~\ref{Eq6} for $x\gg1$). Nevertheless, Eq.~\ref{Eq6} requires the full, exhaustive, and accurate knowledge of all $\sigma_{i,n}$ for all combinations $\{i,n\}$ and over a wide range of energy (that said $\varv_\text{rel.}$) typically from 0.1\,eV to 100-200\,eV. That becomes unpractical as the number of neutral and ion species increases. It may significantly slow down the computation and increase the complexity of the model due to the large number of outcomes for all collisions. In addition, not all cross-sections between the different ions and neutrals are available.

The second approach to approximate Eq.~\ref{Eq3} is :
\begin{equation}
<\!\sigma^\text{tot}_{i,n}(\varv_\text{rel.})\varv_\text{rel.}\!>=\int_{\mathbb{R}^3}\,\sigma^\text{tot}_{i,n}(\vec{V})\Vert\vec{V}\Vert f_{n}(\vec{V})\,\mathrm{d}^3\vec{V}=k_{in}(T_n)
\label{Eq7}
\end{equation}
where $k_{in}(T_n)$ is the ion-neutral reaction rate coefficient. We have decided to use the second approach (see also the discussion in Section~\ref{section41}). The rate coefficients are better known, constrained, and tabulated than the cross-sections needed in the first approach. By only considering kinetic rates from  from \url{https://umistdatabase.net/} \citep[][see Appendix \ref{Appendixreaction}]{Millar2022}, a couple of assumptions are made:
\begin{itemize}
    \item Ion-neutral collisions are treated as if they were occurring at thermal energy (with relative energy $\lesssim 10$\,eV). 
    \item It indirectly means that the considered collisions change the reactants (reactants$\neq$products, elastic and symmetric charge-exchange collisions are not considered) as we primarily focus on the ion composition and ion-neutral chemistry. For example, symmetric charge exchange such as $\text{O}_2^++\text{O}_2\longrightarrow\text{O}_2+\text{O}_2^+$ is ignored. However, \citet{Carnielli2020a} showed that such a process does not affect the total number density in their simulations for the G02 flyby, keeping in mind that their ionosphere was dominated by O$_2^+$.
    \item Only spontaneous reactions in gas phase, that said exothermic and athermic (e.g. O$_2$ + H$_3^+$$\leftrightarrow$O$_2$H$^+$+H$_2$), are considered.
    \item Reactants and products are assumed to be in the ground state. A non-exhaustive list of reactions is missing as they involve excited states. For instance, the reaction {$\text{H}^++\text{H}_2(\nu)\rightarrow\text{H}+\text{H}_2^+$} that is exothermic only when H$_2$ is in an excited vibrational state $\nu\geq 4$ \citep{Huestis2008} is ignored. Similarly, ions may also be left in an excited state following electron-impact ionization that is the main source of ionization within Ganymede's ionosphere, and become more reactive \citep[e.g. $\text{O}_2^{+*}+\text{H}_2\text{O}\rightarrow\text{O}_2+\text{H}_2\text{O}^+$,][]{Turner1968}, being ignored here as well.
\end{itemize}

We fix the neutral temperature to $T_n=100$~K. First, it corresponds to the mean temperature at Ganymede's surface \citep[varying between 80 and 150~K, see][]{Leblanc2017}. Secondly, the lower it is, the higher the rate coefficient is. The simulations will overestimate ions born from chemistry if Ganymede is warmer (putting aside that we have used rate coefficients and not cross sections as discussed in Section~\ref{section41}). As a first application to Ganymede, the main goal is to provide a qualitative understanding of how ion-neutral chemistry may affect the exo-ionosphere of icy moons in the magnetised case.

 At each time step, we draw a random number $p$ following a uniform distribution on the interval $[0,1]$ (i.e. $p\in\mathcal{U}_{[0,1]}$) and compare to $p_\text{collision}$: If $p<p_\text{collision}$, there is collision, none otherwise. The collisionless version of the test particle is equivalent to $p_\text{collision}=0$. In case of a collision, we need to identify the neutral species with which the ion species $i$ has collided. We estimate the probability $p_{i,n}$ for the ion species $i$ to collide with the neutral species $n$ given by:
\begin{equation}
    p_{i,n}=\dfrac{\nu_{i,n}}{\sum_{j}^\text{neutrals} \nu_{i,j}}
\end{equation}
such that $\sum^\text{neutrals}_n p_{i,n}=1$.

Note that for accuracy, $\Delta t$ must obey $\nu_{i,n}\Delta t\lesssim 1$: A maximum of only one collision must occur during $\Delta t$ or, more precisely, the probability to have 2 or more collisions during a timestep must remain low. That approximation holds at Ganymede. A quick way to estimate the maximum ion-neutral collision frequency is to multiply the maximum neutral number density, found in general at the surface, with the largest kinetic rate coefficient to get an upper limit for $\nu_{i,n}=k_{i,n}n_n$. The maximum neutral number density is $\sim 3\times10^{15}$\,m$^{-3}$ \citep[cf. Appendix A in][]{Beth2025} and the maximum rate coefficient is around $\sim 10^{-14}$\,m$^{3}$\,s$^{-1}$ (typical order of magnitude at $T_n=100$~K for reactions with water) such that the ion-neutral collision frequency is $\lesssim30$\,s$^{-1}$. $\Delta t$ is mass-dependent and varies from 1.7$\times 10^{-3}$\,s for $\mathrm{H^+}$ to 5.5$\times 10^{-2}$\,s for $\mathrm{O_2H^+}$. $\nu_{i,n}\Delta t$ is found to be maximal by far for any ion colliding with H$_2$O, up to $\sim 0.53$ at the surface for O$^+$+H$_2$O and therefore still below 1. As the occurrence of collisions follows a Poisson distribution if collisions are independent (though not completely true as the order of the collisions matters and these collisions change the nature of the ions), the probability for no collision, in that case, is $\exp(-0.53)\approx58.8\%$, only 1 collision $31.2\%$, and 2 or more $10\%$ that we found acceptable to not constrain against chemistry. In a less magnetised environment, $\Delta t$ would have to be adapted with respect to $\nu_{i,n}$ instead. Our approach is valid in the limit $\nu_{i,n}\lesssim\omega_{c,i}$, hence perfect for Ganymede where the ion motion is dominated by electromagnetic fields and not collisions owing to Ganymede's own magnetic field. As the ion moves farther away from the moon, the ion-neutral collision frequency decreases with the decrease in the neutral number density. Therefore, $\nu_{i,n}\Delta t$ decreases with altitude as $\Delta t$ is fixed for a given ion. Including collisions increases the computational time compared to the collisionless case: Drawing random numbers requires more time (roughly +50-100\%). To keep the simulation time reasonable, we have reduced the number of simulated macroparticles generated tenfold \citep[that said $N_\text{stat}=10^3$ instead of $10^4$ in][]{Beth2025}. This increases the numerical noise \citep[roughly 3 times larger than those from][as the standard deviation scales with $1/\sqrt{N_\text{stat}}$]{Beth2025} in the number density profiles. The most affected ion species are Jovian H$^+$ and O$^+$ and those with a low statistic, hence number density.

Along with the ion-neutral chemistry and the formation of new ion species, we have also considered the excess energy following the collision. As reactions considered here are exothermic (eventually athermic), the excess energy must be dissipated somehow. In the present model, we assume that the excess energy is only converted into kinetic energy and redistributed between the products. This is described in detail in Appendix~\ref{AppendixPostcoll}.

\section{Results}\label{section3}

In this section, we present the results of our simulations, which include ion-neutral chemistry, for three different Galileo flybys: G01, G07, and G29. We choose these flybys for several reasons. They were all performed in the Alfvén wings, when Ganymede was outside the plasma sheet, and showed the best model-data agreement \citep{Beth2025}. Furthermore, these flybys are associated with different Ganymede's local time. For G01, the Sun illuminated the anti-Jovian side of Ganymede. In contrast, for G07, the trailing hemisphere, slightly towards the Jovian side, was illuminated. G29 flyby was performed when Ganymede passed through Jupiter's shadow, which corresponds to an H$_2$O-deficient exosphere, mainly produced by sputtering and radiolysis, with photoionisation turned off. We summarise the characteristics of these flybys in Table \ref{Table1}. 
First, we have performed simulations along the flybys' trajectories to be compared with results from \citet{Beth2025} (see Section~\ref{flybys}). Next, we have performed 3D cuts of ion number densities for the configurations met during these flybys (see Section~\ref{3D}). All the results are discussed and shown in the GPhiO coordinate system, centred on Ganymede, where $X$ points in the direction of the incident plasma flow, $Y$ points towards Jupiter, and $Z$ completes the orthogonal system.

\begin{table*}
\caption{Configuration of the different flybys considered for our simulations. Adapted from \citet{Beth2025}.}\label{Table1}
\begin{tabular}{cccccccrccc}
    \hline
        Flyby   & \multicolumn{7}{c}{Location}& \multicolumn{3}{c}{$<\vec{\varv}_\text{SC}>$}\\
        \cline{2-8}\cline{9-11}
        & Rel. to& $\angle$ Sun-Jup.-Gan.&Rel. to&Plasma&&CA&&\multicolumn{1}{c}{$<\varv_{\text{SC},x}>$}&\multicolumn{1}{c}{$<\varv_{\text{SC},y}>$}&\multicolumn{1}{c}{$<\varv_{\text{SC},z}>$}\\
        \cline{6-8}
        & the PS&&Ganymede&regions&$r\, [R_G]$&\hphantom{-}lat.&long.&\multicolumn{3}{c}{[km\,s$^{-1}$]}\\
        \hline
        \textbf{G01} &$\uparrow$&  349$^{\circ}$&central wake &&$1.32$&$\hphantom{-}30.6^{\circ}$&$-21.1^{\circ}$&$\hphantom{-}2.08$&$\hphantom{30}7.30\hphantom{7}$&$\hphantom{-}1.12$\\
        \textbf{G07} & $\downarrow$& 115.5$^{\circ}$&mid-lat. downstr.&Alfvén wing&$2.18$&$\hphantom{-}55.6^{\circ}$&$2.9^{\circ}$&$\hphantom{-}0.46$&$\hphantom{.43}{-8.43}\hphantom{-8.}$&$-0.04$\\
        \textbf{G29} & $\uparrow$& 178.5$^{\circ}$&mid-lat. downstr. &&$1.89$&$\hphantom{-}62.4^{\circ}$&$1.4^{\circ}$&$-0.12$&$\hphantom{45}10.45\hphantom{10}$&$-0.07$
    \end{tabular}
    \label{table1}
\end{table*}
\subsection{Along the trajectory of Galileo's flybys: G01, G07, and G29}\label{flybys}

\begin{figure*}
\centering
\includegraphics[width=.49\linewidth]{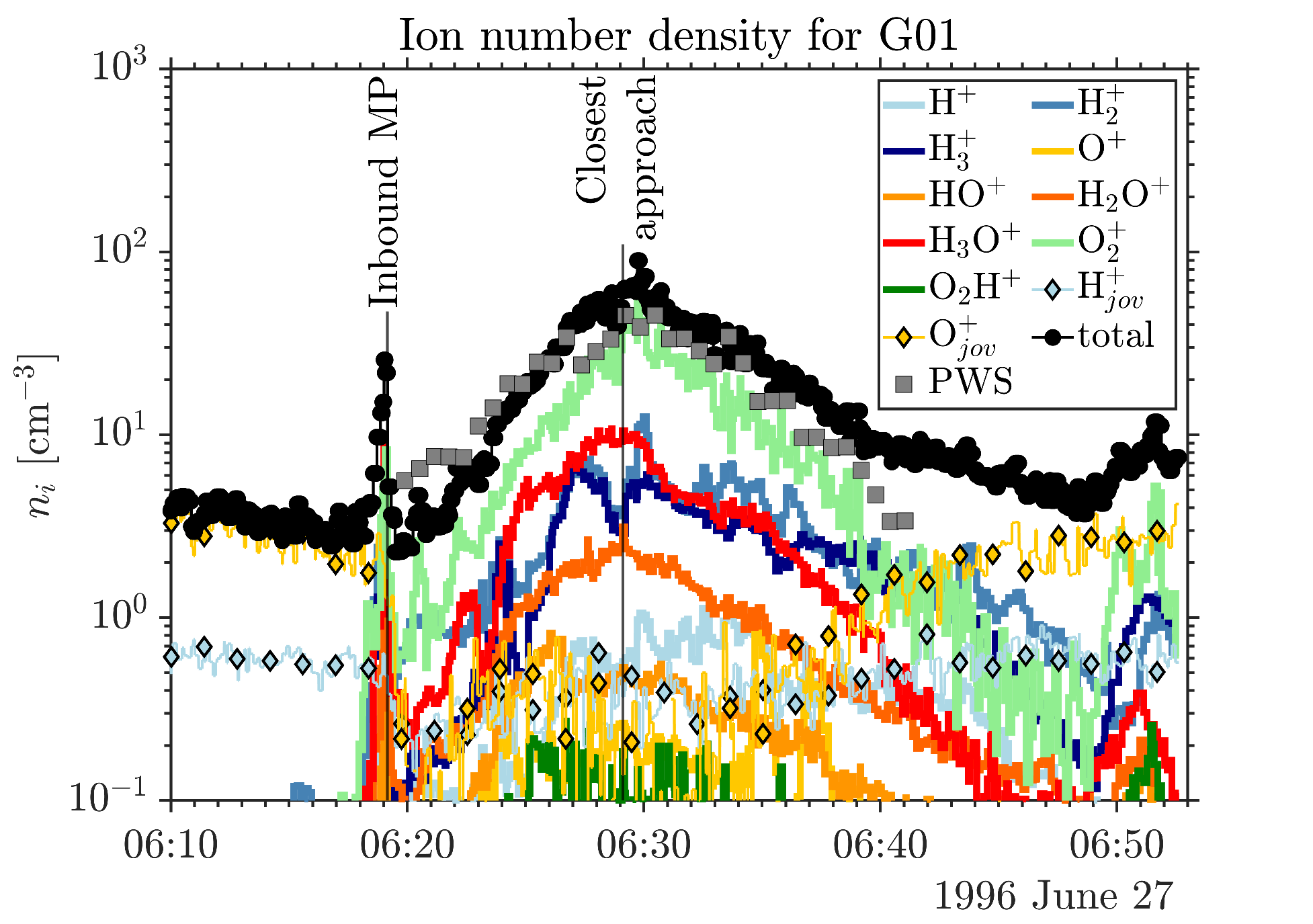}
\includegraphics[width=.49\linewidth]{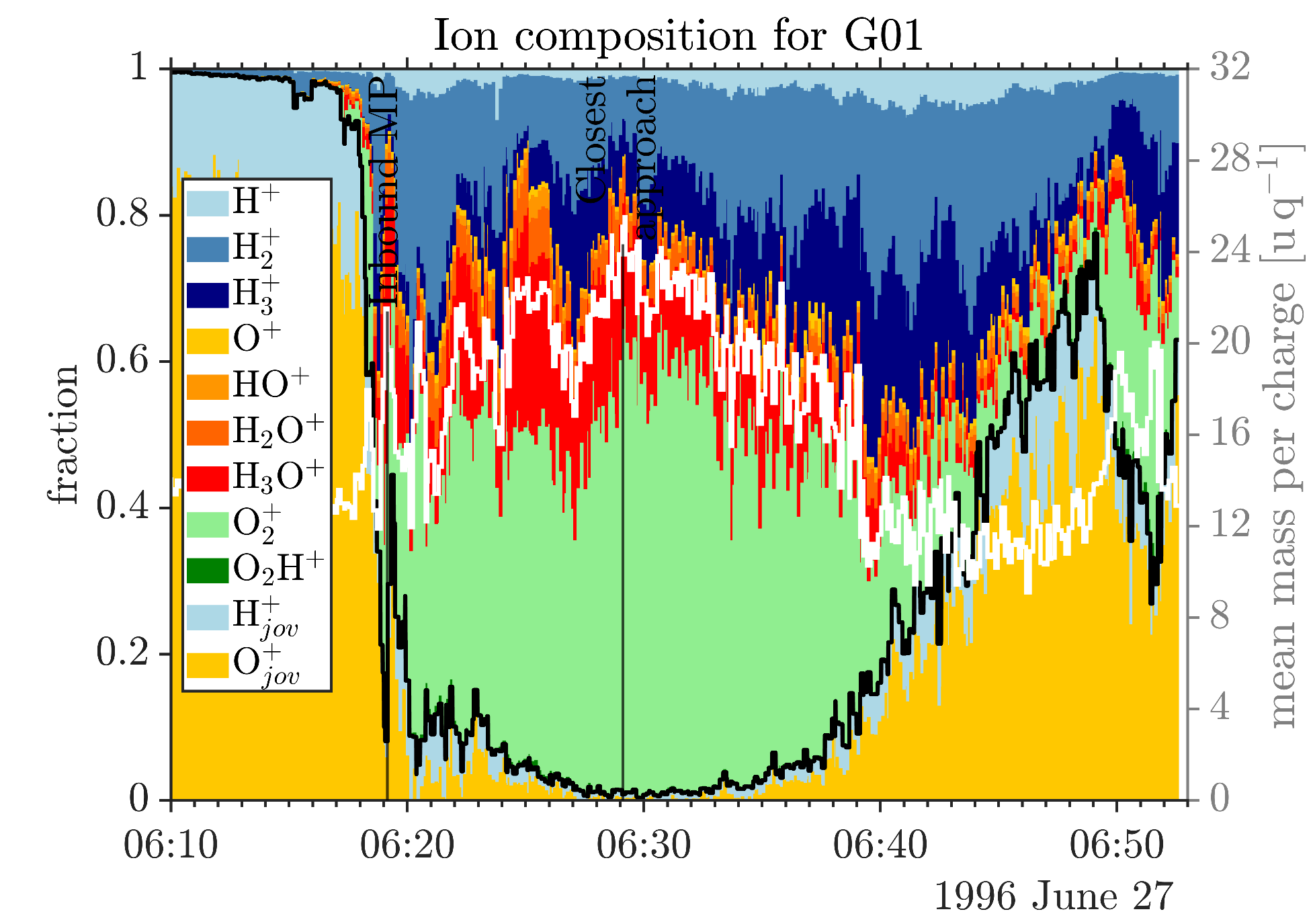}\\
\includegraphics[width=.49\linewidth]{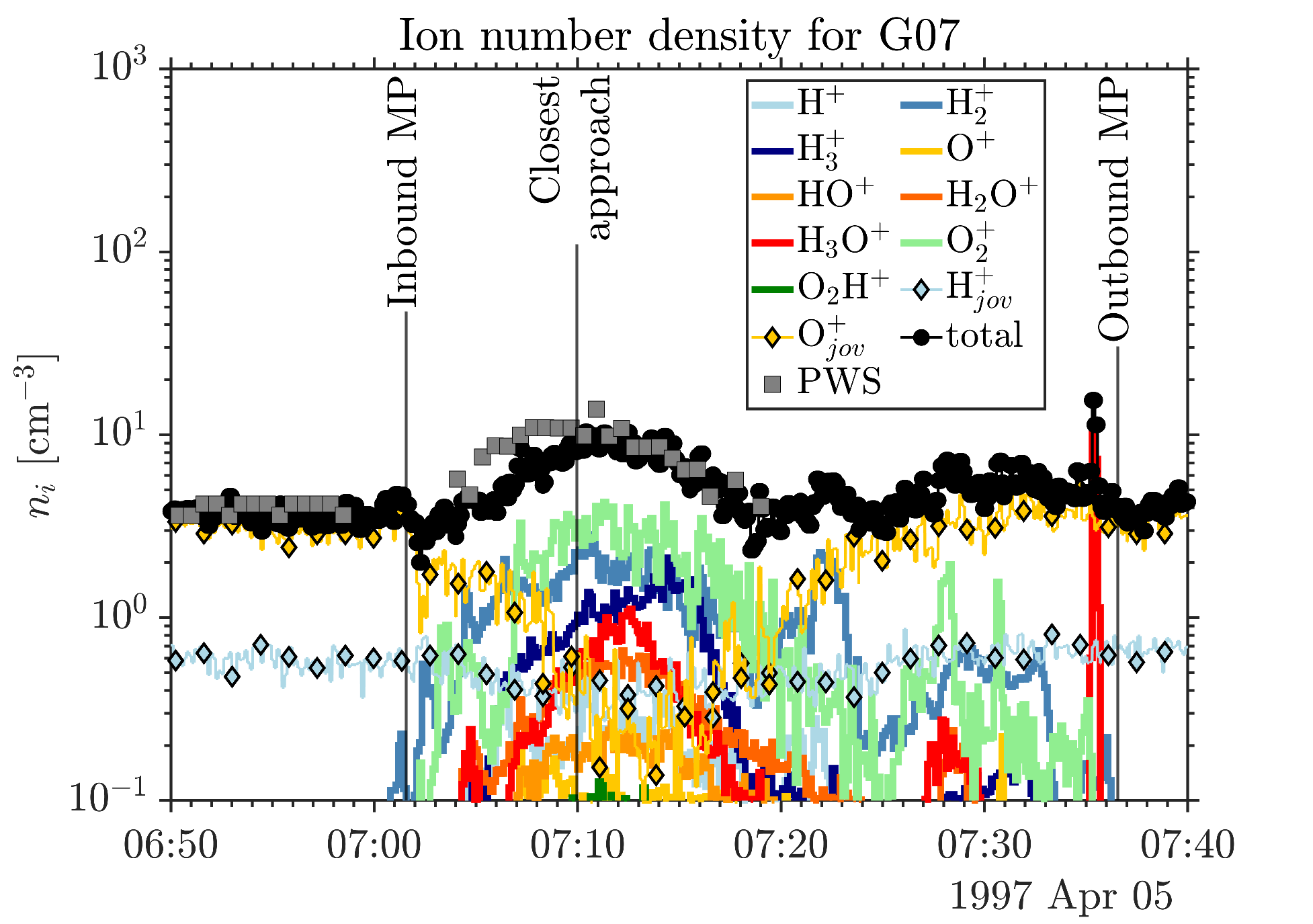}
\includegraphics[width=.49\linewidth]{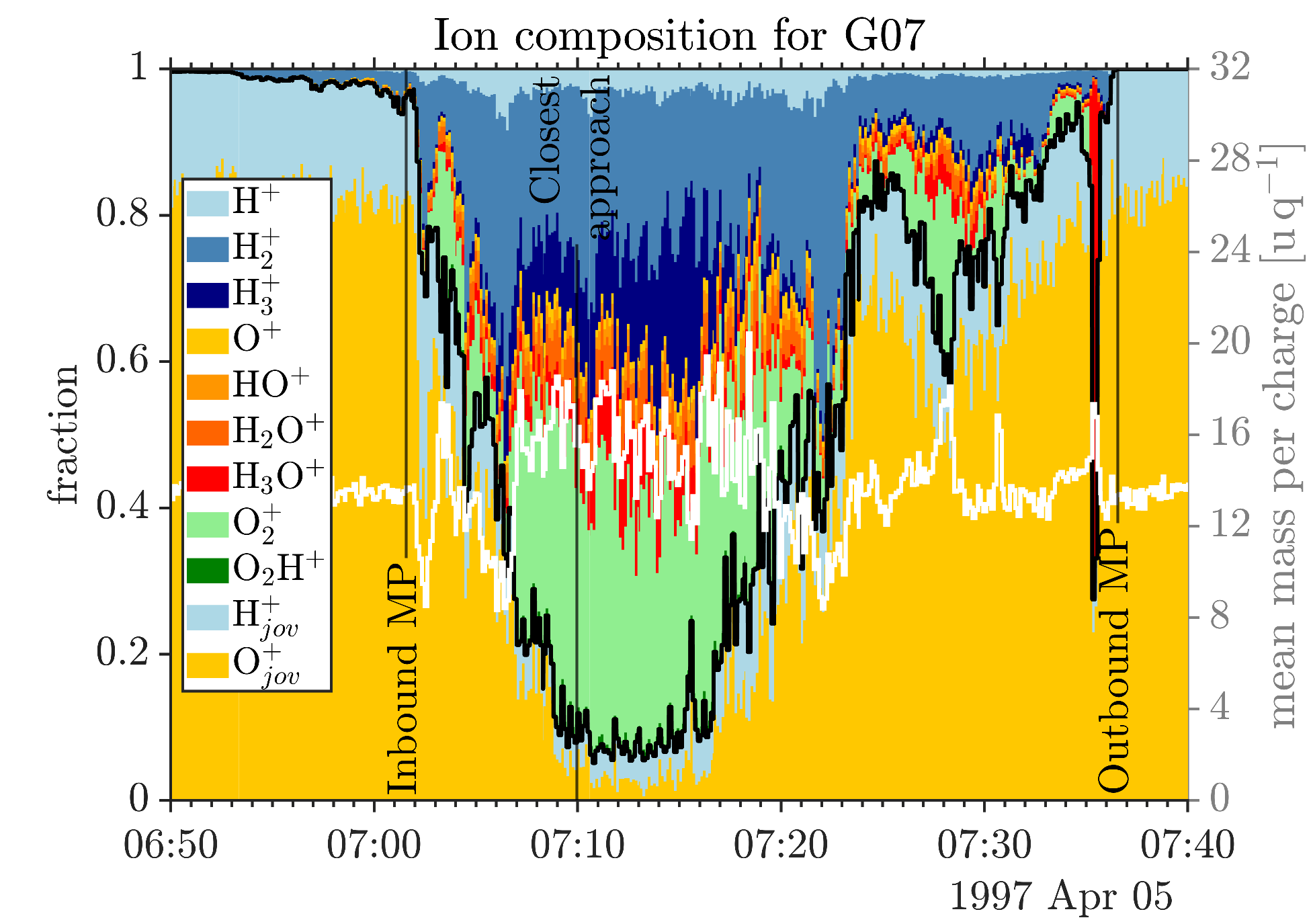}\\
\includegraphics[width=.49\linewidth]{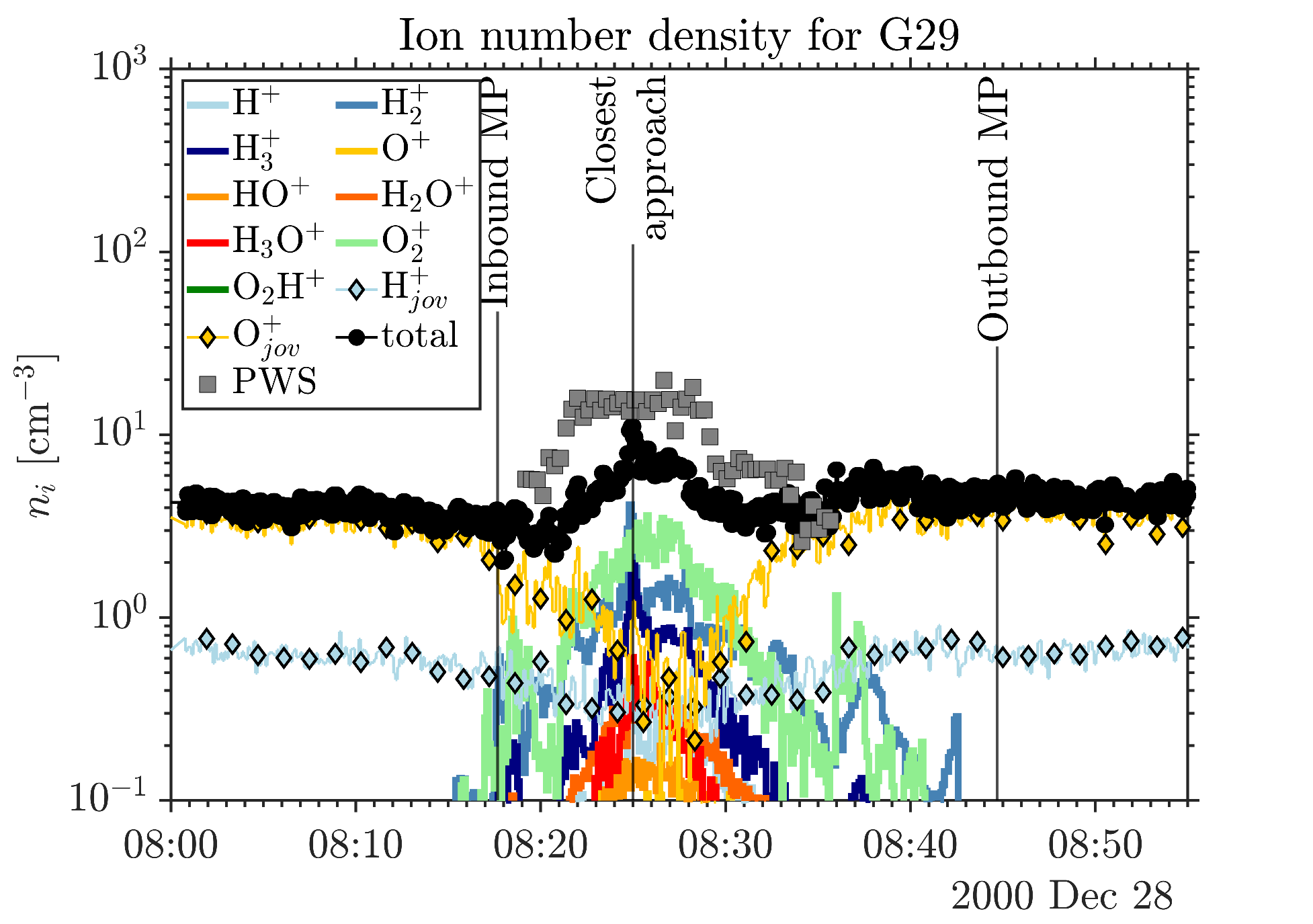}
\includegraphics[width=.49\linewidth]{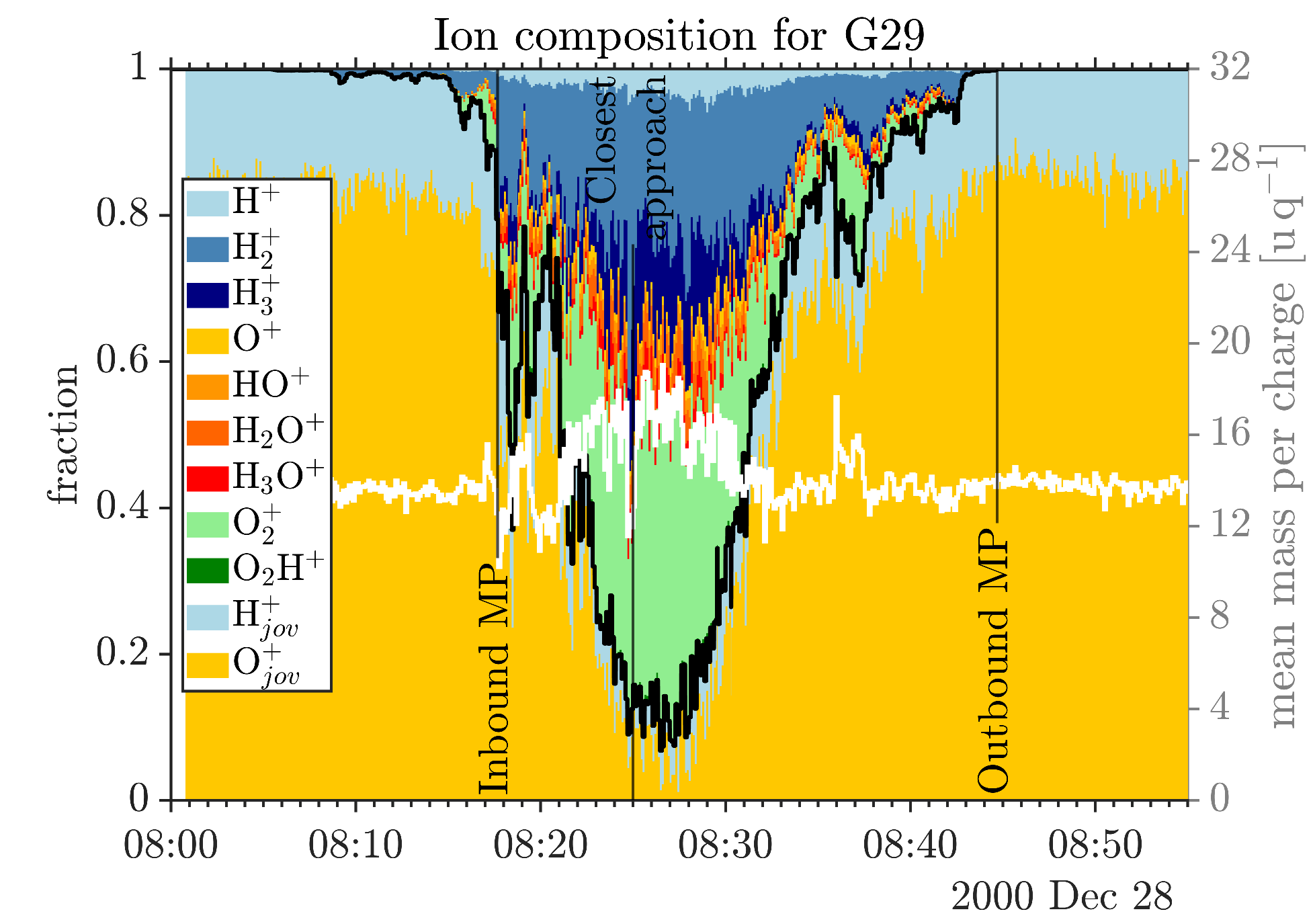}\\
\caption{Simulated ion number densities for Jovian and ionospheric ions compared with in situ PWS electron number density from \citet{KurthPWS} (left column) and simulated ion composition (right column) during G01 (top row), G07 (middle row), and G29 (bottom row) flybys. Vertical, black lines feature the inbound magnetopause crossing, the closest approach, and the outbound magnetopause crossing (if available from the MAG data), respectively. For the ion composition, the black, thick line represents the fraction of Jovian ions. The white, thick line must be read with the right axis and corresponds to the mean ion mass per charge.}
\label{Fig1}
\end{figure*}

Fig.~\ref{Fig1} shows the ion number densities (left) and composition (right) for G01, G07, and G29. These results should be compared with Fig. 5 from \citet{Beth2025} without collisions. In addition to H$^+$, H$_2^+$, O$^+$, HO$^+$, H$_2$O$^+$, and O$_2^+$, three new ion species are included in the new simulations as these species only stem from ion-neutral collisions: H$_3^+$, H$_3$O$^+$, and O$_2$H$^+$ (see Appendix \ref{Appendixreaction}). 

First, ion-neutral chemistry does not modify the total ion number density. The main reason is that ions are primarily all drifting at the same velocity regardless of their mass within Ganymede's ionosphere \citep{Beth2025}. Ionospheric ions drift at the same perpendicular velocity perpendicular to $\vec{B}$ (i.e. $\vec{E}\times\vec{B}/B^2$ where $\vec{E}$ and $\vec{B}$ are the electric and magnetic fields from the MHD simulation) as the simulations assume ideal MHD \citep{Jia2008,Jia2009}. Ion parallel velocities (negligible at the surface) increase as ions move away from Ganymede's surface and get closer to the magnetopause: The parallel velocity is mass-dependent, hence attributed to finite Larmor radius effects \citep{Beth2025}. Therefore, the total ion number density is expected to change only where ion velocities depart from each other; that said, around the magnetopause or the edges of the Alfvén wings, due to ion-neutral chemistry. We do not observe such a change in our simulations as the mean mass (white solid line) at the magnetopause (MP) is barely changed compared with the collisionless case (Fig.~\ref{Fig1}, right column). Indeed, the Jovian plasma density is not yet depleted at the MP, O$_2^+$ number density is the same as for the collisionless case, and ionospheric ions (H$_2$O$^+$, HO$^+$, O$^+$, and a fraction of H$_2^+$), as they react, are mainly turned into H$_3$O$^+$ that has a mass close to the mean one in the collisionless case (going from 10 to 20\,u, with an average of 14\,u for the Jovian plasma). The main deviation of the mean ion mass between the collisionless case and the collision case is observed before the G01 closest approach. Dominated by H$_2^+$ and O$_2^+$ in the collisionless case, it is now dominated by O$_2^+$ and water-group ions: Part of H$_2^+$ has been converted either directly to H$_2$O$^+$/H$_3$O$^+$ or indirectly through the intermediate formation of H$_3^+$.

For all flybys, before inbound and after outbound MP crossings, the Jovian ion number density is unperturbed by ion-neutral collisions. Although our neutral exosphere extends up to $5\,R_G$, the neutral number density remains low \citep[$\lesssim 10^{11}$\,m$^{-3}$,][]{Beth2025}; hence, the collision probability mainly driven by H$_2$ at these altitudes is four orders of magnitude lower compared with that at the surface.

For G01 within Ganymede's magnetosphere, the O$_2^+$ number density does not change either, as this ion species does not react with any neutral species included in the model: It is a terminal ion.  In our simulation, it can only be lost through transport, as dissociative recombination is ignored as it is negligible. Indeed, compared with ion-neutral collision frequency, for dissociative recombination to be of the same order of magnitude would require a plasma number density to be 10$^{7}$~cm$^{-3}$ at least, whereas a posteriori it is less than 10$^{5}$~cm$^{-3}$ from the simulations. Although a few ion-neutral reactions, namely charge exchange, lead to O$_2^+$ production, they remain far less efficient than the direct ionisation of O$_2$. The production of O$_2^+$, through charge exchange or ionisation, relies on O$_2$. This neutral species is confined near the surface with the lowest scale height amongst neutral species, around 20~km. Although produced only near the surface, O$_2^+$ is efficiently transported from the surface on the trailing side, through the whole ionosphere (cf. streamlines in O$_2^+$ panel) and contributes mostly to the ion composition. In addition, H$_3^+$ and H$_3$O$^+$ contribute significantly to the ion composition (see Fig.~\ref{Fig1}, top panel). The main chemical pathways for producing these ions are:
\begin{align}
\text{H}_2^+\phantom{O_3}+\text{H}_2\phantom{O}&\longrightarrow \text{H}_3^+\phantom{O_3}+\text{H}\label{reaction1}\\
\text{H}_3^+\phantom{O_3}+\text{H}_2\text{O}&\longrightarrow \text{H}_3\text{O}^++\text{H}_2\label{reaction2}\\
\text{H}_2\text{O}^++\text{H}_2\text{O}&\longrightarrow \text{H}_3\text{O}^++\text{HO}\label{reaction3}
\end{align}
Interestingly, H$_2$O$^+$ number density is equal to or even slightly higher than in the collisionless case. This suggests that the primary path for forming H$_3$O$^+$ is not through the loss of H$_2$O$^+$ but through that of H$_3^+$, and that an ion-neutral reaction leads to the production of H$_2$O$^+$. Given the neutral composition (mainly O$_2$, H$_2$, and H$_2$O), number densities, and by looking at the initial `identity' of the macroparticle at the start of the simulation, we determined that H$_2$O$^+$ is also produced through charge exchange
\begin{align}
\text{H}_2^+\phantom{O_3}+\text{H}_2\text{O}&\longrightarrow \text{H}_2\text{O}^++\text{H}_2\label{reaction4}
\end{align}
Therefore, H$_2^+$ number density decreases to the benefit of H$_3^+$ and H$_2$O$^+$ formation.

As H$_2$ is relatively uniform around Ganymede (i.e. with a roughly spherical symmetry), H$_2^+$ production and H$_2^+$ loss/H$_3^+$ production are also expected to be uniform around the moon. Unlike H$_2$, H$_2$O is primarily produced from the sublimation of the water ice on the surface: It is mainly present on the sunlit side. Galileo crossed the terminator plane from the sunlit side to the dark one around 06:33 \citep[cf. Fig. 12 in][]{Beth2025}, H$_3^+$ reacts with H$_2$O to become H$_3$O$^+$. As illustrated in Fig.~\ref{Fig1} (top row), beside O$_2^+$ which dominates around CA, H$_3^+$ is more present on the dark side owing to the lack of H$_2$O whereas H$_3$O$^+$ is mainly formed on the sunlit side. 

For G07 and G29 (cf. Fig.~\ref{Fig1}, middle and bottom rows), the change in ion composition is even less dramatic as these flybys were performed much farther away from Ganymede (resp. 2.18 and 1.89\,$R_G$ to compare with 1.32\,$R_G$ for G01 at closest approach). For G07, H$_2^+$ number density is mainly reduced between 07:05 and 07:15 compared with the profiles from \citet{Beth2025} as it is turned into H$_3^+$ (directly through proton transfer with H$_2$), H$_2$O$^+$ (through charge exchange with H$_2$O), and H$_3$O$^+$ (either directly through proton transfer with H$_2$O, or indirectly through two successive proton transfers, first with H$_2$ and then H$_2$O). We observed an asymmetry in the H$_3^+$ ion number density. While most ion number densities peak at closest approach, H$_3^+$ number density peaks slightly later around 07:12-07:15, interestingly corresponding to local midnight in terms of `illumination' in Ganymede's frame \citep[cf. Fig. 12 in][]{Beth2025}. In addition, right before the G07 outbound MP crossing, the plasma peak is dominated by H$_3$O$^+$ instead of H$_2$O$^+$ in the collisionless case \citep[to be compared with Fig. 5 in][]{Beth2025}. Although the outbound MP crossing occurs on the dark side \citep[cf. Fig. 12 from][]{Beth2025} where H$_2$O number density is much lower, hence water-group ions are less likely to be produced, G07 was the only flyby going from the sub-Jovian to the anti-Jovian flank (i.e. away from Jupiter), and therefore in the direction/hemisphere where ionospheric ions are picked up, along the convective electric field $\vec{E}_\text{conv.}=-\vec{\varv}_\text{jov}\times \vec{B}$ that points towards $-y$ ($\vec{B}$ mainly points towards $-z$ and the Jovian plasma velocity $\vec{\varv}_\text{jov}$ points towards $+x$ such that $\vec{E}_\text{conv.}$ points towards $-y$). Therefore, the presence of H$_3$O$^+$ is consistent with the fact that H$_2$O$^+$ (from which H$_3$O$^+$ originates here) is produced on the dayside \citep[Galileo crossed the terminator plane before the inbound MP crossing, cf. Fig. 12 from][]{Beth2025}, then transported upstream (i.e. in the $-x$ direction towards the trailing side of the magnetopause) inside Ganymede's magnetosphere, and finally deflected either along the MP preferentially along the anti-Jovian flank and along/within the Alfvén wings. That is shown in more detail in Section~\ref{3D}.

G29 is the least affected flyby by ion-neutral chemistry. As Ganymede was within Jupiter's shadow, H$_2$O is much lower and only produced through ice sputtering. With low H$_2$O number density, H$_2^+$ reacts either with H$_2$ to produce H$_3^+$, or with O$_2$ to be turned into O$_2^+$ or O$_2$H$^+$. On the one hand, the lack of H$_2$O minimises the loss of H$_2^+$/H$_3^+$ (and O$_2$H$^+$ though negligible). On the other hand, the lack of photoionisation due 
 to eclipse drastically impedes the production of H$_2^+$, the precursor for forming H$_3^+$ and O$_2$H$^+$ (see Section~\ref{section43}). H$_3^+$ and O$_2$H$^+$ are less likely to be produced but also less likely to be lost. However, in the case of G29, ionisation only relies on electron impact, which is poorly constrained.

For all flybys, the most affected ion species by ion-neutral chemistry is O$^+$. For example, for G01, the simulated ionospheric O$^+$ number density reaches $\sim~10$\,cm$^{-3}$ at closest approach in the collisionless case \citep{Beth2025} whereas it only reaches $\sim~0.5-0.7$\,cm$^{-3}$ once ion-neutral collisions are considered. O$^+$ reacts with all the different neutral species present in Ganymede's exosphere, while it is mainly produced through ionisation of H$_2$O and O$_2$. This deficiency found for O$^+$ has strong implications on the interpretation of previous flybys regarding ion composition (see Section \ref{section42}).

\subsection{3D spatial distribution}\label{3D}
\begin{figure*}
        \centering
        \fcolorbox{Hp}{white}{\parbox{.467\linewidth}{\stackinset{l}{0pt}{b}{87pt}{\colorbox{Hp}{\contour{black}\bf\color{white}H$^+$+H$_\text{jov}^+$}}{\includegraphics[width=\linewidth,clip,trim=0.8cm 8cm 2cm 2.8cm]{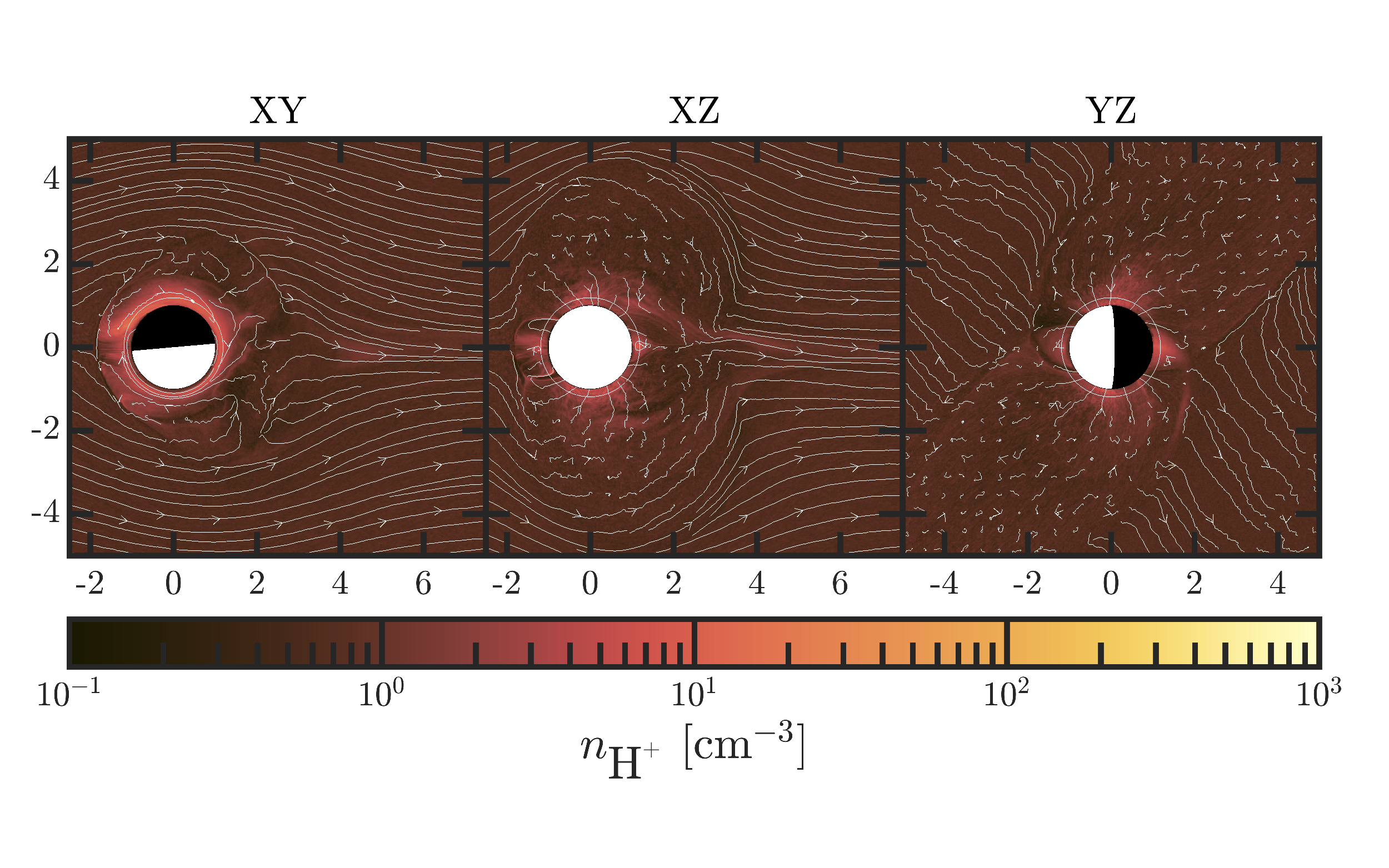}}}}\!
        \fcolorbox{Op}{white}{\parbox{.467\linewidth}{\stackinset{l}{0pt}{b}{87pt}{\colorbox{Op}{\contour{black}\bf\color{white}O$^+$+O$_\text{jov}^+$}}{\includegraphics[width=\linewidth,clip,trim=0.8cm 8cm 2cm 2.8cm]{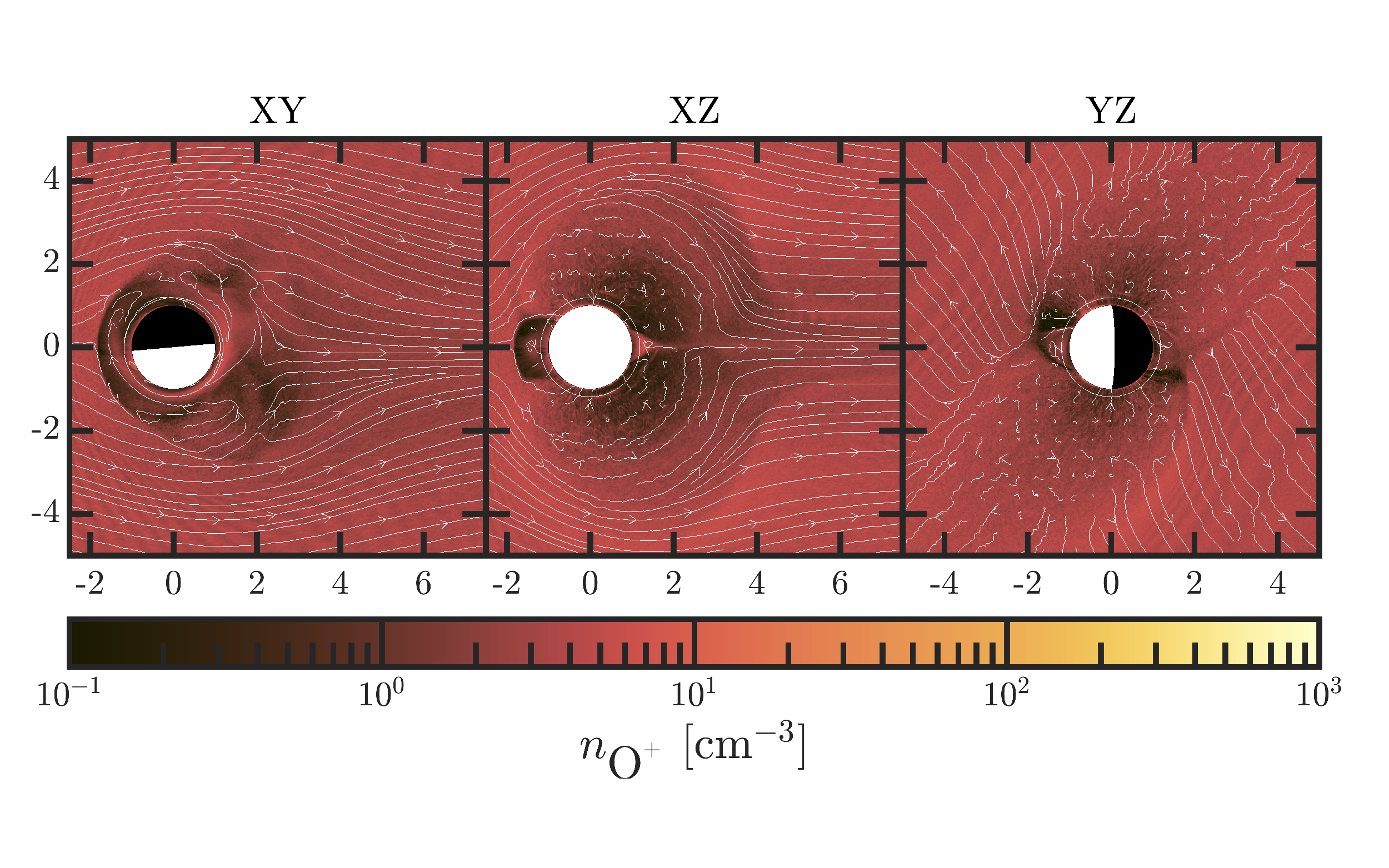}}}}\\
        \fcolorbox{H2p}{white}{\parbox{.467\linewidth}{\stackinset{l}{0pt}{b}{87pt}{\colorbox{H2p}{\contour{black}\bf\color{white}H$_2^+$}}{\includegraphics[width=\linewidth,clip,trim=0.8cm 8cm 2cm 2.8cm]{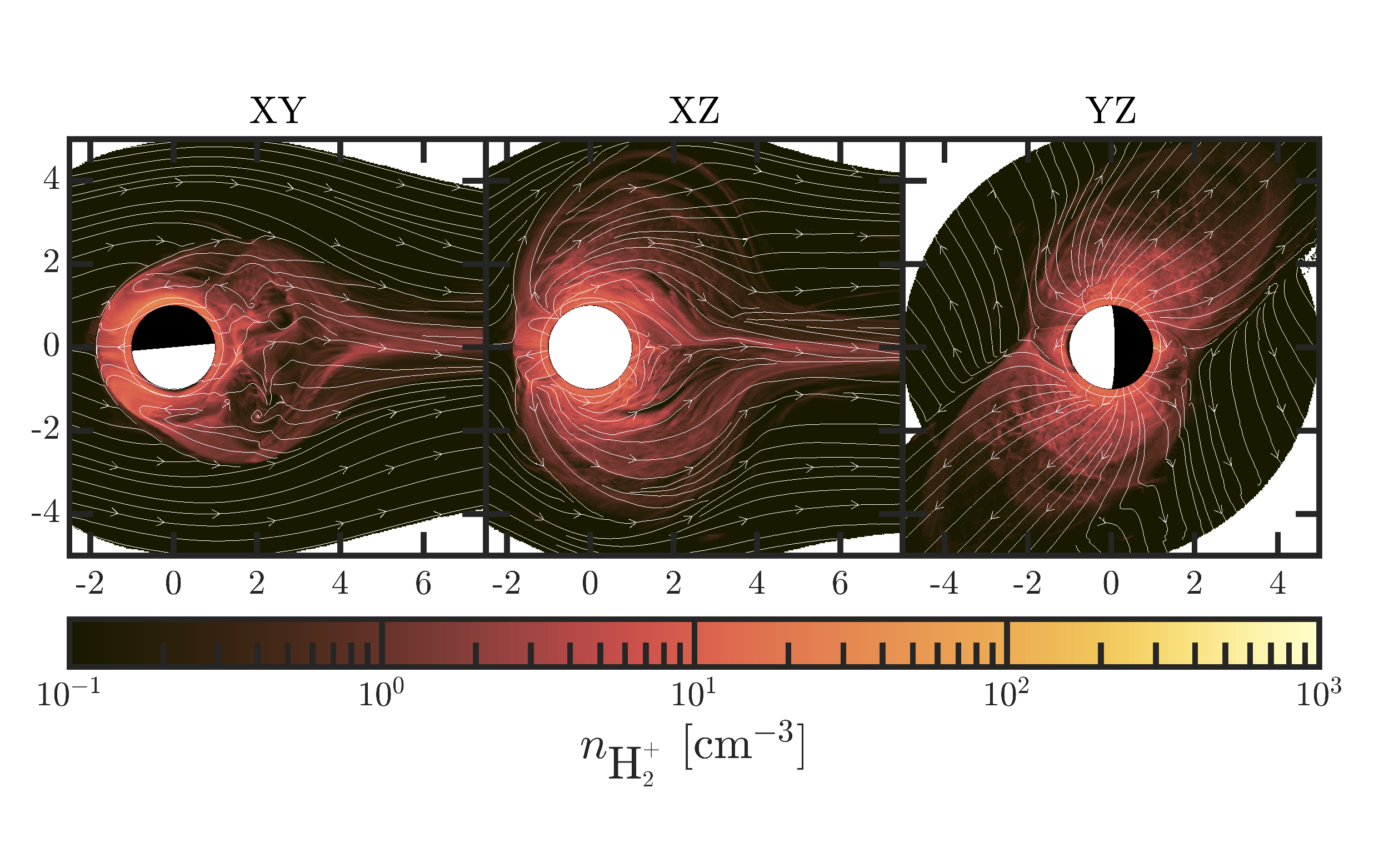}}}}\!
        \fcolorbox{HOp}{white}{\parbox{.467\linewidth}{\stackinset{l}{0pt}{b}{87pt}{\colorbox{HOp}{\contour{black}\bf\color{white}HO$\vphantom{O_2^+}^+$}}{\includegraphics[width=\linewidth,clip,trim=0.8cm 8cm 2cm 2.8cm]{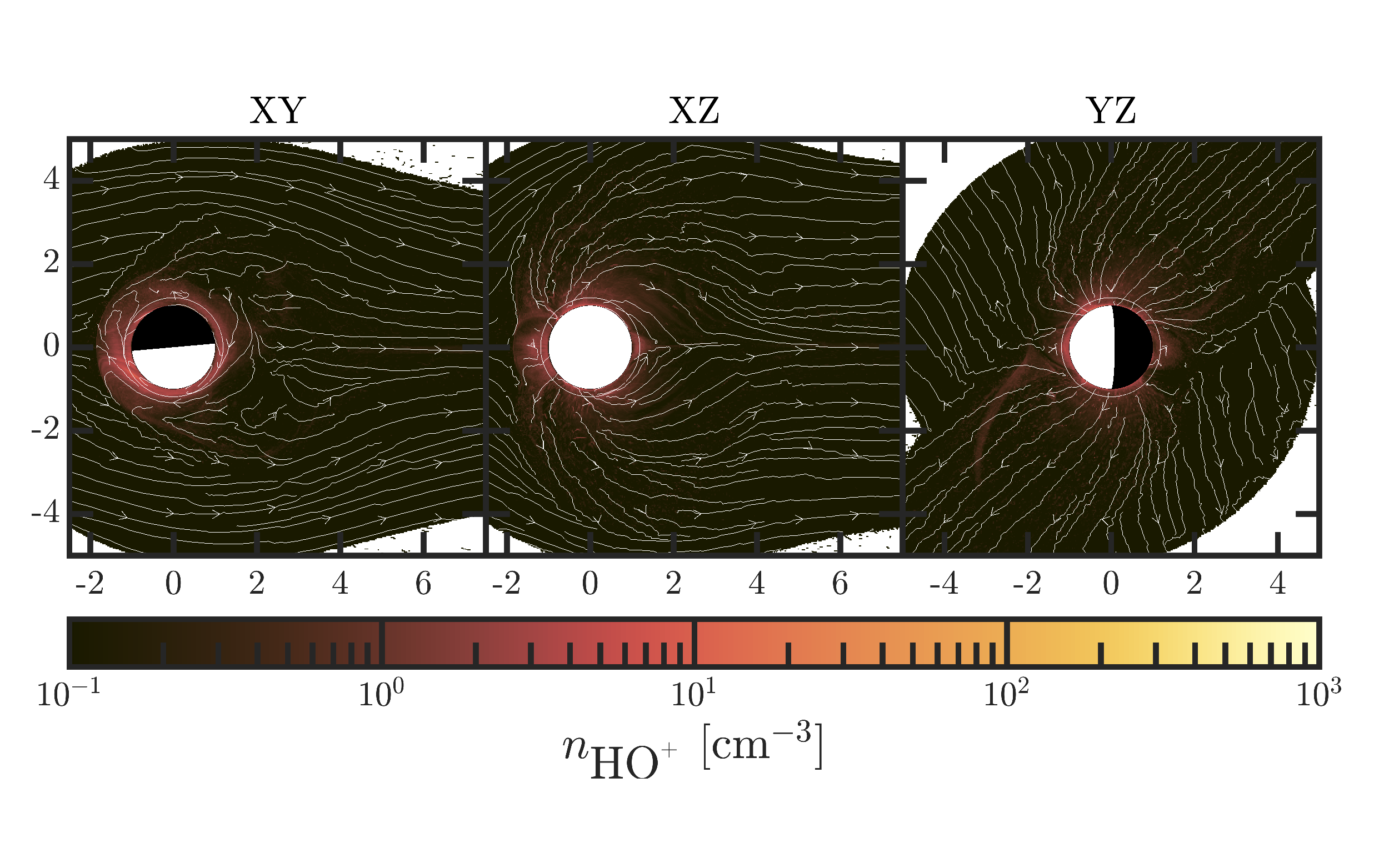}}}}\\
        \fcolorbox{H3p}{white}{\parbox{.467\linewidth}{\stackinset{l}{0pt}{b}{87pt}{\colorbox{H3p}{\contour{black}\bf\color{white}H$_3^+$}}{\includegraphics[width=\linewidth,clip,trim=0.8cm 8cm 2cm 2.8cm]{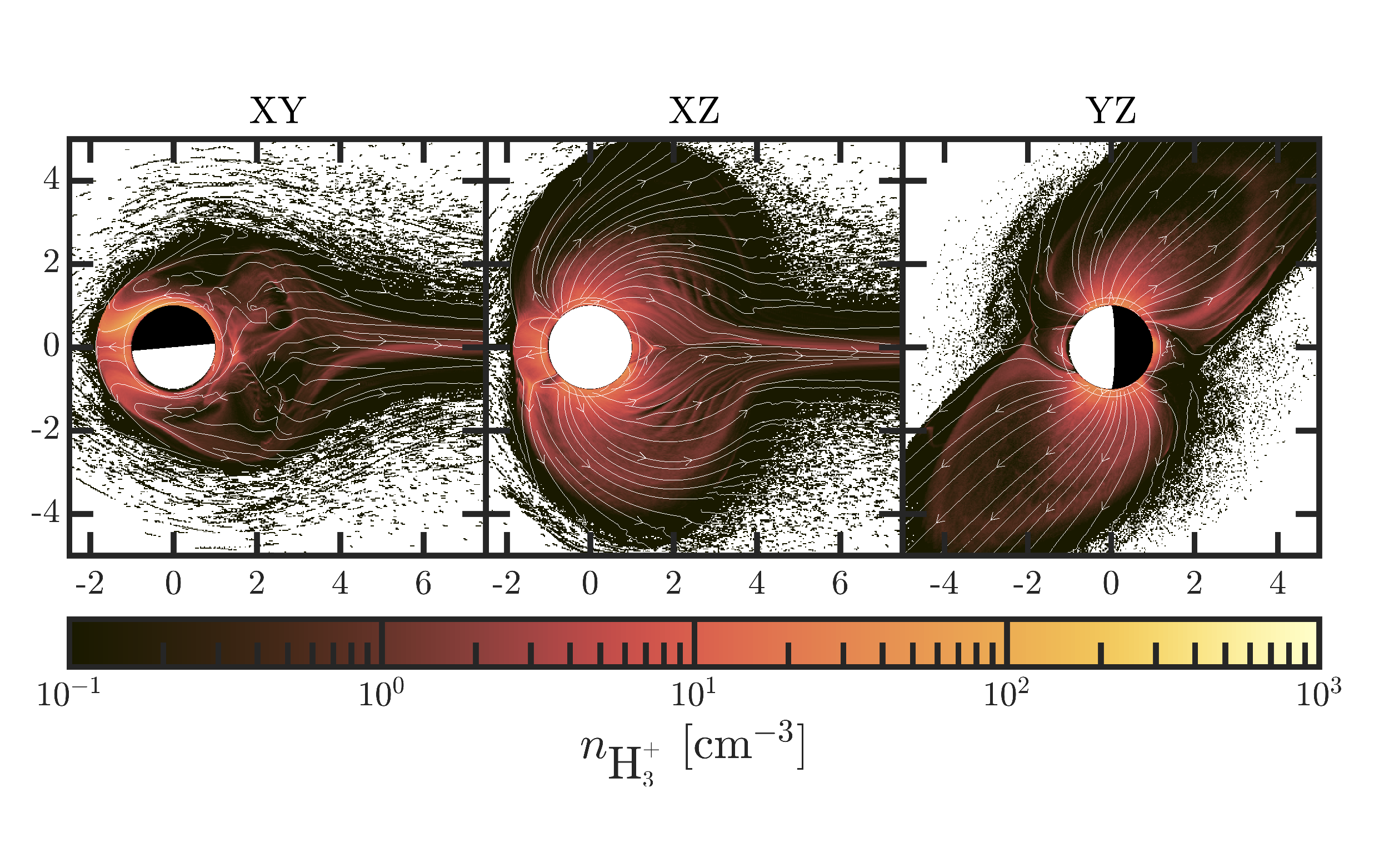}}}}\!
        \fcolorbox{H2Op}{white}{\parbox{.467\linewidth}{\stackinset{l}{0pt}{b}{87pt}{\colorbox{H2Op}{\contour{black}\bf\color{white}H$_2$O$\vphantom{O_2^+}^+$}}{\includegraphics[width=\linewidth,clip,trim=0.8cm 8cm 2cm 2.8cm]{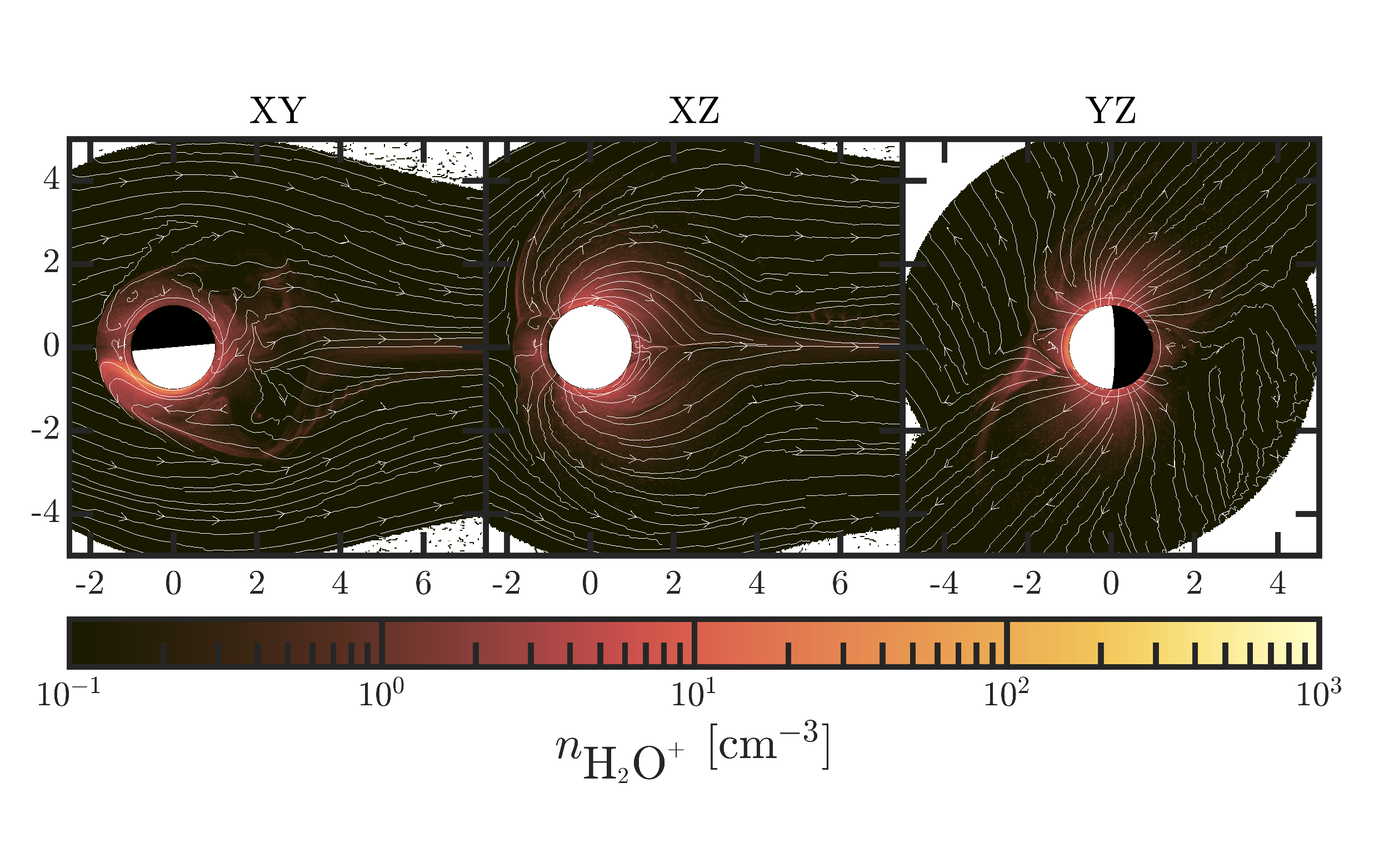}}}}\\
        \fcolorbox{O2p}{white}{\parbox{.467\linewidth}{\stackinset{l}{0pt}{b}{87pt}{\colorbox{O2p}{\contour{black}\bf\color{white}O$_2^+$}}{\includegraphics[width=\linewidth,clip,trim=0.8cm 8cm 2cm 2.8cm]{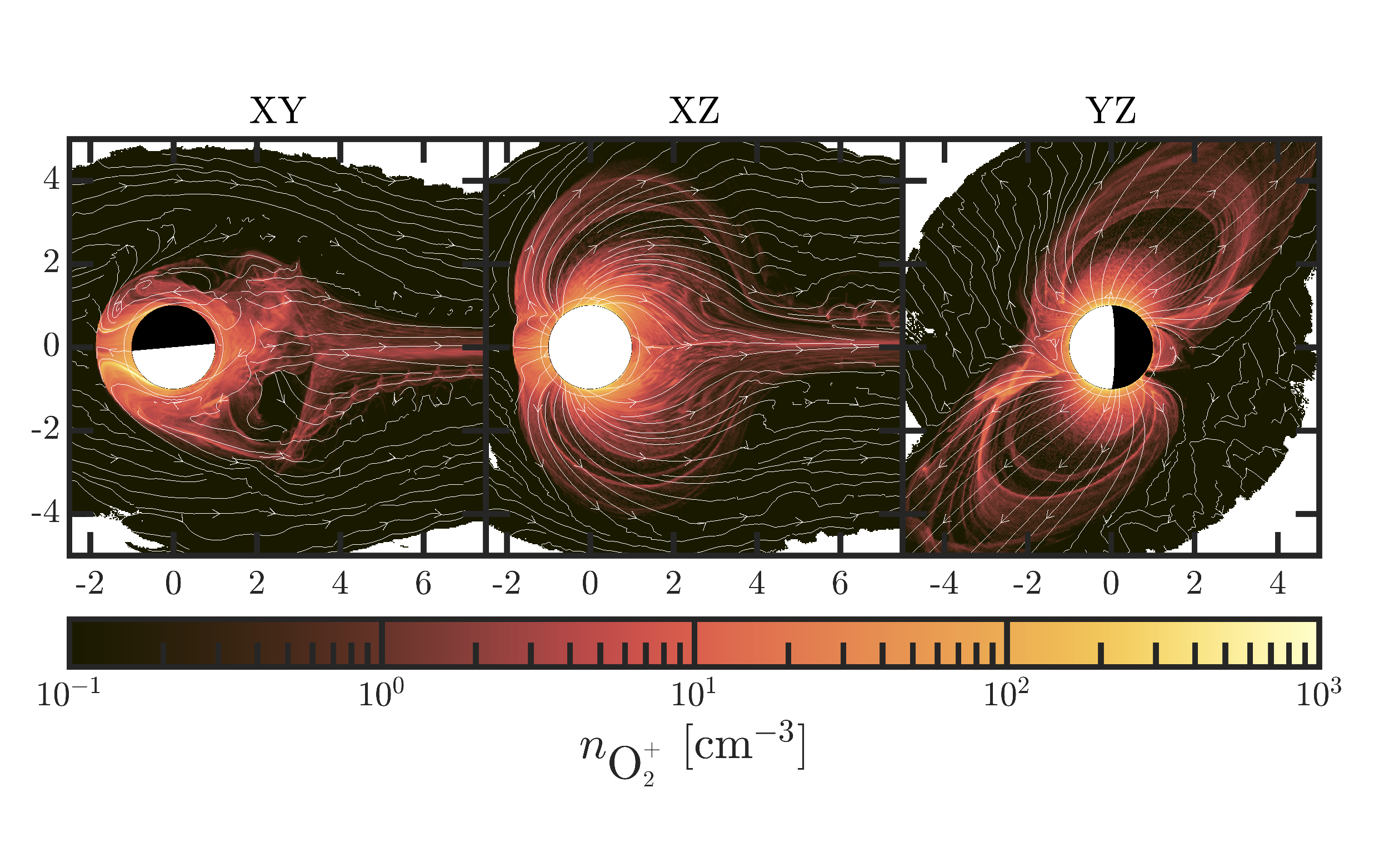}}}}\!
        \fcolorbox{H3Op}{white}{\parbox{.467\linewidth}{\stackinset{l}{0pt}{b}{87pt}{\colorbox{H3Op}{\contour{black}\bf\color{white}H$_3$O$\vphantom{_3^+}^+$}}{\includegraphics[width=\linewidth,clip,trim=0.8cm 8cm 2cm 2.8cm]{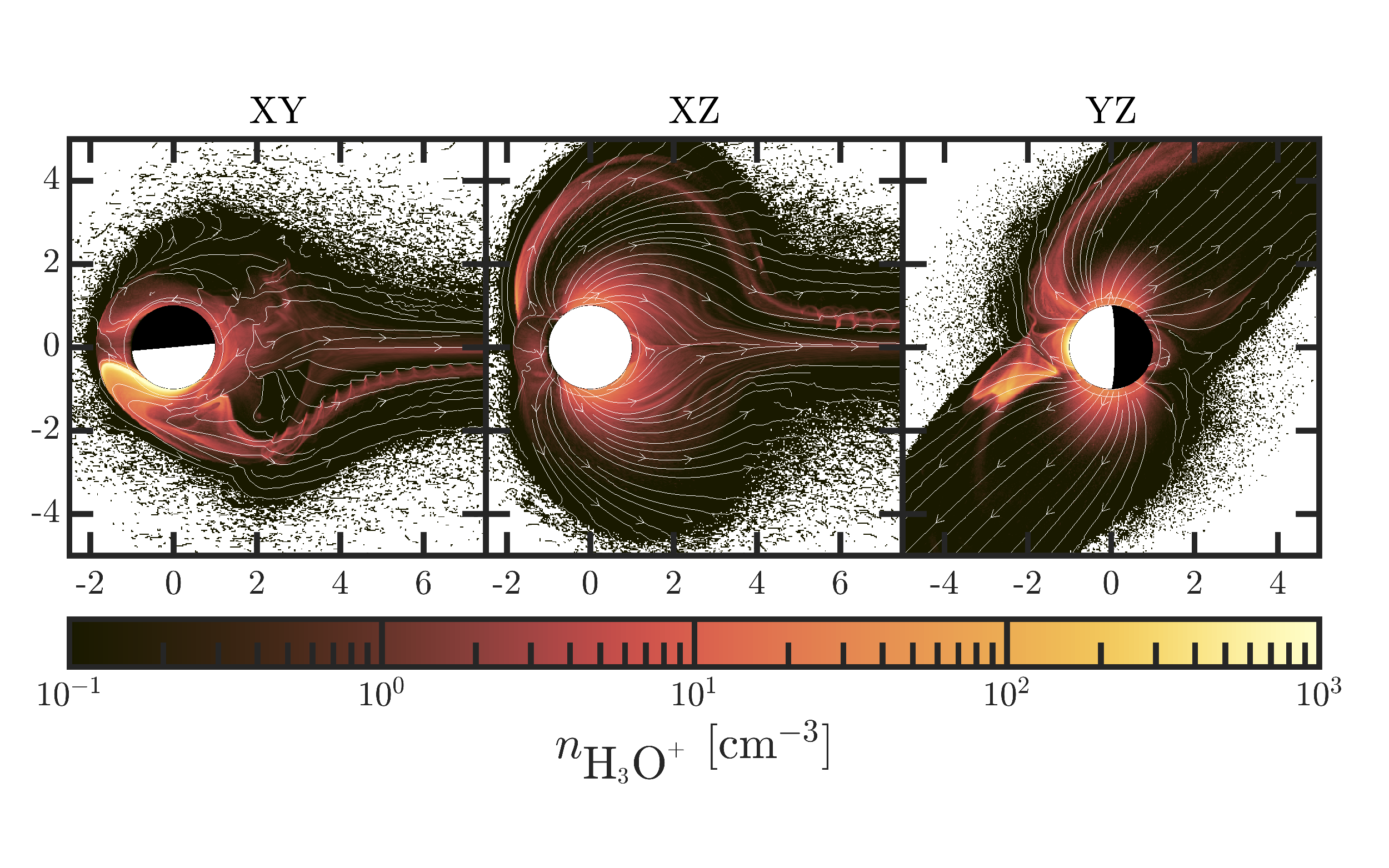}}}}\\
        \fcolorbox{O2Hp}{white}{\parbox{.467\linewidth}{\stackinset{l}{0pt}{b}{87pt}{\colorbox{O2Hp}{\contour{black}\bf\color{white}O$_2$H$\vphantom{_2^+}^+$}}{\includegraphics[width=\linewidth,clip,trim=0.8cm 8cm 2cm 2.8cm]{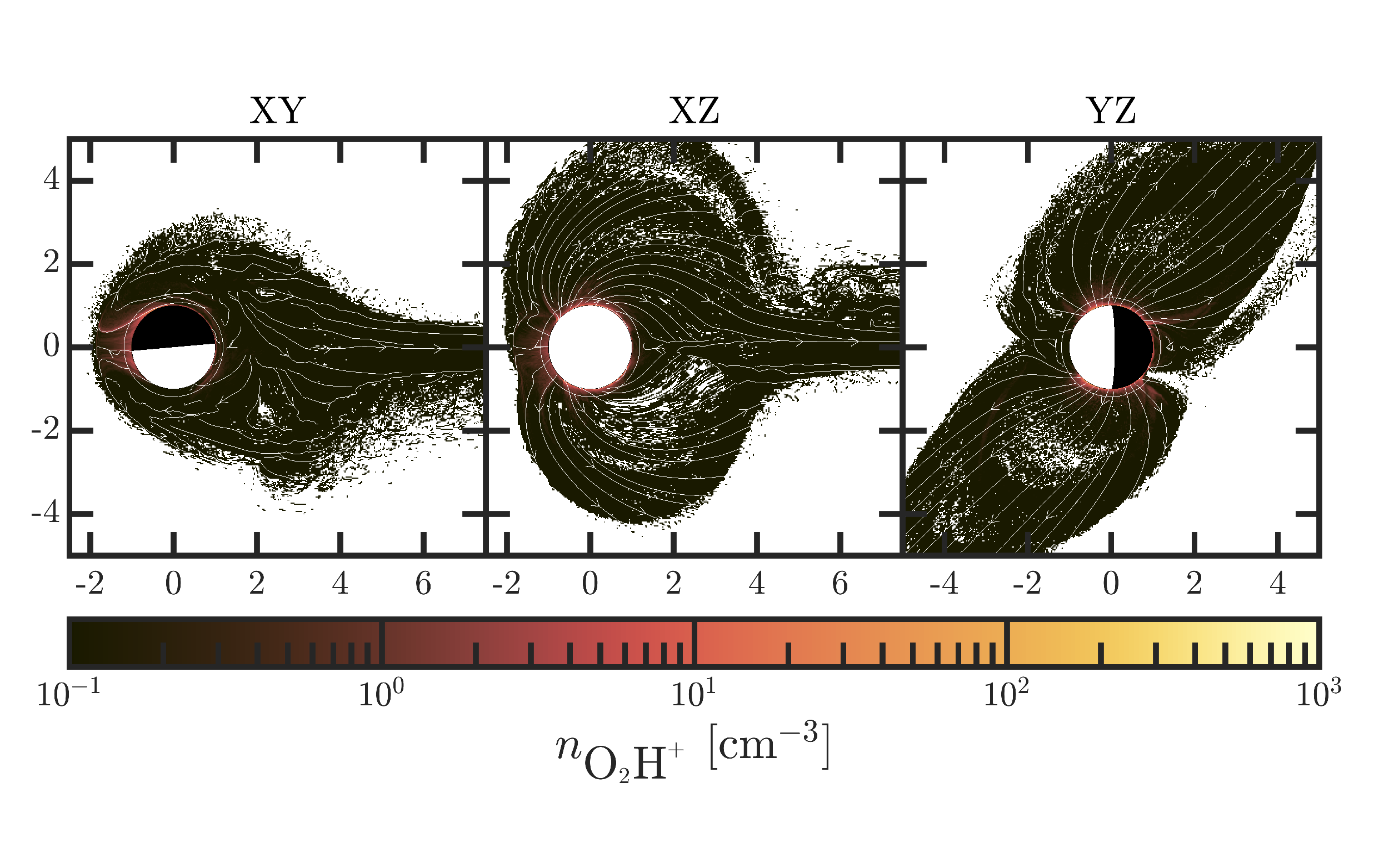}}}}\!
        \fcolorbox{black}{white}{\parbox{.467\linewidth}{\stackinset{l}{0pt}{b}{87pt}{\colorbox{black}{\contour{black}\bf\color{white}$\vphantom{O_2^+}$\bf SUM}}{\includegraphics[width=\linewidth,clip,trim=0.8cm 8cm 2cm 2.8cm]{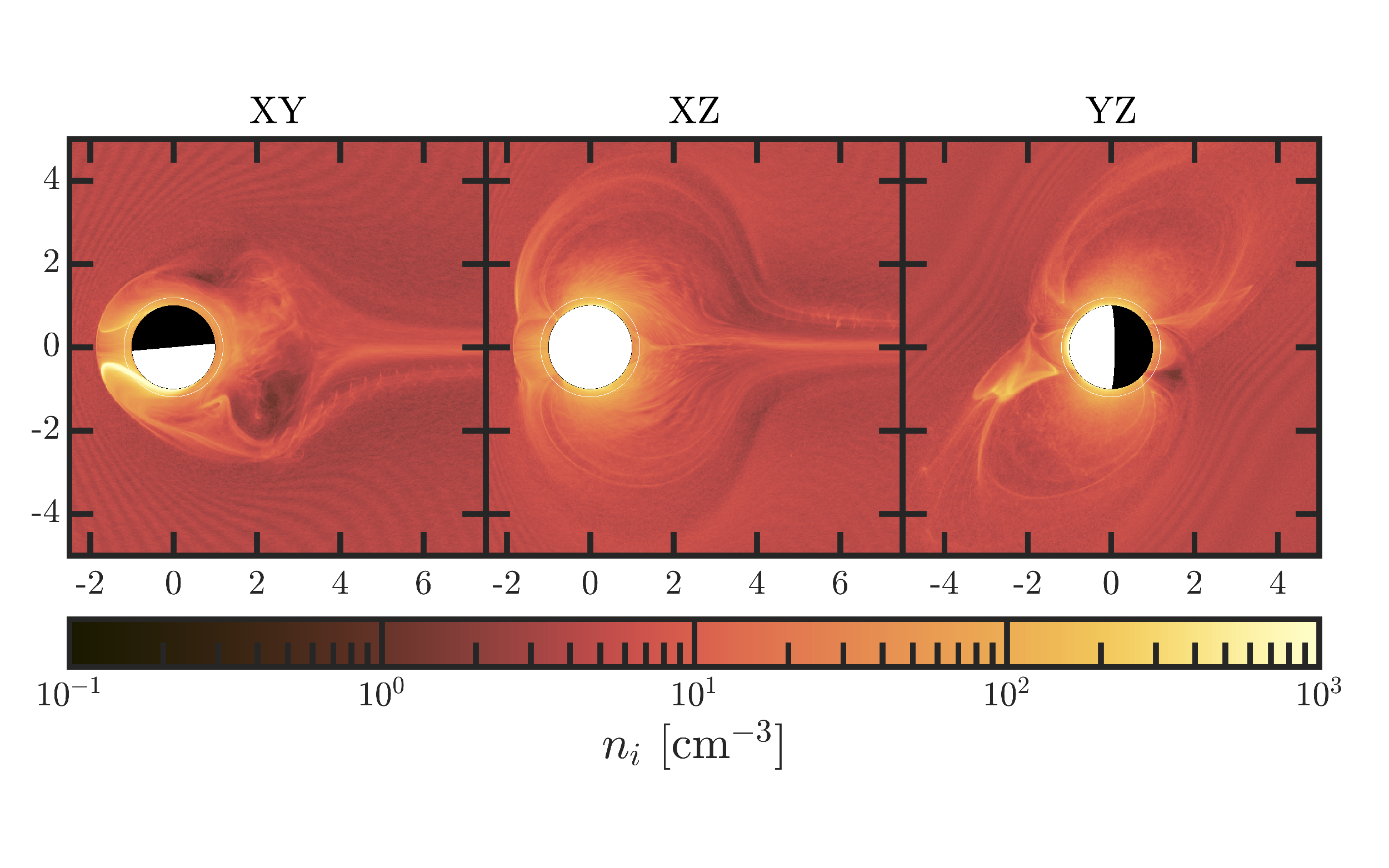}}}}\\
        \includegraphics[width=\linewidth,clip,trim=1.2cm 3.0cm 1.2cm 18.5cm]{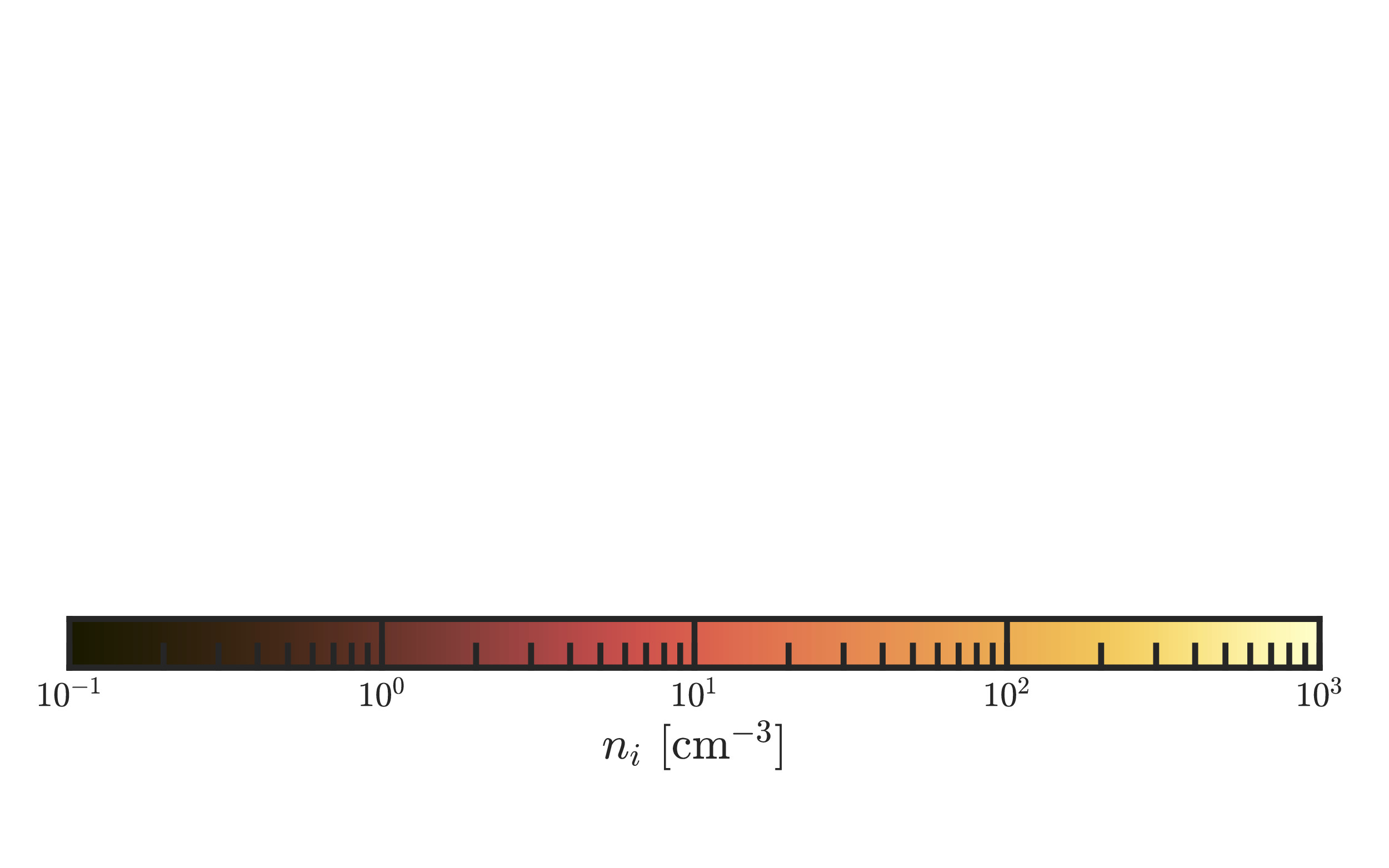}\\
        \caption{Cuts of ion number densities in the different planes around Ganymede for G01. The cuts are made in the XY, YZ, and YZ planes in the GPhiO coordinate system. The axes' unit is in Ganymede's radius. The three upper left panels correspond to light ions: H$^+$ (both jovian and ionospheric), H$_2^+$, and H$_3^+$. The four upper right panels correspond to intermediate-mass (or water-group) ions: O$^+$ (both jovian and ionospheric), HO$^+$, H$_2$O$^+$, and H$_3$O$^+$. The two bottom left panels correspond to heavy ions: O$_2^+$ and O$_2$H$^+$, H$_2^+$. The bottom right panel corresponds to the total ion number density. Ion species are indicated in the top left corner of each frame. The central disk corresponds to Ganymede. The white part of the disk is the dayside. Any white area outside Ganymede corresponds to regions that macroparticles have not reached or because their weight was 0 (they are launched from a cell where the neutral number density is 0). The thin white circle around Ganymede is located at 500\,km which corresponds to one of the two radii of JUICE circular orbits. We plotted the streamlines for each ion species, with arrows indicating the flow direction.} \label{Fig2}
    \end{figure*}
    \begin{figure*}
        \centering
        \fcolorbox{Hp}{white}{\parbox{.467\linewidth}{\stackinset{l}{0pt}{b}{87pt}{\colorbox{Hp}{\contour{black}\bf\color{white}H$^+$+H$_\text{jov}^+$}}{\includegraphics[width=\linewidth,clip,trim=0.8cm 8cm 2cm 2.8cm]{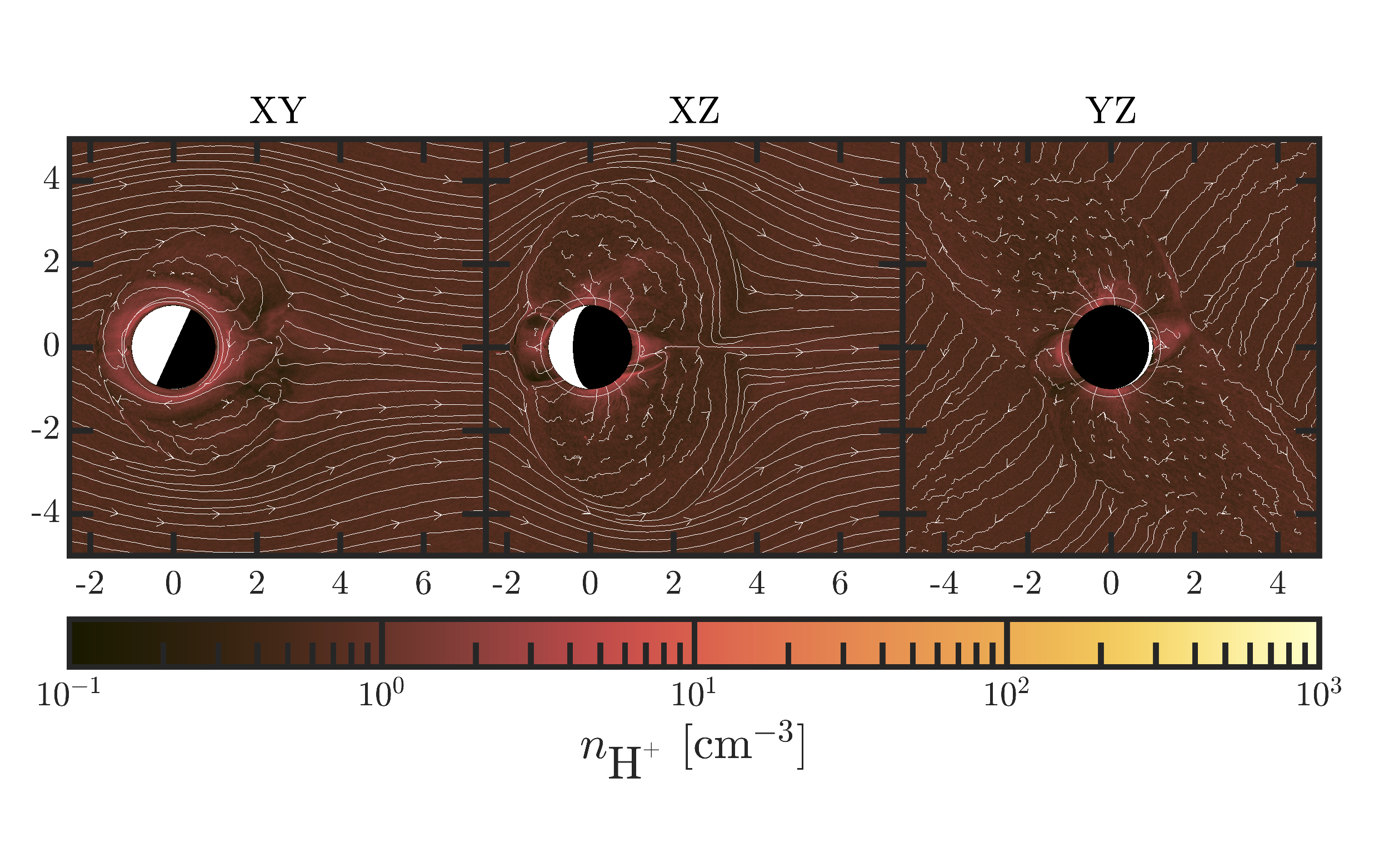}}}}\!
        \fcolorbox{Op}{white}{\parbox{.467\linewidth}{\stackinset{l}{0pt}{b}{87pt}{\colorbox{Op}{\contour{black}\bf\color{white}O$^+$+O$_\text{jov}^+$}}{\includegraphics[width=\linewidth,clip,trim=0.8cm 8cm 2cm 2.8cm]{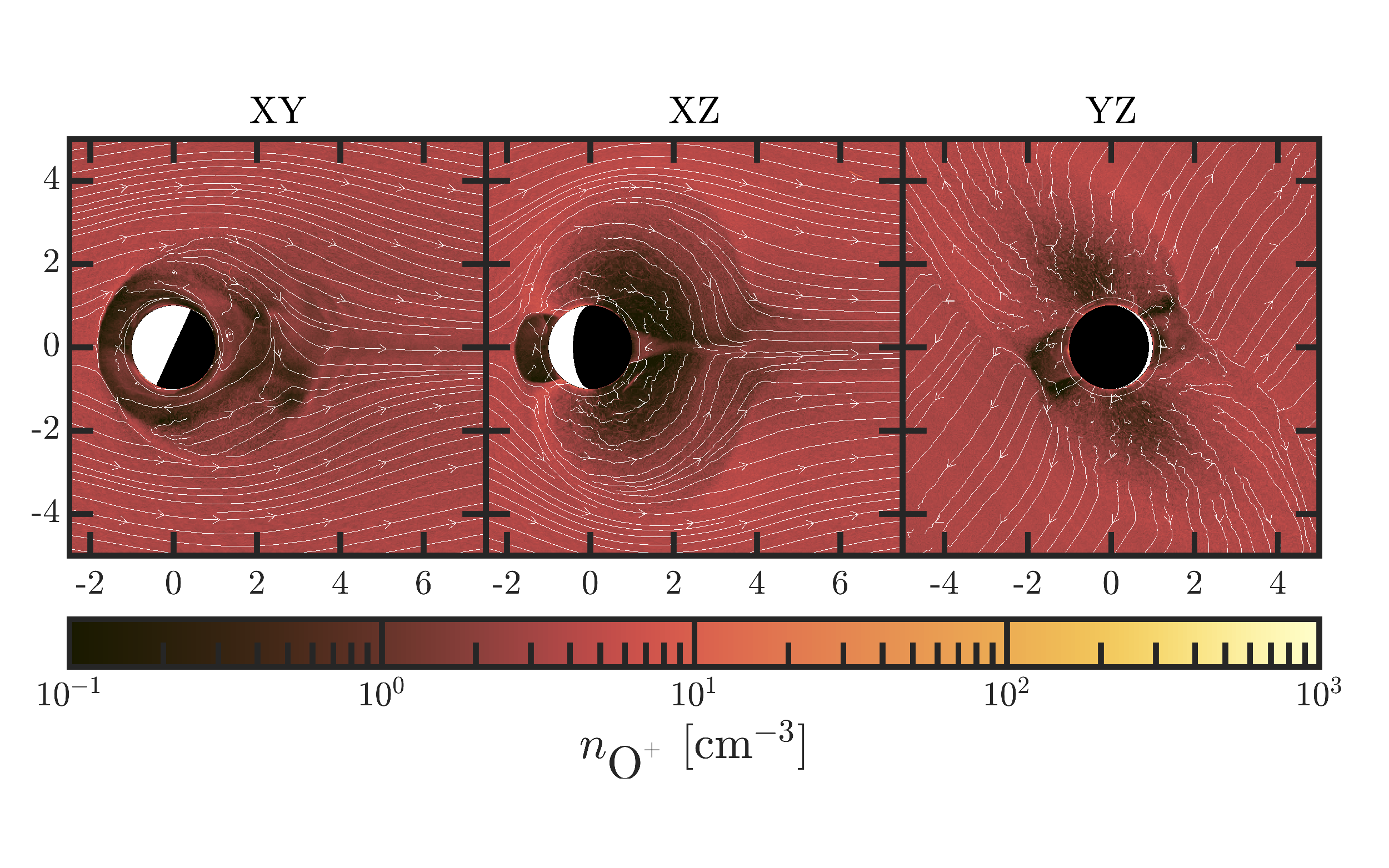}}}}\\
        \fcolorbox{H2p}{white}{\parbox{.467\linewidth}{\stackinset{l}{0pt}{b}{87pt}{\colorbox{H2p}{\contour{black}\bf\color{white}H$_2^+$}}{\includegraphics[width=\linewidth,clip,trim=0.8cm 8cm 2cm 2.8cm]{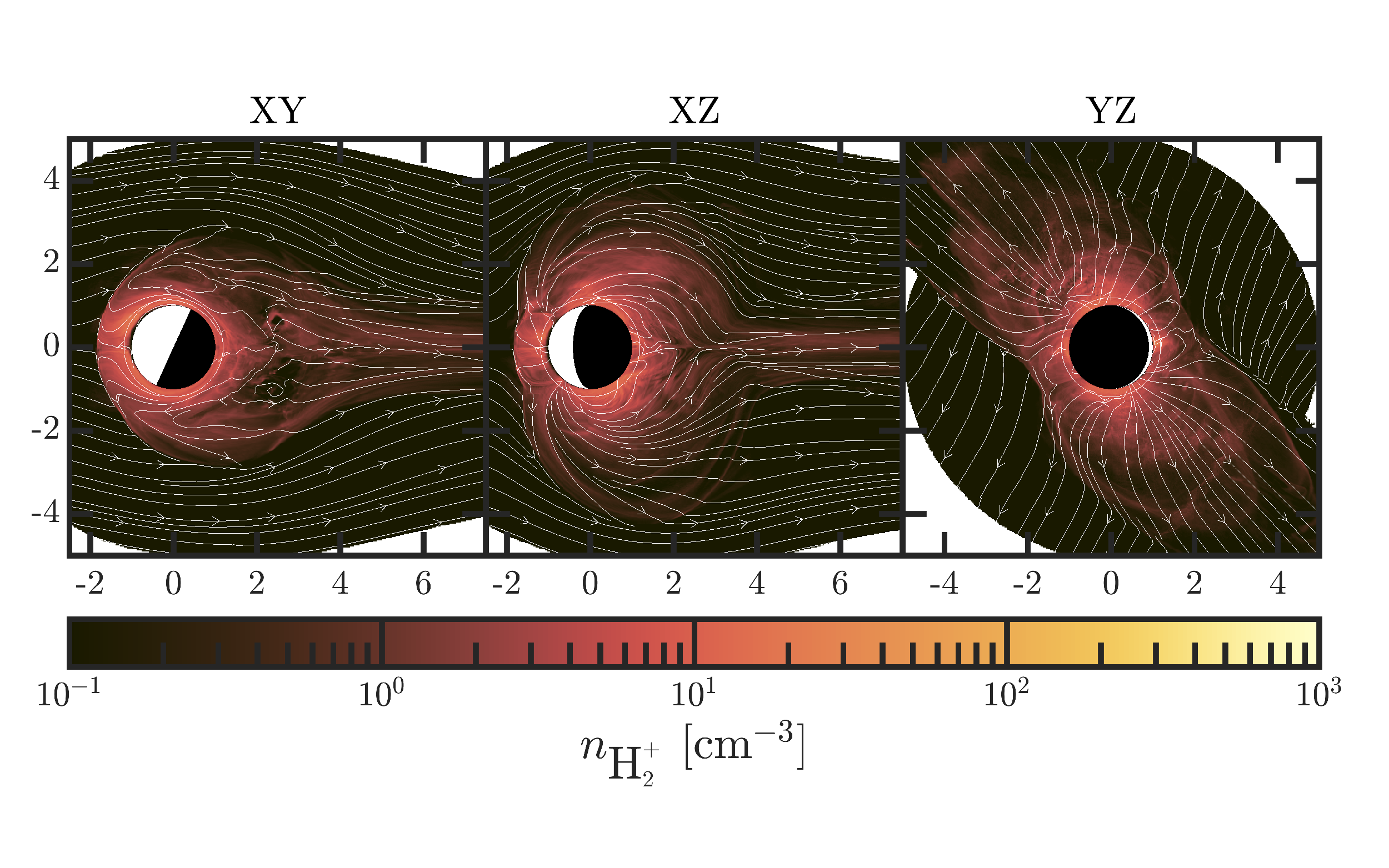}}}}\!
        \fcolorbox{HOp}{white}{\parbox{.467\linewidth}{\stackinset{l}{0pt}{b}{87pt}{\colorbox{HOp}{\contour{black}\bf\color{white}HO$\vphantom{O_2^+}^+$}}{\includegraphics[width=\linewidth,clip,trim=0.8cm 8cm 2cm 2.8cm]{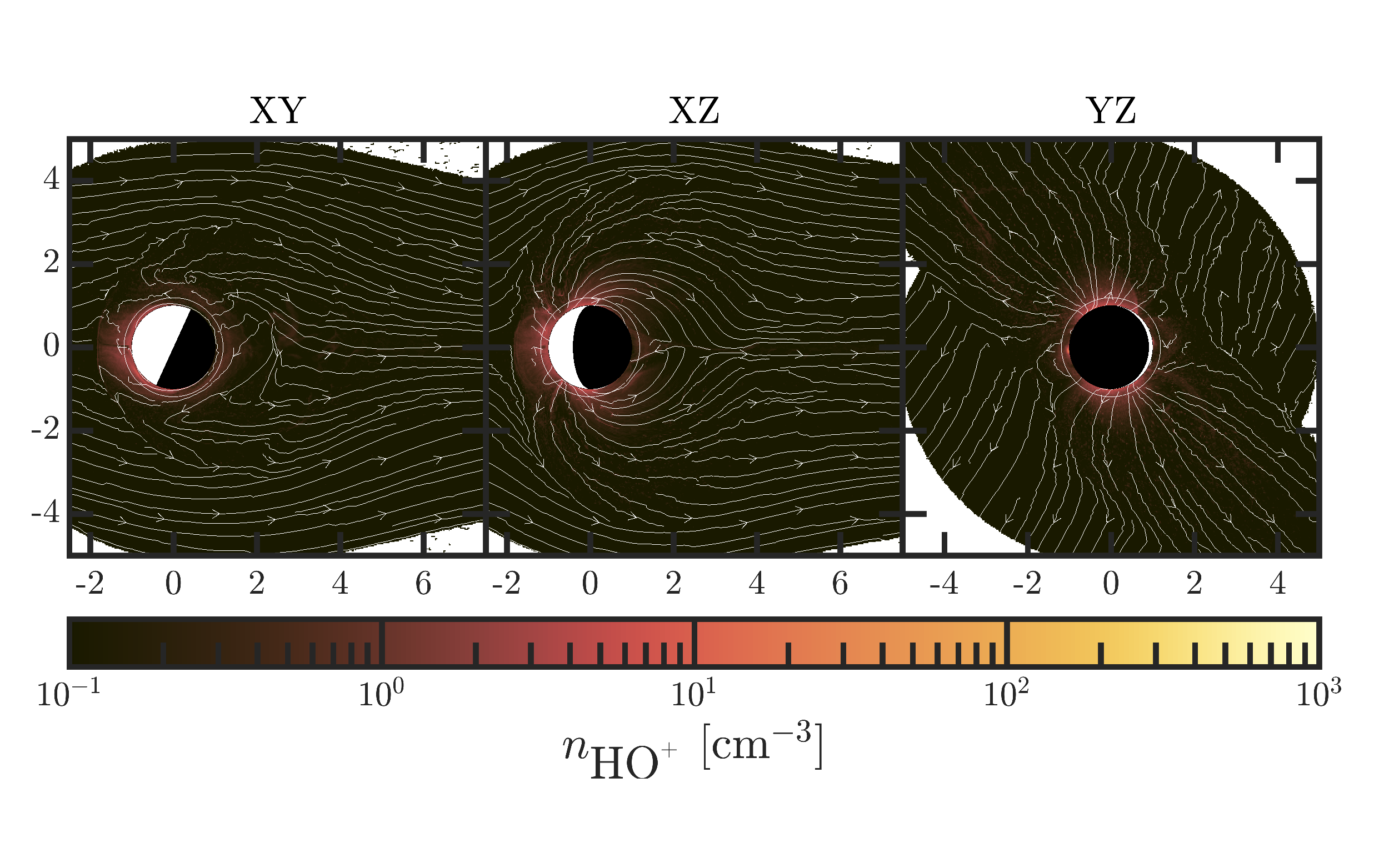}}}}\\
        \fcolorbox{H3p}{white}{\parbox{.467\linewidth}{\stackinset{l}{0pt}{b}{87pt}{\colorbox{H3p}{\contour{black}\bf\color{white}H$_3^+$}}{\includegraphics[width=\linewidth,clip,trim=0.8cm 8cm 2cm 2.8cm]{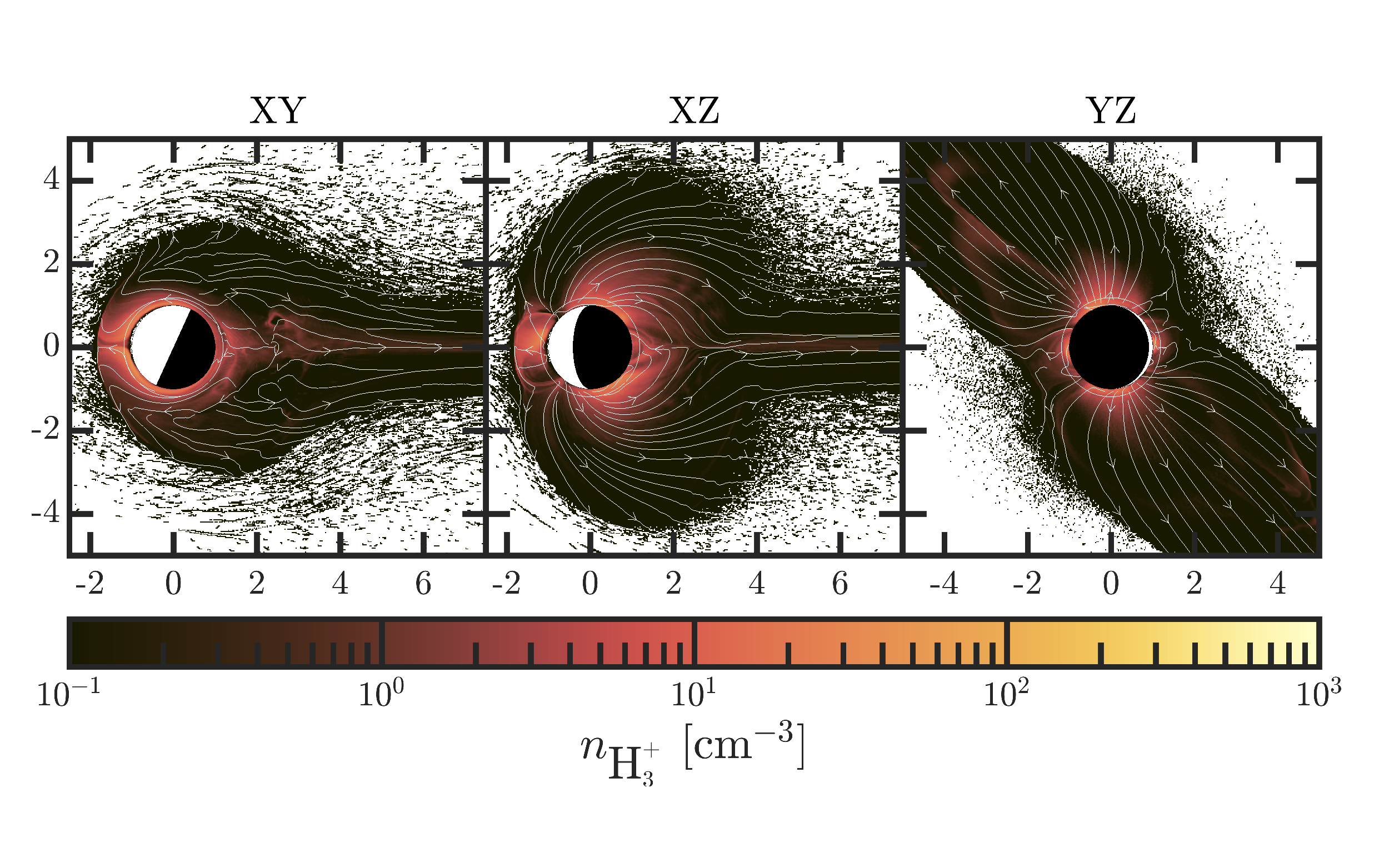}}}}\!
        \fcolorbox{H2Op}{white}{\parbox{.467\linewidth}{\stackinset{l}{0pt}{b}{87pt}{\colorbox{H2Op}{\contour{black}\bf\color{white}H$_2$O$\vphantom{O_2^+}^+$}}{\includegraphics[width=\linewidth,clip,trim=0.8cm 8cm 2cm 2.8cm]{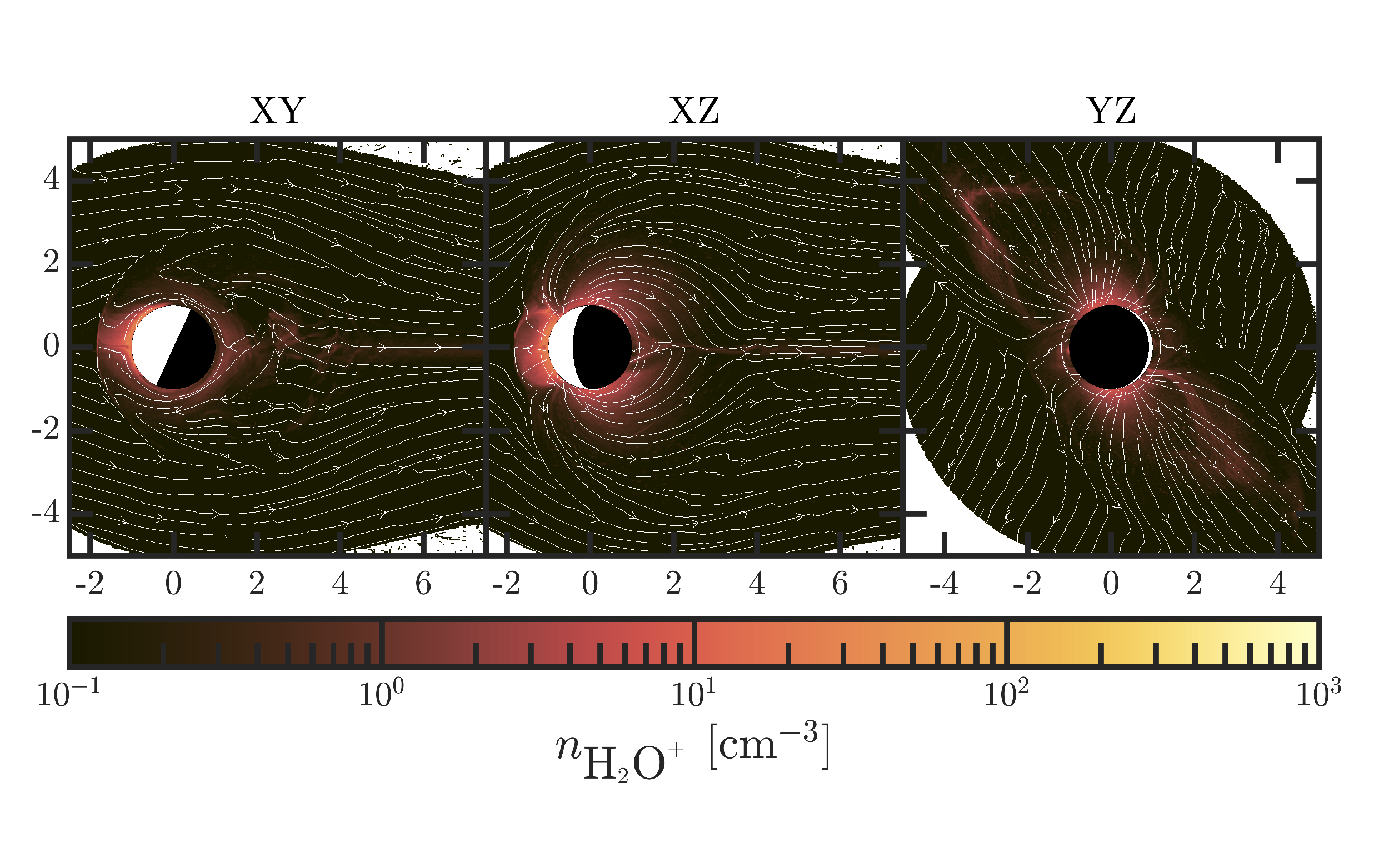}}}}\\
        \fcolorbox{O2p}{white}{\parbox{.467\linewidth}{\stackinset{l}{0pt}{b}{87pt}{\colorbox{O2p}{\contour{black}\bf\color{white}O$_2^+$}}{\includegraphics[width=\linewidth,clip,trim=0.8cm 8cm 2cm 2.8cm]{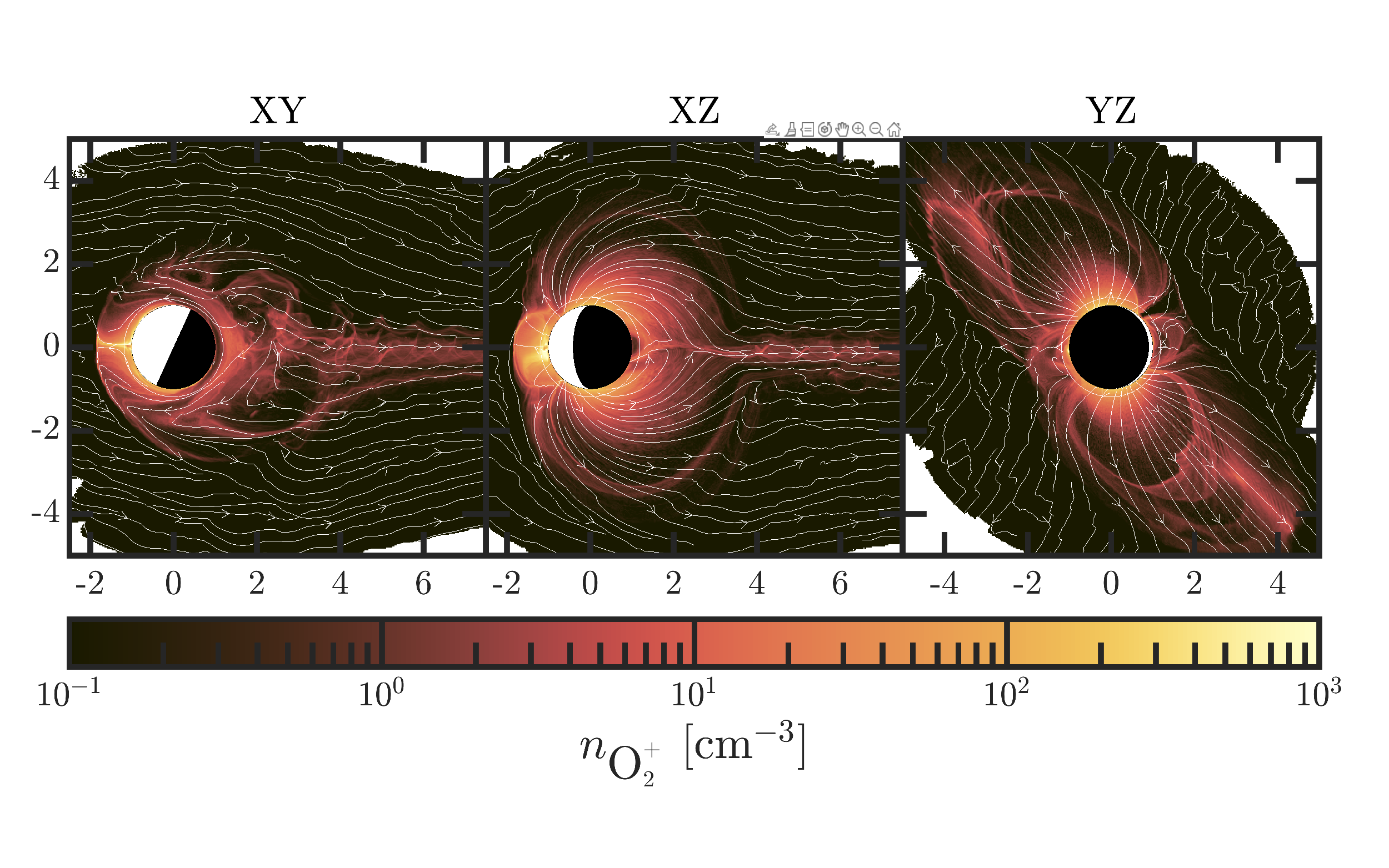}}}}\!
        \fcolorbox{H3Op}{white}{\parbox{.467\linewidth}{\stackinset{l}{0pt}{b}{87pt}{\colorbox{H3Op}{\contour{black}\bf\color{white}H$_3$O$\vphantom{_3^+}^+$}}{\includegraphics[width=\linewidth,clip,trim=0.8cm 8cm 2cm 2.8cm]{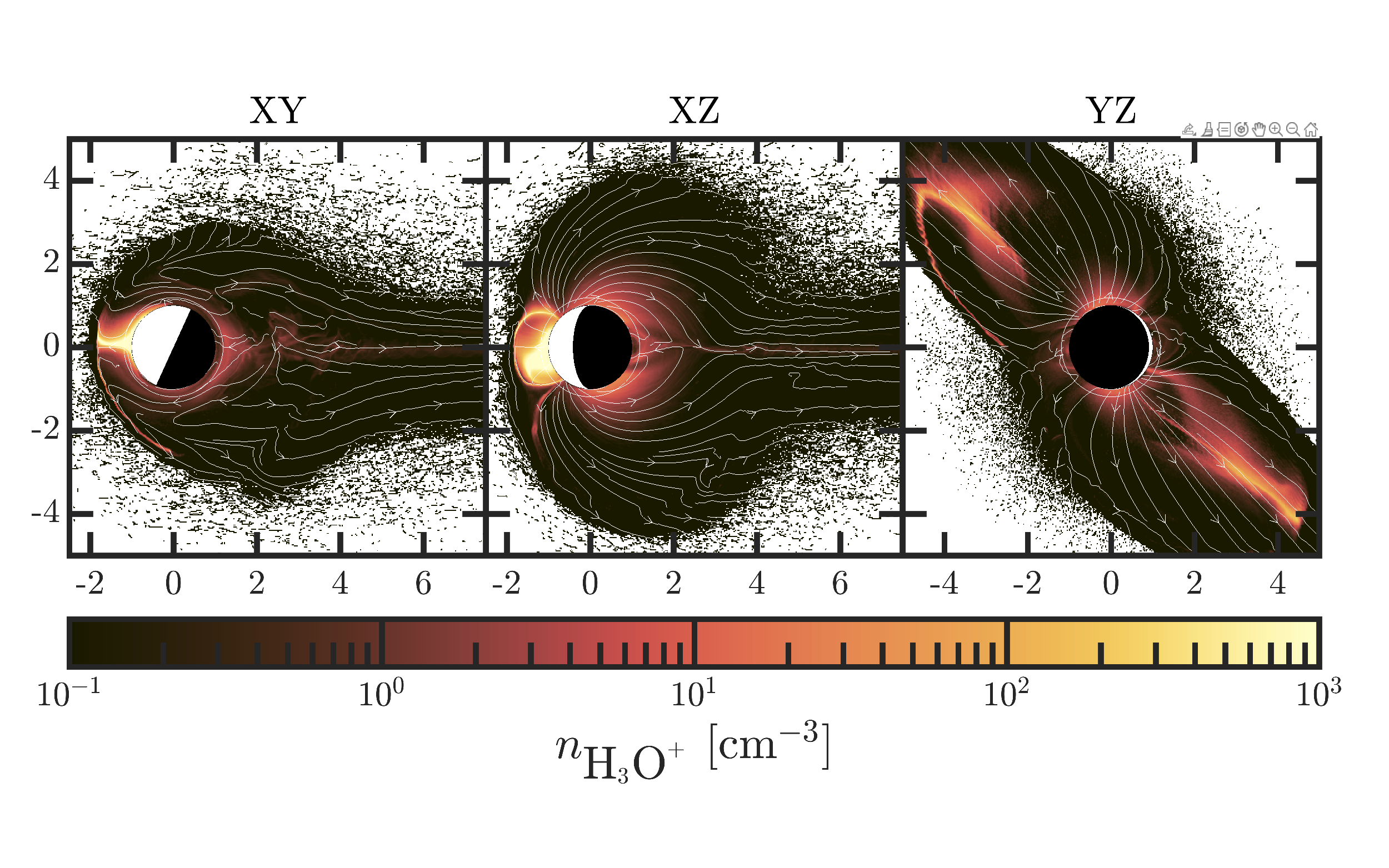}}}}\\
        \fcolorbox{O2Hp}{white}{\parbox{.467\linewidth}{\stackinset{l}{0pt}{b}{87pt}{\colorbox{O2Hp}{\contour{black}\bf\color{white}O$_2$H$\vphantom{_2^+}^+$}}{\includegraphics[width=\linewidth,clip,trim=0.8cm 8cm 2cm 2.8cm]{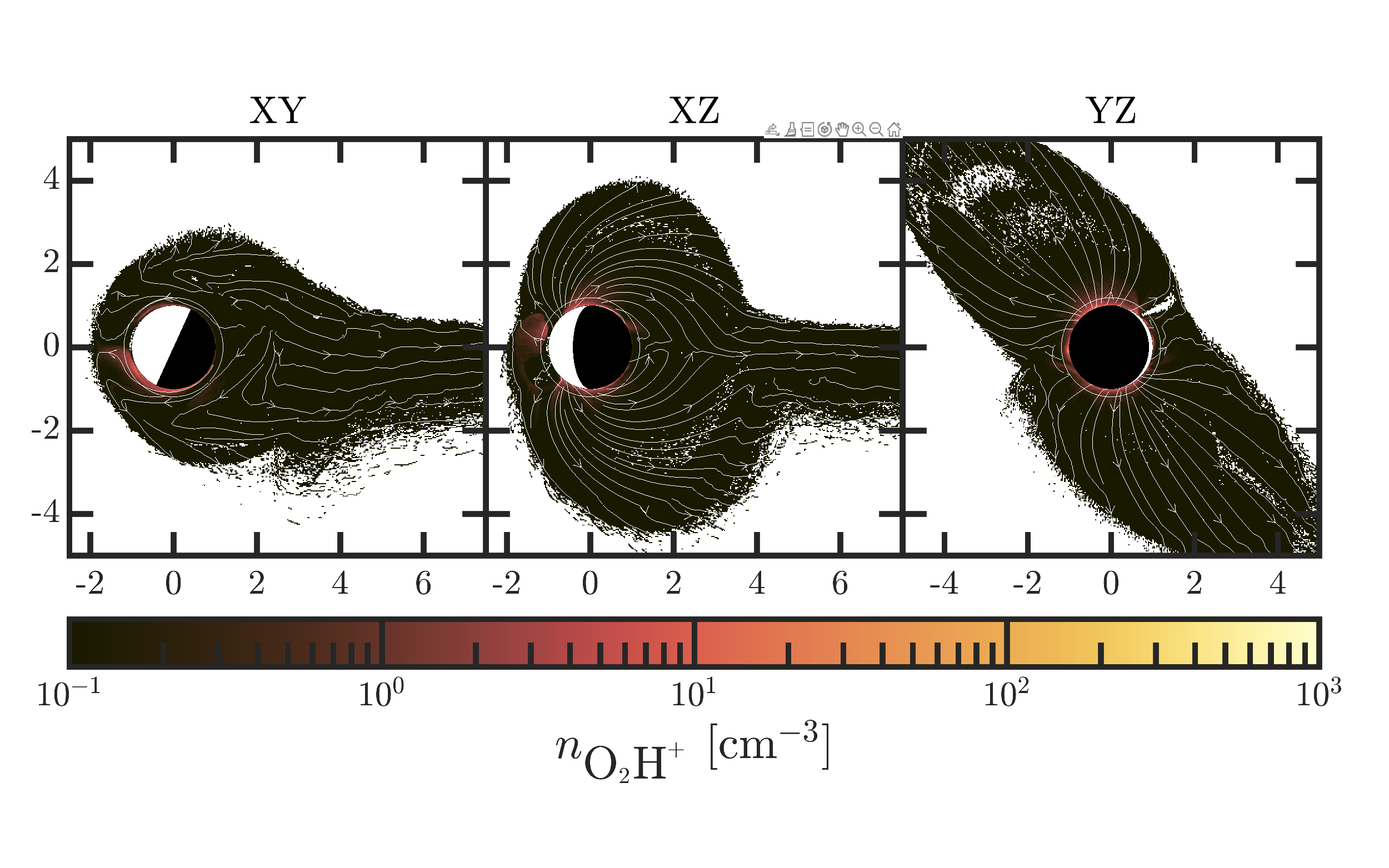}}}}\!
        \fcolorbox{black}{white}{\parbox{.467\linewidth}{\stackinset{l}{0pt}{b}{87pt}{\colorbox{black}{\contour{black}\bf\color{white}$\vphantom{O_2^+}$\bf SUM}}{\includegraphics[width=\linewidth,clip,trim=0.8cm 8cm 2cm 2.8cm]{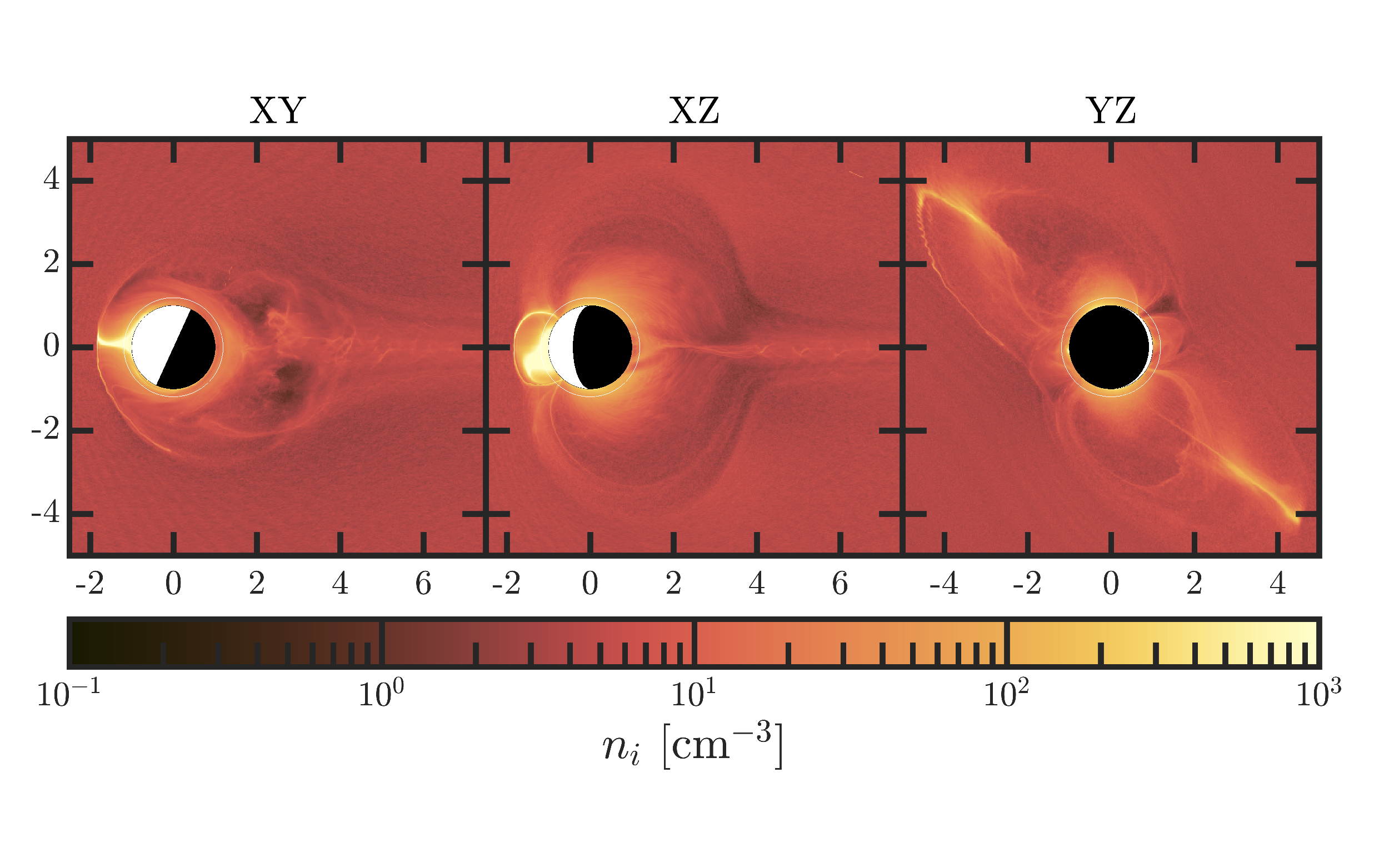}}}}\\
        \includegraphics[width=\linewidth,clip,trim=1.2cm 3.0cm 1.2cm 18.5cm]{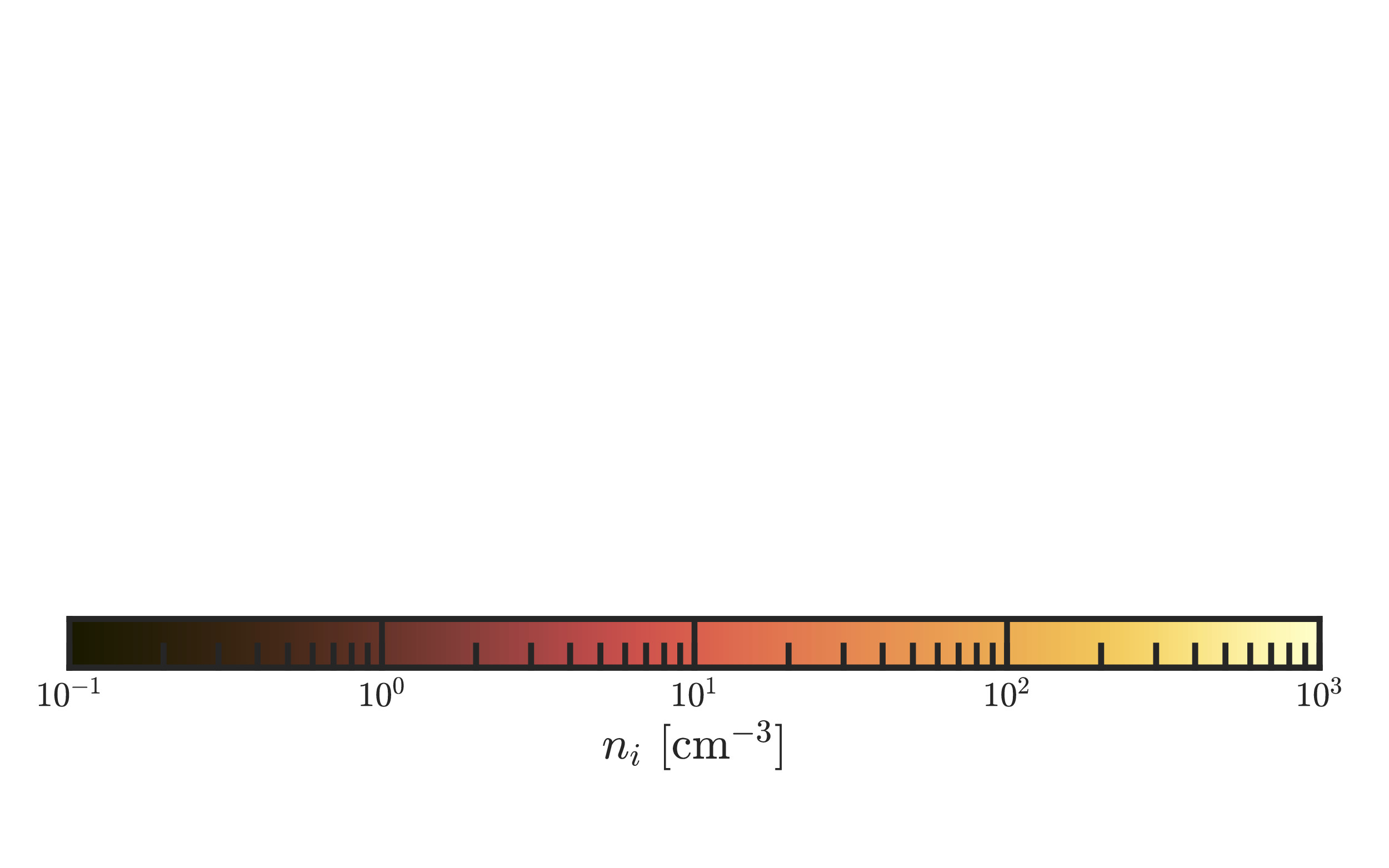}\\
        \caption{Similar to Fig.~\ref{Fig2} for G07}
        \label{Fig3}
    \end{figure*}
    \begin{figure*}
        \centering
        \fcolorbox{Hp}{white}{\parbox{.467\linewidth}{\stackinset{l}{0pt}{b}{87pt}{\colorbox{Hp}{\contour{black}\bf\color{white}H$^+$+H$_\text{jov}^+$}}{\includegraphics[width=\linewidth,clip,trim=0.8cm 8cm 2cm 2.8cm]{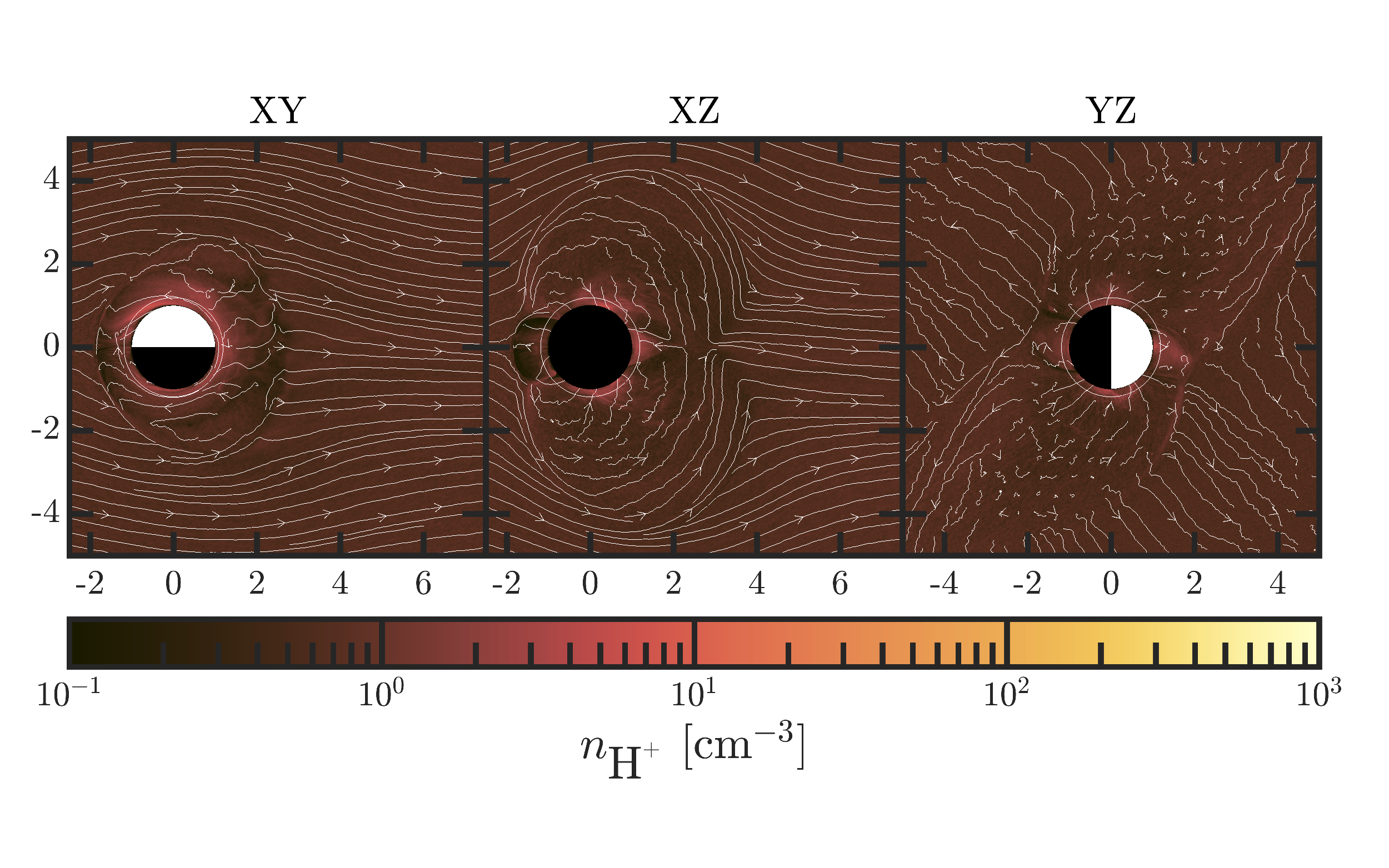}}}}\!
        \fcolorbox{Op}{white}{\parbox{.467\linewidth}{\stackinset{l}{0pt}{b}{87pt}{\colorbox{Op}{\contour{black}\bf\color{white}O$^+$+O$_\text{jov}^+$}}{\includegraphics[width=\linewidth,clip,trim=0.8cm 8cm 2cm 2.8cm]{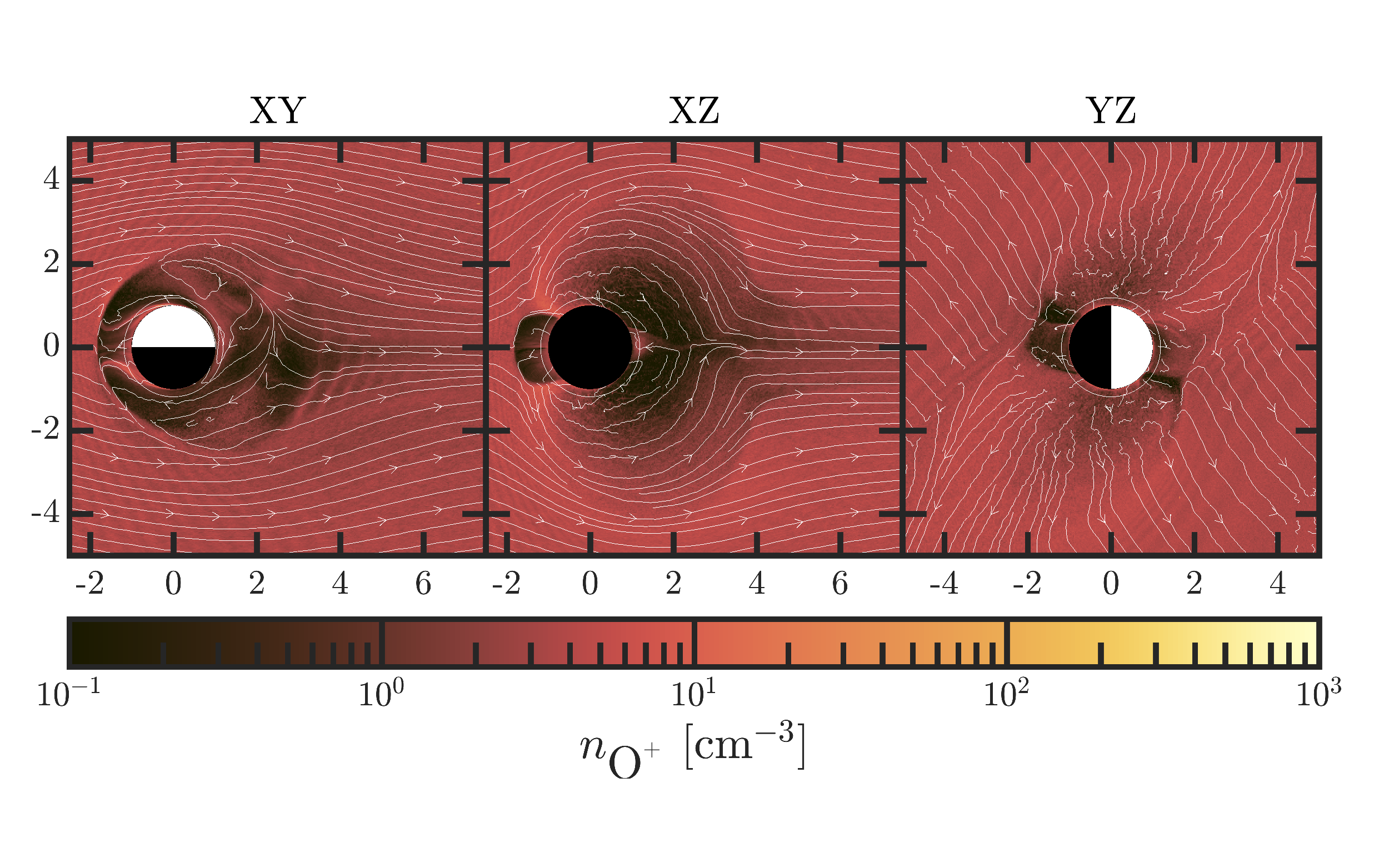}}}}\\
        \fcolorbox{H2p}{white}{\parbox{.467\linewidth}{\stackinset{l}{0pt}{b}{87pt}{\colorbox{H2p}{\contour{black}\bf\color{white}H$_2^+$}}{\includegraphics[width=\linewidth,clip,trim=0.8cm 8cm 2cm 2.8cm]{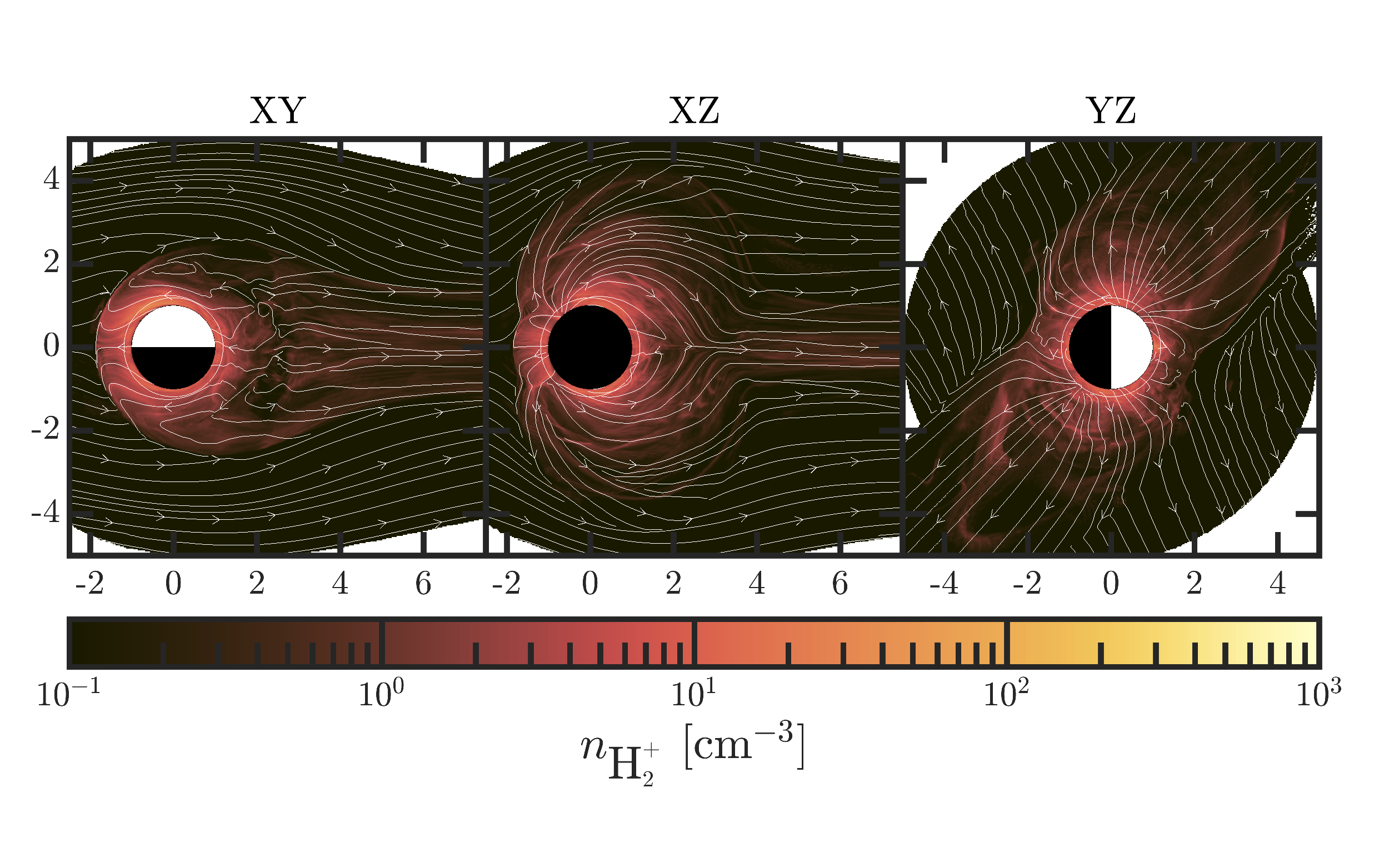}}}}\!
        \fcolorbox{HOp}{white}{\parbox{.467\linewidth}{\stackinset{l}{0pt}{b}{87pt}{\colorbox{HOp}{\contour{black}\bf\color{white}HO$\vphantom{O_2^+}^+$}}{\includegraphics[width=\linewidth,clip,trim=0.8cm 8cm 2cm 2.8cm]{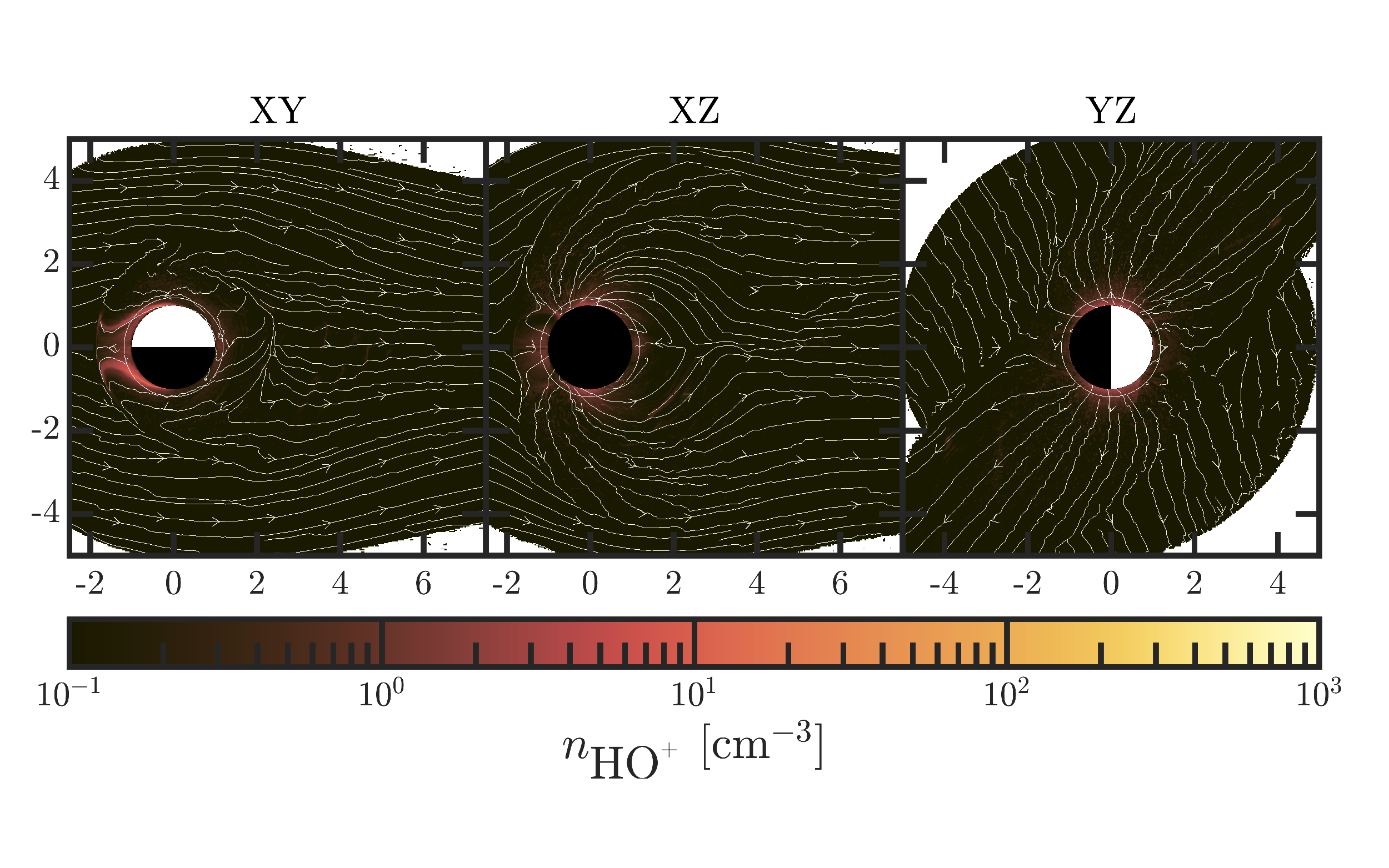}}}}\\
        \fcolorbox{H3p}{white}{\parbox{.467\linewidth}{\stackinset{l}{0pt}{b}{87pt}{\colorbox{H3p}{\contour{black}\bf\color{white}H$_3^+$}}{\includegraphics[width=\linewidth,clip,trim=0.8cm 8cm 2cm 2.8cm]{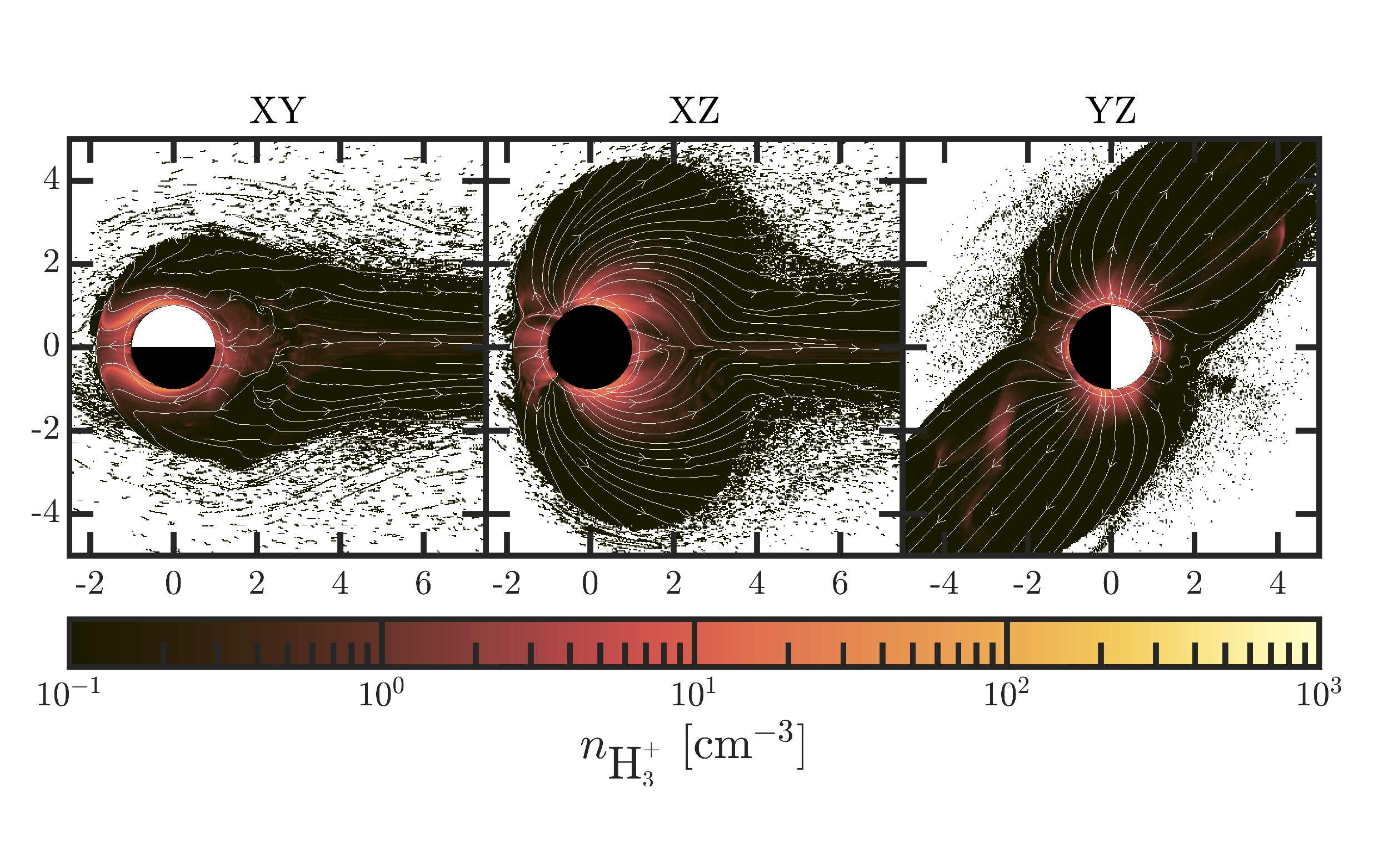}}}}\!
        \fcolorbox{H2Op}{white}{\parbox{.467\linewidth}{\stackinset{l}{0pt}{b}{87pt}{\colorbox{H2Op}{\contour{black}\bf\color{white}H$_2$O$\vphantom{O_2^+}^+$}}{\includegraphics[width=\linewidth,clip,trim=0.8cm 8cm 2cm 2.8cm]{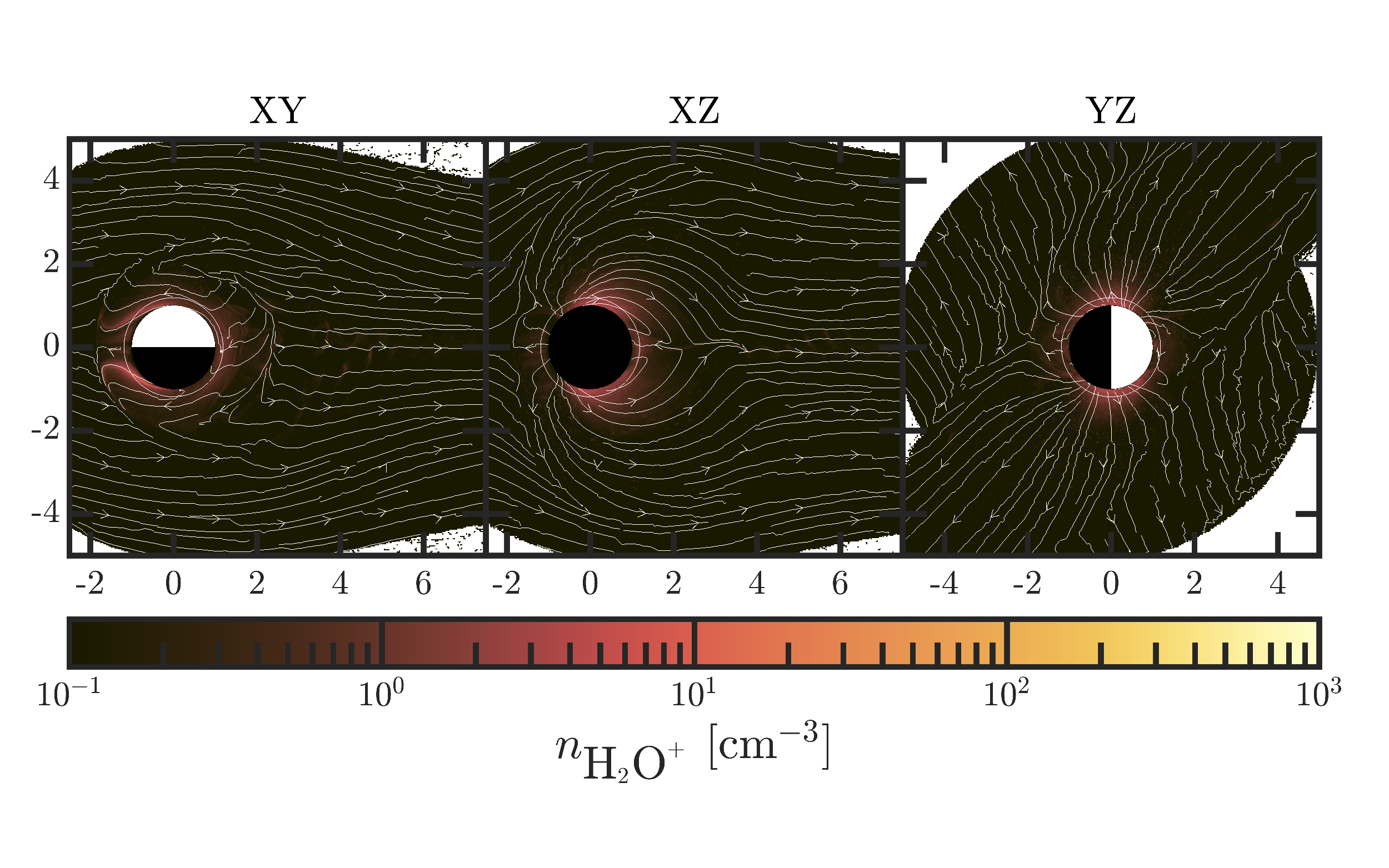}}}}\\
        \fcolorbox{O2p}{white}{\parbox{.467\linewidth}{\stackinset{l}{0pt}{b}{87pt}{\colorbox{O2p}{\contour{black}\bf\color{white}O$_2^+$}}{\includegraphics[width=\linewidth,clip,trim=0.8cm 8cm 2cm 2.8cm]{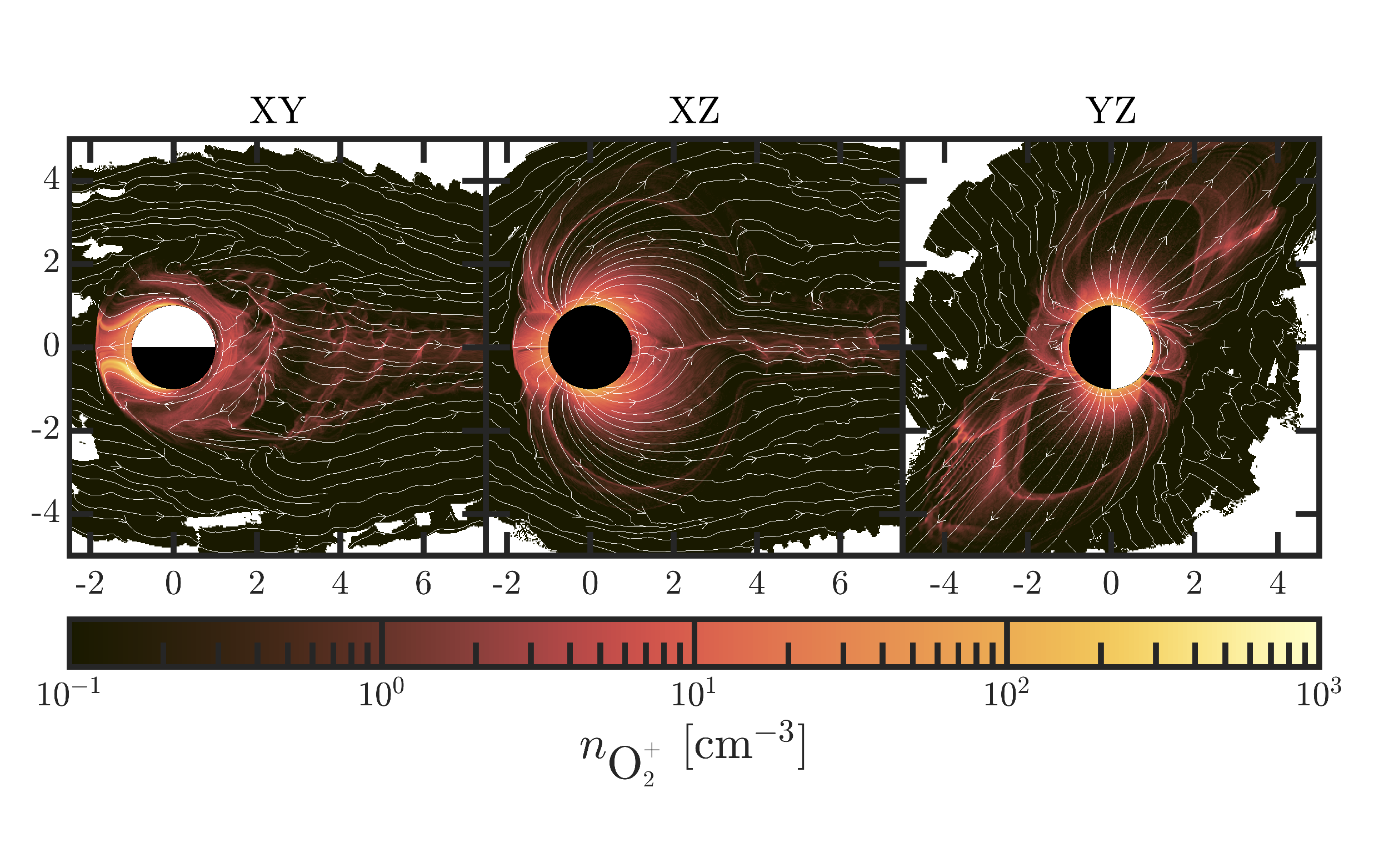}}}}\!
        \fcolorbox{H3Op}{white}{\parbox{.467\linewidth}{\stackinset{l}{0pt}{b}{87pt}{\colorbox{H3Op}{\contour{black}\bf\color{white}H$_3$O$\vphantom{_3^+}^+$}}{\includegraphics[width=\linewidth,clip,trim=0.8cm 8cm 2cm 2.8cm]{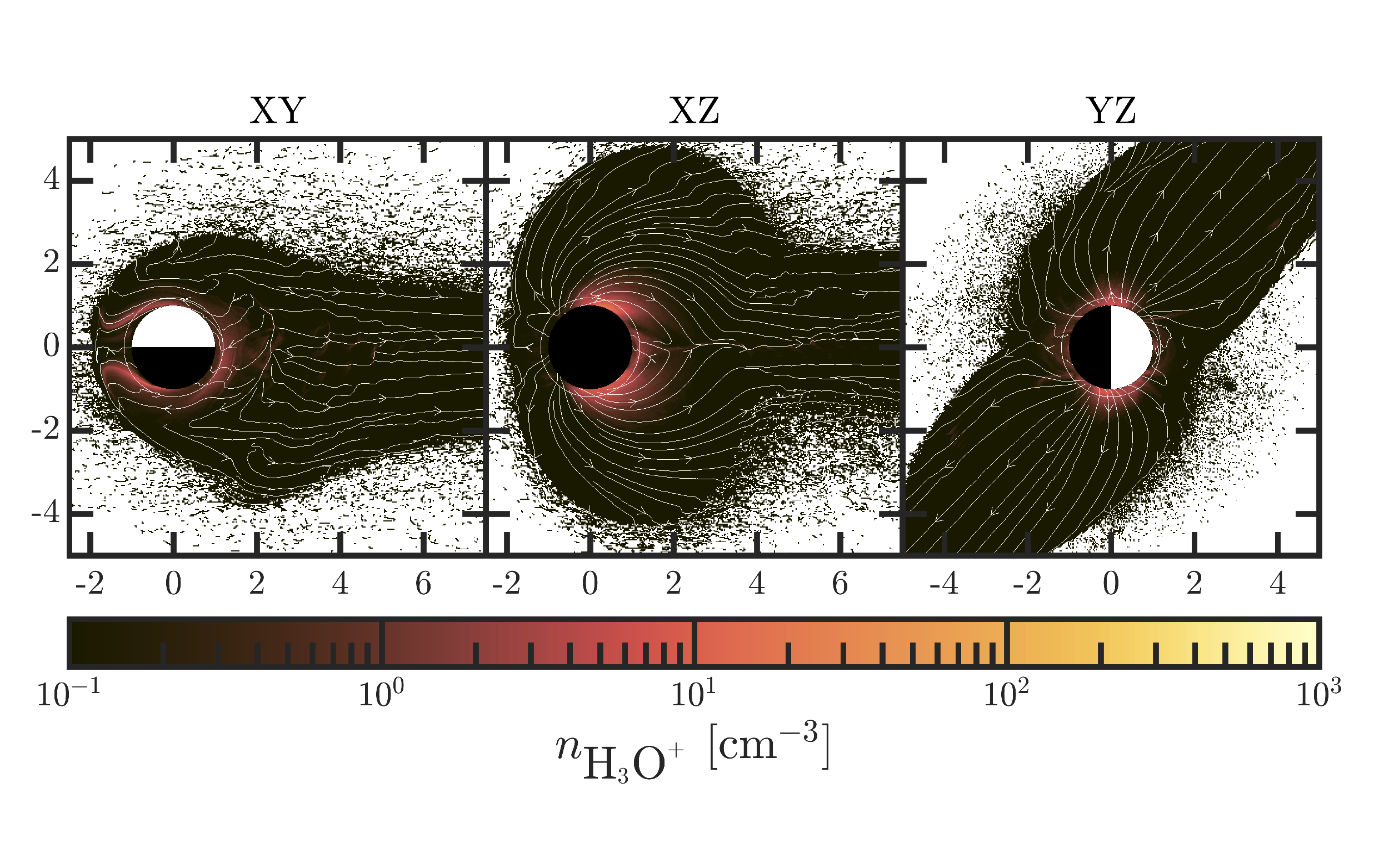}}}}\\
        \fcolorbox{O2Hp}{white}{\parbox{.467\linewidth}{\stackinset{l}{0pt}{b}{87pt}{\colorbox{O2Hp}{\contour{black}\bf\color{white}O$_2$H$\vphantom{_2^+}^+$}}{\includegraphics[width=\linewidth,clip,trim=0.8cm 8cm 2cm 2.8cm]{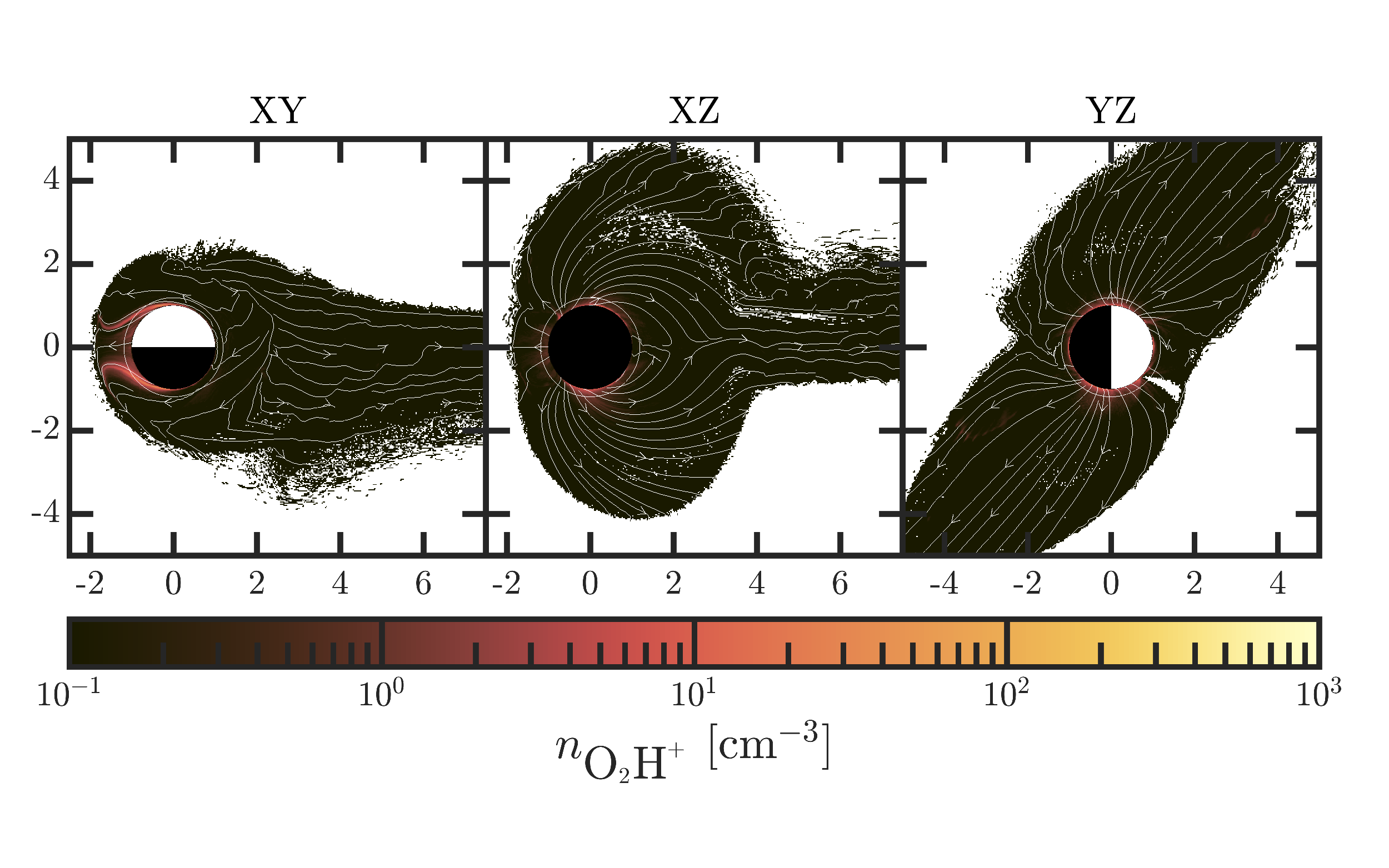}}}}\!
        \fcolorbox{black}{white}{\parbox{.467\linewidth}{\stackinset{l}{0pt}{b}{87pt}{\colorbox{black}{\contour{black}\bf\color{white}$\vphantom{O_2^+}$\bf SUM}}{\includegraphics[width=\linewidth,clip,trim=0.8cm 8cm 2cm 2.8cm]{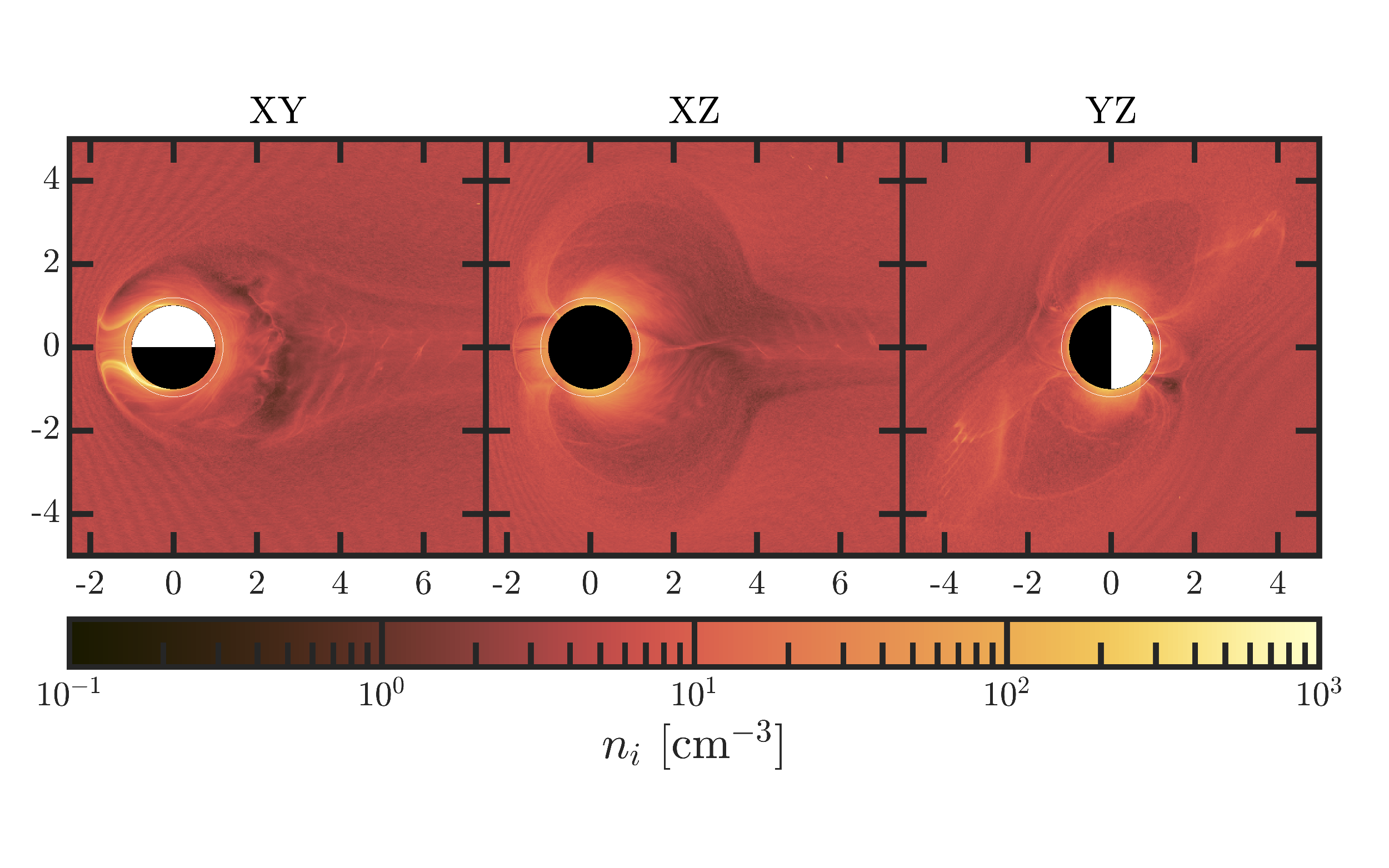}}}}\\
        \includegraphics[width=\linewidth,clip,trim=1.2cm 3.0cm 1.2cm 18.5cm]{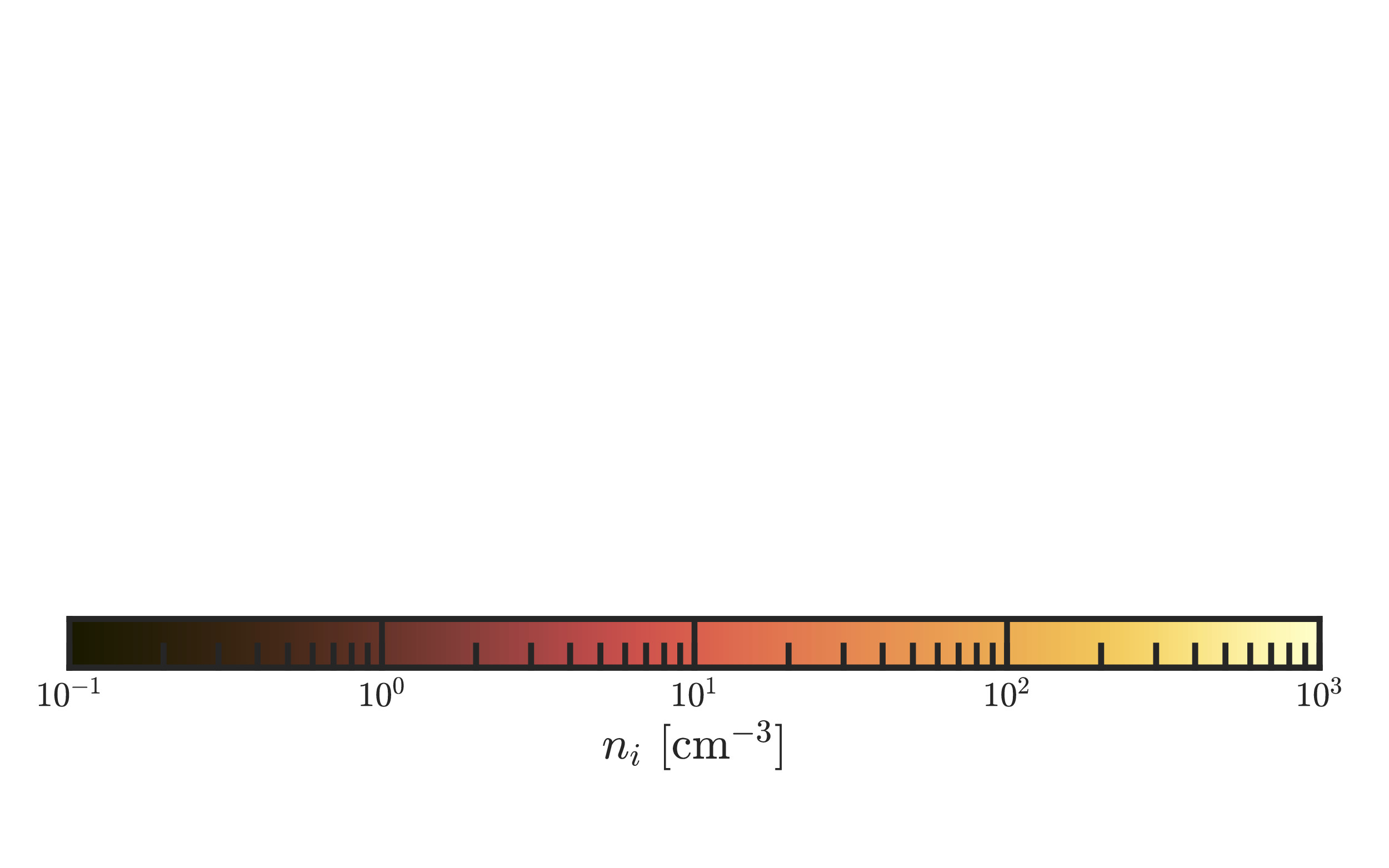}\\
        \caption{Similar to Fig.~\ref{Fig2} for G29. Ganymede was in Jupiter's shadow, though we still indicate the expected dayside.}
        \label{Fig4}
    \end{figure*}

Fig.~\ref{Fig2}, \ref{Fig3}, and \ref{Fig4} show different cuts (in the XY, XZ, and YZ planes in the GPhiO coordinate system centred on Ganymede) of number densities of the different ion species for the three flybys considered, G01, G07, and G29, respectively. These cuts help to reveal different asymmetries: dayside/nightside (except for G29 eclipsed by Jupiter) and Jovian/anti-Jovian. 

A common feature of G01 and G07 is the absence of H$_2^+$ and H$_3^+$ (and of O$_2$H$^+$ and O$^+$ to a lesser extent) near the subsolar point in the few tens of kilometres above the surface (seen as a faint white crescent, in the case of a lack of macroparticles, or a black one, in the case of very low number densities, present at the edge of the disk that defines Ganymede). As H$_2^+$ macroparticles, which are also the sole ``parent'' of H$_3^+$, are supposed to be produced homogeneously through the simulation box (regardless of the neutral number density that only affects their statistical weight), this low amount (black crescent) of H$_2^+$ means that the macroparticle is readily turned into another, but which one? A priori, near the surface, the main loss mechanisms for H$_2^+$ are charge exchange and proton transfer with O$_2$, producing O$_2^+$ and O$_2$H$^+$, respectively, as O$_2$ is much denser than H$_2$ at the surface. This loss of H$_2^+$ and production of O$_2$H$^+$ (only produced through ion-neutral chemistry, O$_2^+$ being mainly yielded through O$_2$ ionisation) must occur homogeneously around Ganymede, as O$_2$ and H$_2$ do not exhibit strong spatial dependency. However, near the subsolar point, there is the extra presence of sublimated H$_2$O. This large amount of H$_2$O, larger than those of O$_2$ and H$_2$ \citep{Beth2025}, increases the loss of H$_2^+$, yielding H$_2$O$^+$ and H$_3$O$^+$. This also causes H$_2$O$^+$ to be under photo-chemical equilibrium near the subsolar point, reaching a number density of $\sim 15-20$\,cm$^{-3}$ (the crescent has the corresponding colour).
As a matter of fact, H$_2$O reacts with all ions except O$_2^+$ and H$_3$O$^+$, driving their loss over the subsolar point to ultimately become H$_3$O$^+$ through successive ion-neutral reactions (see Appendix \ref{Appendixreaction} and Section~\ref{section43}).    

Still in the case of G01 and G07, we also found that ion-neutral chemistry contributes non-negligibly to H$_2$O$^+$ production rate via charge-exchange with H$_2$O. In the collisionless case, H$_2$O$^+$ is solely produced through ionisation of H$_2$O. 

G29 is a case of its own: Ganymede passed in Jupiter's shadow, inhibiting the presence of H$_2$O and photoionisation. Consequently, it allows us to focus solely on the Jovian/anti-Jovian asymmetry. Overall, ion number densities are lower owing to the absence of photoionisation and of sublimated water when comparing with the other flybys on the dayside. Removing H$_2$O from the chemical network leads to a few changes as well. H$_3$O$^+$ production is still possible but unlikely (it can be formed from O$^+$ and successive proton transfer with H$_2$), H$_3^+$ and O$_2$H$^+$ become terminal ions (see Section \ref{section43}). Therefore, we anticipate that the ion composition may change during the JUICE escorting phase while passing behind Jupiter. Although O$_2$H$^+$ survives longer in these conditions, it remains confined at the surface where O$_2$ resides. Afar from Ganymede, O$_2$H$^+$ is turned into H$_3^+$ through proton transfer with H$_2$.      

In all cases, Ganymede's plasma tail is asymmetric and more extended on the anti-Jovian side (cf. XY cut, left and centre panels for each species), populated by heavy ions. As ionospheric ions escape Ganymede's magnetosphere upstream (cf. the streamlines), they are picked up in the direction of the convective electric field. Ions are transported, surfing along the magnetopause before leaving the system downstream, forming a tail in front of the moon. When the subsolar point, where the water ice sublimates the most, is located on the far side of Jupiter, as for G01, the plasma is loaded with water ions (H$_2$O$^+$ and H$_3$O$^+$). It is unclear if `mass-loading' would be the most appropriate term here as the Jovian plasma has a mean mass-to-charge ratio of $m/z=14$\,u\,q$^{-1}$ while water ions do not have a mass drastically different. However, it may affect the plasma dynamics as these ions are injected at a much lower speed than the ambient plasma flow: It would be then `momentum-(un)loading'. For instance, for G01, in Fig. \ref{Fig2}, for O$_2^+$, H$_2$O$^+$ and H$_3$O$^+$, we can identify the cycloidal motion of these heavy ions downstream along the wake: If one wants to include ionospheric ions in the simulation (in particular heavy ions with a large gyroradius), one must consider simulating them kinetically with a hybrid approach. This partly explains why MHD simulations might struggle to accurately capture the magnetopause crossings, in particular along the anti-Jovian flank of Ganymede's magnetosphere, in the wake, and on the dayside \citep[e.g.][]{Duling2022}, as the plasma may be momentum-loaded by ionospheric heavy ions with large gyroradii effects. 

\begin{figure*}
        \centering
        {\includegraphics[width=\linewidth,clip,trim=0cm 4cm 1cm 0cm]{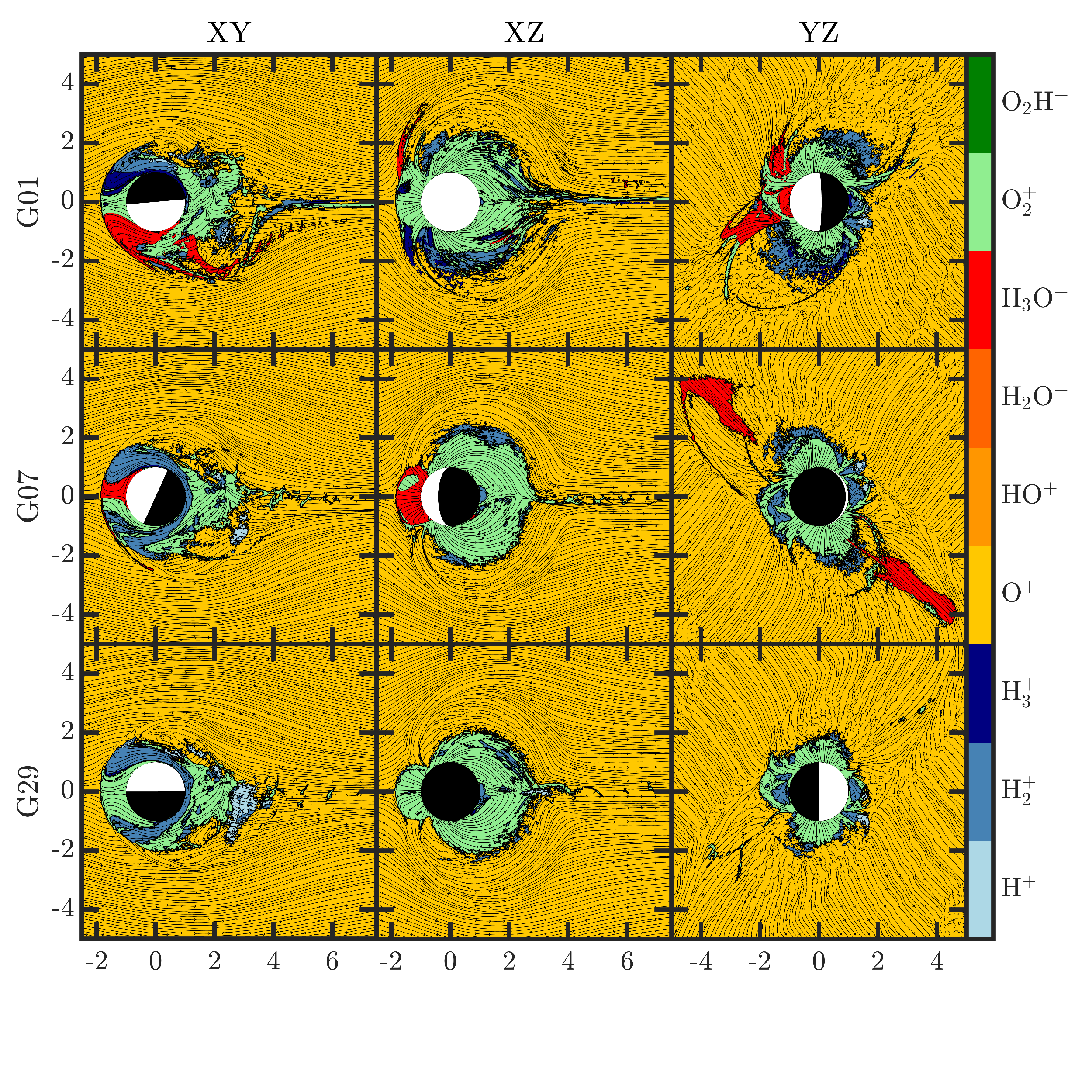}}
        \caption{Cuts for G01, G07, and G29, as Figs.~\ref{Fig2}, \ref{Fig3}, and \ref{Fig4}, coloured as a function of the dominant ion species. Note that Ganymede is within Jupiter's shadow for G29. Where O$^+$ and H$^+$ dominate, this is solely caused by Jovian contribution. Streamlines of the total ion momentum are superimposed.}
        \label{Fig5}
    \end{figure*}

Another way to summarise our findings is to look at the dominant ion species around Ganymede. Fig.~\ref{Fig5} shows cuts of the dominant ion species for the different flybys.  Therein, we define regions around Ganymede depending on the ion species with the largest number density for the three analysed configurations, G01 (top row), G07 (middle row), and G29 (bottom row). This reveals the extent of Ganymede's ionosphere. Although the ionosphere does not have a strict upper boundary, it is appropriate to say that it extends as far as the plasma of exospheric origin dominates over that of magnetospheric origin. For O$^+$ and H$^+$, we considered together ions of both ionospheric and magnetospheric origins. Outside of Ganymede's magnetosphere, Jovian O$^+$ dominates the plasma composition. Even though the Jovian magnetospheric plasma may still access part of Ganymede's magnetosphere, even the closed-field-line region (extending roughly up to $\sim2 R_G$ from the centre of Ganymede in its equatorial plane) via the wake, the ionospheric plasma dominates within $\sim 1 R_G$ from the surface. During G29, Ganymede's ionosphere shrinks in size compared to other flybys due to the absence of sublimation and photoionisation.

Fig.~\ref{Fig5} emphasises the importance of the location of the subsolar point compared to the overall magnetospheric configuration of Ganymede (i.e. depending on Ganymede's position relative to Jupiter's dipole). Depending on the location of the subsolar point (e.g. G01 vs G07), H$_3$O$^+$ may be the dominant ion species around the subsolar point within Ganymede's magnetosphere. H$_3^+$ is expected to be more present on the dark side (cf. G01). However, in the case of G07, the dark side and leading hemispheres coincide. The plasma within Ganymede's magnetosphere flows in the -$x$ direction around the moon, sputtering the leading hemisphere \citep{Carnielli2020b}. H$_3^+$ is therefore convected towards the shadowed surface, where it reacts with O$_2$: A thin layer of O$_2$H$^+$ is seen in Fig.~\ref{Fig3} on the nightside. 

Our simulations agree with the findings of \citet{Eviatar2001} on the fact that O$_2^+$ dominates the composition over the polar regions (cf. XZ and YZ cuts). However, while \citet{Eviatar2001} concluded that O$^+$ dominates at low latitudes and flows along the open flux tubes, we found that H$_3$O$^+$ appears to be a better candidate at low latitudes.

\section{Discussion}\label{section4}

In this section, we discuss a few aspects regarding our modelling, its reliability, the assumptions made, and the new questions raised by this study that must be addressed in the future.

\subsection{Coefficient rates versus cross-sections}\label{section41}
In this paper, we treat collisions using kinetic rates instead of cross-sections in part due to a lack of laboratory measurements (see Section \ref{section22}). Considering kinetic coefficient rates instead of cross sections might overestimate the effect of ion-neutral chemistry, but this might not be drastically the case in our case study. Let's take the example of H$_2^+$+H$_2$ presented in Fig. 4 of \citet{Phelps1990}. Below a laboratory ion energy $E_\text{lab.}$ of 5~eV ($E_\text{rel.}<2.5$ eV, cf. Appendix \ref{AppendixPostcoll}), the formation of H$_3^+$ dominates the process, and the ion-neutral collision can be well approximated by the Langevin model \citep[that describes the interaction between an ion and a non-polar molecule, see e.g.]{Fox2015}, meaning that using a coefficient rate is appropriate (i.e. ${<\!\sigma \varv_\text{rel.}\!>_{\text{H}_2^+-\text{H}_2}}(E_\text{lab.}<\!5~\text{eV})\sim \text{const.} \approx 2.08\times 10^{-15}\,\text{m}^{3}\,\text{s}^{-1}$). Above 5~eV, symmetric charge exchange dominates the interaction by forming a slow H$_2^+$ and a fast H$_2$. Because the cross-section becomes roughly constant at these energies ($\sigma_{\text{H}_2^+-\text{H}_2}(E_\text{rel.}\!>\!5~\text{eV})\sim 10^{-19}$~m$^{-2}$), ${<\!\sigma \varv_\text{rel.}\!>}$ must be calculated using Eq.~\ref{Eq6} instead and hence it increases with energy, linearly with respect to the relative speed (as the relative speed is well above the thermal neutral speed): ${<\!\sigma \varv_\text{rel.}\!>_{\text{H}_2^+-\text{H}_2}\sim 10^{-19} \Vert\vec{\varv}_\text{rel.}\Vert\,\text{m}^{3}\,\text{s}^{-1}}$ (quantities given in SI units). As shown in \citet{Beth2025} and Fig.~\ref{Figapp5}, within Ganymede's magnetosphere, light ion species remain below $\sim 20$~eV, this limit being reached around the MP. Nevertheless, at the MP, H$_2$ number density has already decreased by two orders of magnitude compared with that at the surface. Therefore, it would require H$_2^+$ to be around 40\,keV at the MP (i.e. 10 times faster than the Jovian magnetospheric flow) to have the same collision probability than that at the surface. Even if we virtually underestimate cross-sections at large energies, the likelihood of ion-neutral collisions at these energies is still negligible owing to drastically lower neutral number densities. Most of the ion-neutral chemistry occurs near the surface where neutral number densities are large and ions are slow (that said $E_\text{ref.}\!<\!5$~\text{eV}). Therefore, we anticipate that the use of the reaction rate coefficient remains valid with the electromagnetic fields used in these simulations.

As pointed out by \citet{Beth2025}, using the electromagnetic fields from \citet{Jia2008,Jia2009}, there are still discrepancies between simulated and observed ion energies \citep[observed ion energies exceed those simulated, see Fig. 7 in][and Appendix \ref{Figapp5}]{Beth2025}, suggesting that ions, in particular heavy ones, such as water ions and O$_2^+$, may exceed this threshold of 5\,eV even near the surface of Ganymede. Processes at higher energies (i.e. not thermal and/or endothermic such as charge-exchange, dissociation, and excitation), above tens of eVs, are not considered here as they tend to be endothermic and require cross-sections. The ion species most likely to exceed these threshold energies are the heaviest because all species tend to drift at similar speeds with these ideal MHD background fields \citep{Beth2025} such that their kinetic energy is only a function of their mass. For example, let's look at reactions that involve heavy ions, such as O$_2^+$+H$_2$: O$_2^+$ is heavy and H$_2$ has the most extended exosphere with the largest scale height. These species do not react together at thermal energy in the ground state (hence not included in our chemical network) but do at much larger energies \citep{Irvine1997}. In addition, one has also to consider the energy states of the reactants. For example, the excited metastable O$_2^+$ is more reactive than O$_2^+$ in the ground state (namely $X_2\Pi_g$): The former has a charge-exchange cross section with H$_2$ $\sim30$ times larger at 100\,eV than the latter \citep{Irvine1997}. Likewise, O$_2^+$ may be left in an excited state following electron impact \citep{Turner1968}, the primary ionisation source at Ganymede \citep{Carnielli2019}. Therefore, not only the state in which ions are born, ground or excited, depending on the ionisation process (photon or electron impact), but also the electromagnetic fields, may affect the ion-neutral chemistry and the outcome of our simulations. This must not be forgotten. On the one hand, little is known about the formation of excited states and branching ratios (see also Section \ref{section43}). On the other hand, we still miss proper constraints of the 3D electromagnetic fields (the electric field being unknown) around Ganymede.   

\subsection{Insight on Juno PJ34 flyby}\label{section42}
Although the current work does not focus on the Juno flyby (that is left for future studies), our modelling includes the production of H$_3^+$ that was detected during the PJ34 Juno flyby. Therefore, our simulations can help us interpret the findings of \citet{Valek2022}. One finding was the inbound-outbound asymmetry in the H$_3^+$ number density. During PJ34, inbound occurred on the nightside, whereas outbound occurred on the dayside. Consequently, Ganymede's exosphere is expected to be denser and wetter (more H$_2$O) along the outbound leg. That may explain the H$_3^+$ asymmetry (i.e. a steeper decrease of the number density as a function of the altitude): H$_3^+$ is lost and turned into H$_3$O$^+$ as it reacts with H$_2$O on the sunlit side as the spacecraft approached the subsolar point. 

A second finding was the ``O$^+$'' asymmetry. It would be more appropriate to refer to it as water-group ion asymmetry as the instrument lacks mass precision. The number density of water-group ions is higher on the dayside. This might support our finding that H$_2$O$^+$ and H$_3$O$^+$ are significantly produced on the dayside where H$_2$O is present, misidentifying these ions as O$^+$.

As shown in Section \ref{flybys}, O$^+$ number density (both Jovian and ionospheric) is strongly depleted near Ganymede compared with the collisionless case \citep[cf.][]{Beth2025}. It has been reported in the literature that O$^+$ was the dominant ion within Ganymede's ionosphere during Galileo's flybys \citep{Vasyliunas2000} and the main contributor to the range 16-19\,u\,q$^{-1}$ (i.e. water-group ions) during Juno flyby \citep[][]{Allegrini2022,Valek2022}. However, the ion identification from observations remains inconclusive owing to the inability to separate ion masses (e.g. Galileo) or due to a mass resolution that is too low to separate water-group ions (e.g. Juno). By including ion-neutral chemistry, we find that O$^+$ number density is significantly reduced: O$^+$ reacts with all neutral species, mainly through charge exchange (see Appendices \ref{Appendixreaction}, \ref{AppendixPAEI}, and Fig.~\ref{network}). Aside from the ionization of neutrals, it can only be produced through collision between H$^+$ and O (almost reversible as H and O have close ionisation energy, see Appendix \ref{AppendixPAEI}), which is a negligible chemical path due to the low level of hydrogen. O$^+$ is destroyed and barely replenished through ion-neutral collisions compared with the collisionless case. Therefore, it is unlikely within Ganymede's magnetosphere that ionospheric O$^+$ is the dominant species amongst the water-group ions at low energy, especially on the dayside due to the additional presence of H$_2$O.

Within the context of the Juno flyby, \citet{Waite2024} attempted to simulate and retrieve H$_2^+$, H$_3^+$, and O$_2^+$ under severe assumptions. The electron number density obtained from \citet{Buccino2022} is prescribed for the estimation of H$_2^+$ and H$_3^+$ number densities (i.e. not self-consistently calculated). Ions are assumed to be in photochemical equilibrium, neglecting transport. Applying these assumptions, including dissociative recombination, to G01 conditions for O$_2^+$ over the polar caps, which dominates the ion composition in this region, we infer a number density of $\sim 8\times 10^3$\,cm$^{-3}$. In contrast, in our simulations, O$_2^+$ ion number density reaches $\sim 1\times10^3$\,cm$^{-3}$. This demonstrates that our model, which includes transport, is best suited for interpreting Juno observations.

\subsection{Comprehensive chemical network at Ganymede}\label{section43}
A summary of chemical pathways is shown in Fig.~\ref{network}, where two cases are addressed: one with H$_2$O, representative of dayside, and one without, representative of nightside, in Jupiter's shadow (like G29) or farther away from the subsolar point as the scale height for the sublimated H$_2$O is small. H$_3$O$^+$ is expected to be a terminal ion in the presence of H$_2$O, whereas O$_2$H$^+$ and H$_3^+$ are terminal ions only in its absence. In both cases, O$_2^+$ is always a terminal ion. While Fig.~\ref{network} exhibits similarities with \citet{Shematovich2008}, there are differences. For example, we did not include the possibility to convert O$_2^+$ to O$_2$H$^+$: Such a reaction was not found in either UMIST \citep{Millar2022} or KIDA \citep{Wakelam2012} databases. However, O$_2^+$+H$_2$ might produce O$_2$H$^+$ and H$_2$O$^+$ if O$_2^+$ is an excited state \citep{Ajello1974,Weber1993}. This suggests that we might be able to infer whether O$_2^+$ is excited depending on the amount of O$_2$H$^+$ (and maybe H$_2$O$^+$) that Jupiter Icy Moons Explorer (JUICE) will ever detect. Times when Ganymede will be in Jupiter's shadow must be of particular interest: The ionisation is driven solely by electron impact and H$_2$O number density is highly reduced, hindering the ability to produce H$_2$O$^+$ from H$_2$O through ionisation. H$_2$O$^+$ will then be only produced through ion-neutral chemistry and these aforementioned chemical pathways. This again highlights the importance of considering the energy state of ions.

The recent discovery of a localised CO$_2$ gas patch by \citet{Bockelee2024}, though low in terms of number density, may affect the ion composition and our findings. First, including CO$_2$ in our simulation will increase the loss of H$_2^+$, H$_3^+$ and O$_2$H$^+$, through proton transfer as CO$_2$ has a higher proton affinity than H, H$_2$, and O$_2$. However, the newborn ion, HCO$_2^+$ (45\,u\,q$^{-1}$), reacts with H$_2$O to produce H$_3$O$^+$: The former ion is a terminal ion in the absence of H$_2$O and would likely be found over the CO$_2$ patch when in shadow. In contrast, CO$_2^+$ from CO$_2$ ionisation is reacting with all other main neutral species either through charge-exchange or hydrogen transfer (e.g. $\text{CO}_2^++\text{H}_2\longrightarrow \text{HCO}_2^++\text{H}$). Therefore, we do not expect CO$_2$, CO$_2^+$, and HCO$_2^+$ to affect our findings drastically. Nevertheless, CO$_2$ may be dissociated into C and CO or ionised into C$^+$ and CO$^+$: The chemical network becomes more complex owing to C and C$^+$. The ion-neutral reaction involving C leads to CH$^+$ and, by cascade effect, to hydrocarbon cations: Mass spectrometers onboard JUICE (Barabash et al., in prep), may be able to detect them. Indeed, even with a limited mass resolution, any signal detected around 12 and 13\,u\,q$^{-1}$ will be unambiguously attributed to resp. C$^+$ and CH$^+$ as long as the mass separation is $\sim1$ and neglecting isotopologues.

\section{Conclusion}\label{section5}

By including ion-neutral collisions, by means of kinetic rate coefficients, in our kinetic test-particle model, we have tentatively assessed their impact on the ionosphere of an icy moon, with the application here to Ganymede. While ion-neutral chemistry barely affects the total ion number density, except in very specific locations such as the magnetopause, the ion composition, in contrast, is strongly affected, especially near the moon. We show that not only new ion species are produced, namely H$_3^+$, H$_3$O$^+$, and O$_2$H$^+$, and others are destroyed, such as H$_2^+$ and O$^+$, but also the ion composition significantly depends on the location of the subsolar point with respect to Ganymede's global magnetospheric configuration (which varies with the moon's local time with respect to Jupiter), whether located around the trailing hemisphere (e.g G07), the leading one, the nearside, or the farside (e.g. G01) of Jupiter. 

From a modelling perspective, we provide preliminary results and a simple framework to implement ion-neutral chemistry for computational models, particularly adequate for multi-fluid MHD and hybrid models. As we treat ions as test particles, therefore kinetically, the approach exposed here, described in Appendices \ref{Algo} and \ref{AppendixPostcoll} combined with a better parametrisation of the neutral atmosphere \citep[cf.][]{Beth2025}, is the best course of action for those who are interested in accounting for ion-neutral chemistry in their hybrid models \citep[e.g. those of][]{Fatemi2022,Stahl2023} and may compare with our results.

From an observational perspective, this work has strong implications on the future JUICE in-situ measurements. At the current altitude of the spacecraft orbit (500 or 200\,km altitude), we anticipate that JUICE will observe H$_2^+$, H$_3^+$, H$_3$O$^+$, and O$_2^+$, the dominant ion species in our simulations. However, we cast doubt on the ability of the JUICE Particle Environment Package (PEP, Barabash et al., in prep.) to separate and distinguish these ions appropriately. Indeed, two PEP sensors are dedicated to ion measurements: Neutral Gas and Ion Mass (NIM) and Jovian Plasma Dynamics and Composition (JDC). While NIM has a sufficient mass resolution ($m/\Delta m\gtrsim 750$), its energy acceptance is limited below 5-10\,eV, excluding deceleration/acceleration caused by the spacecraft potential. The lighter the ion species is, the more likely NIM will detect it: a 10~eV H$_2^+$ goes at 30\,km\,s$^{-1}$ whereas a 10~eV O$_2^+$ goes at 7\,km\,s$^{-1}$. \citet{Beth2025} showed that a good proxy of the ion velocity based on these MHD simulations is $\vec{E}\times\vec{B}/B^2$. Near the surface at an altitude of $\sim 130$\,km, the location of the inner boundary of the MHD simulation, the ion speed reaches already $\sim 10$ km\,s$^{-1}$ depending on the conditions/flybys \citep[cf. Fig 12 in][]{Beth2025}, such that H$_2^+$ would be detectable, H$_2$O$^+$ maybe (at the detection limit), but not O$_2^+$. The lowest speeds are found above the magnetic poles, within the cusps, and near the magnetopause such that these locations will be the best places to separate the different water-group ions. In contrast, JDC has an energy acceptance much wider, from 1\,eV to 41\,keV, but a much lower mass resolution ($m/\Delta m\sim 20$), meaning that water-group ions will be challenging to separate from each other. To interpret the Juno and JUICE ion composition measurements around Ganymede, kinetic models including chemistry, such as the one presented here, are critical. 
\begin{figure}

\centering
\begin{tikzpicture}[scale=3.5]
        \node[align=center] (title) at (0.35,0.2) {Including H$_2$O};
        \node[circle,draw,align=center] (H2p) at (0,0) {\,H$_2^+$};
        \node[align=center] (H2Op) at (0.7,-0.504) {\,H$_2$O$^+$};
        \node[rectangle,draw,align=center] (O2p) at (1.4,0) {\,O$_2^+$};
        \node[align=center] (H3p) at (0,-0.808) {\phantom{\,H$_3^+$}};
        \node[rectangle,align=center] (HOp) at (1.4,-0.808) {\,HO$^+$};
        \node[circle,draw,align=center] (Hp) at (0.7,-0.16) {\,H$^+$};
        \node[circle,draw,align=center] (Op) at (1.2,-0.404) {\,O$^+$};
        \node[rectangle,draw,align=center] (H3Op) at (0.7,-1.21) {\,H$_3$O$^+$};
        \node[align=center] (O2Hp) at (-0.7,-1.21) {\,O$_2$H$^+$};
        \draw [-{Stealth[length=3mm, width=2mm]},draw=cH2,densely dotted,pos=0.4,line width=0.5mm] (H2p) -- (H3p) node[pos=0.4,sloped,above=-2pt]{+H$_2$};
        \draw [-{Stealth[length=3mm, width=2mm]},draw=cH2O,line width=0.5mm] (H2p) -- (H2Op) node[pos=0.4,sloped,above=-2pt]{+H$_2$O};
        \begin{scope}[transform canvas={xshift=-.075em}]
        \draw [-{Stealth[length=3mm, width=2mm,right]},draw=cH2,densely dotted,line width=0.5mm] (H2Op.south) -- (H3Op.north) ;
        \end{scope}
        \begin{scope}[transform canvas={xshift=+.075em}]
        \draw [-{Stealth[length=3mm, width=2mm,left]},draw=cH2O,line width=0.5mm] (H2Op.south) -- (H3Op.north) ;
        \end{scope}
        \draw [-{Stealth[length=3mm, width=2mm]},draw=cH2O,line width=0.5mm] (H3p) -- (H3Op.north west) ;
        \draw [-{Stealth[length=3mm, width=2mm]},draw=cH2O,line width=0.5mm] (H2p) -- (H3Op) ;
        \draw [-{Stealth[length=3mm, width=2mm]},draw=cO2,dash dot,line width=0.5mm] (H2p) -- (O2p) node[pos=0.5,above=-2pt]{+O$_2$\phantom{$^+$}};
        \draw [-{Stealth[length=3mm, width=2mm]},draw=cO2,dash dot,line width=0.5mm] (H2p) -- (O2Hp) ;
        \begin{scope}[transform canvas={xshift=-.0em,yshift=.095em}]
        \draw [-{Stealth[length=3mm, width=2mm,left]},draw=cH2,densely dotted,line width=0.5mm] (O2Hp) -- (H3p) ;
        \end{scope}
        \begin{scope}[transform canvas={xshift=.0em,yshift=-.07em}]
        \draw [-{Stealth[length=3mm, width=2mm,left]},draw=cO2,dash dot,line width=0.5mm] (H3p) -- (O2Hp) ;
        \end{scope}
        \node[align=center] at (0,-0.808) {\,H$_3^+$};
        \draw [-{Stealth[length=3mm, width=2mm]},draw=cH2O,line width=0.5mm] (O2Hp) -- (H3Op) ;
        \draw [-{Stealth[length=3mm, width=2mm]},draw=cO2,dash dot,line width=0.5mm] (H2Op) -- (O2p.south west) ;
        \draw [-{Stealth[length=3mm, width=2mm]},draw=cO2,dash dot,line width=0.5mm] (HOp) -- (O2p) ;
        \draw [-{Stealth[length=3mm, width=2mm]},draw=cH2O,line width=0.5mm] (HOp.south west) -- (H3Op.north east) ;
        \begin{scope}[transform canvas={xshift=-0.2mm,yshift=-0.2mm}]
        \draw [-{Stealth[length=3mm, width=2mm,left]},draw=cH2O,line width=0.5mm] (HOp.north west) -- (H2Op.south east) ;
        \end{scope}
        \begin{scope}[transform canvas={xshift=0mm,yshift=0.29mm}]
        \draw [-{Stealth[length=3mm, width=2mm,right]},draw=cH2,densely dotted,line width=0.5mm] (HOp.north west) -- (H2Op.south east) ;
        \end{scope}
        \draw [-{Stealth[length=3mm, width=2mm]},draw=cH2,densely dotted,line width=0.5mm] (Op) -- (HOp) ;
        \draw [-{Stealth[length=3mm, width=2mm]},draw=cH2O,line width=0.5mm] (Op) -- (H2Op) ;
        \draw [-{Stealth[length=3mm, width=2mm]},draw=cO2,dash dot,line width=0.5mm] (Op) -- (O2p) ;
        \draw [-{Stealth[length=3mm, width=2mm]},draw=cH2O,line width=0.5mm] (Hp) -- (H2Op) ;
        \draw [-{Stealth[length=3mm, width=2mm]},draw=cO2,dash dot,line width=0.5mm] (Hp) -- (O2p) ;
    \end{tikzpicture}
    \vspace{0.5em}
    \hrule
    \vspace{0.5em}
    \begin{tikzpicture}[scale=3.5]
        \node[align=center] (title) at (0.35,0.2) {Neglecting H$_2$O};
        \node[circle,draw,minimum size=2pt,align=center] (H2p) at (0,0) {\,H$_2^+$};
        \node[rectangle,align=center] (H2Op) at (0.7,-0.504) {\,H$_2$O$^+$};
        \node[rectangle,draw,align=center] (O2p) at (1.4,0) {O$_2^+$};
        \node[rectangle,draw,align=center] (H3p) at (0,-0.808) {\phantom{\,H$_3^+$}};
        \node[align=center] (HOp) at (1.4,-0.808) {HO$^+$};
        \node[circle,draw,align=center] (Hp) at (0.7,-0.16) {\,H$^+$};
        \node[circle,draw,align=center] (Op) at (1.2,-0.404) {\,O$^+$};
        \node[rectangle,draw,align=center] (H3Op) at (0.7,-1.21) {\,H$_3$O$^+$};
        \node[rectangle,draw,align=center] (O2Hp) at (-0.7,-1.21) {O$_2$H$^+$};
        \draw [-{Stealth[length=3mm, width=2mm]},draw=cH2,densely dotted,pos=0.4,line width=0.5mm] (H2p) -- (H3p) node[pos=0.4,sloped,above=-2pt]{+H$_2$};
        \draw [-{Stealth[length=3mm, width=2mm]},draw=cH2,densely dotted,line width=0.5mm] (H2Op.south) -- (H3Op.north) ;
        \draw [-{Stealth[length=3mm, width=2mm]},draw=cO2,dash dot,line width=0.5mm] (H2p) -- (O2p) node[pos=0.5,above=-2pt]{+O$_2$\phantom{$^+$}};
        \draw [-{Stealth[length=3mm, width=2mm]},draw=cO2,dash dot,line width=0.5mm] (H2p) -- (O2Hp) ;
        \begin{scope}[transform canvas={xshift=-.0em,yshift=.095em}]
        \draw [-{Stealth[length=3mm, width=2mm,left]},draw=cH2,densely dotted,line width=0.5mm] (O2Hp) -- (H3p) ;
        \end{scope}
        \begin{scope}[transform canvas={xshift=.0em,yshift=-.07em}]
        \draw [-{Stealth[length=3mm, width=2mm,left]},draw=cO2,dash dot,line width=0.5mm] (H3p) -- (O2Hp) ;
        \end{scope}
        \node[rectangle,draw,align=center] at (0,-0.808) {\,H$_3^+$};
        \draw [-{Stealth[length=3mm, width=2mm]},draw=cO2,dash dot,line width=0.5mm] (H2Op) -- (O2p.south west) ;
        \draw [-{Stealth[length=3mm, width=2mm]},draw=cO2,dash dot,line width=0.5mm] (HOp) -- (O2p) ;
        \draw [-{Stealth[length=3mm, width=2mm]},draw=cH2,densely dotted,line width=0.5mm] (HOp) -- (H2Op) ;
        \draw [-{Stealth[length=3mm, width=2mm]},draw=cH2,densely dotted,line width=0.5mm] (Op) -- (HOp) ;
        \draw [-{Stealth[length=3mm, width=2mm]},draw=cO2,dash dot,line width=0.5mm] (Op) -- (O2p) ;
        \draw [-{Stealth[length=3mm, width=2mm]},draw=cO2,dash dot,line width=0.5mm] (Hp) -- (O2p) ;
    \end{tikzpicture}
    \caption{Chemical pathway for ion-neutral collision at Ganymede considering O$_2$, H$_2$, with (top) or without (bottom) H$_2$O as the main neutral species.  Arrows' styles and/or colour indicate the neutral species with which ions react: dotted red for H$_2$, solid blue for H$_2$O, and dash-dotted green for O$_2$. Circles identify species only produced through ionisation and lost through at least one ion-neutral reaction. Squares identify species that cannot be lost through ion-neutral collisions (excluding processes that are endothermic or involve excited states). Loss through ion-electron dissociative recombination is disregarded here.}\label{network}
\end{figure}
\section*{Acknowledgements}
AB acknowledges Ronan Modolo for the fruitful interactions and comments on the study. Work at Imperial College London was supported by the Science and Technology Facilities Council (STFC) of the UK under ST/S000364/1 and ST/W001071/1. Test-particle simulations were performed at the Imperial College Computing Service (DOI: \url{10.14469/hpc/2232}). The research performed by FL was funded by the project FACOM (ANR-22-CE49-0005-01 ACT). XJ acknowledges support by the National Aeronautics and Space Administration (NASA) through Early Career Fellow Startup Grant \#80NSSC20K1286 and JUICE/PEP contract \#183512 with Johns Hopkins University Applied Physics Laboratory. AB acknowledges Fabio Crameri and his scientific colourmaps (used for Figs. \ref{Fig2}, \ref{Fig3}, and \ref{Fig4}) available at \url{10.5281/zenodo.1243862} and presented in \citet{Crameri2020}.

\section*{Data availability}
The data underlying this article will be shared on reasonable request to the corresponding author. MAG \citep{KivelsonG} and PWS \citep{KurthPWS} data are available on the PDS.

\bibliographystyle{mnras}
\bibliography{example}

\begin{thebibliography}{}
\makeatletter
\relax
\def\mn@urlcharsother{\let\do\@makeother \do\$\do\&\do\#\do\^\do\_\do\%\do\~}
\def\mn@doi{\begingroup\mn@urlcharsother \@ifnextchar [ {\mn@doi@} {\mn@doi@[]}}
\def\mn@doi@[#1]#2{\def\@tempa{#1}\ifx\@tempa\@empty \href {http://dx.doi.org/#2} {doi:#2}\else \href {http://dx.doi.org/#2} {#1}\fi \endgroup}
\def\mn@eprint#1#2{\mn@eprint@#1:#2::\@nil}
\def\mn@eprint@arXiv#1{\href {http://arxiv.org/abs/#1} {{\tt arXiv:#1}}}
\def\mn@eprint@dblp#1{\href {http://dblp.uni-trier.de/rec/bibtex/#1.xml} {dblp:#1}}
\def\mn@eprint@#1:#2:#3:#4\@nil{\def\@tempa {#1}\def\@tempb {#2}\def\@tempc {#3}\ifx \@tempc \@empty \let \@tempc \@tempb \let \@tempb \@tempa \fi \ifx \@tempb \@empty \def\@tempb {arXiv}\fi \@ifundefined {mn@eprint@\@tempb}{\@tempb:\@tempc}{\expandafter \expandafter \csname mn@eprint@\@tempb\endcsname \expandafter{\@tempc}}}

\bibitem[\protect\citeauthoryear{Adams \& Smith}{Adams \& Smith}{1984}]{Adams1984}
Adams N.,  Smith D.,  1984, \mn@doi [Chemical Physics Letters] {10.1016/0009-2614(84)85665-1}, 105, 604

\bibitem[\protect\citeauthoryear{Ajello, Huntress, Lane, LeBreton  \& Williamson}{Ajello et~al.}{1974}]{Ajello1974}
Ajello J.~M.,  Huntress W.~T.,  Lane A.~L.,  LeBreton P.~R.,   Williamson A.~D.,  1974, \mn@doi [The Journal of Chemical Physics] {10.1063/1.1681184}, 60, 1211

\bibitem[\protect\citeauthoryear{Allegrini et~al.,}{Allegrini et~al.}{2022}]{Allegrini2022}
Allegrini F.,  et~al., 2022, Geophysical Research Letters, 49, e2022GL098682

\bibitem[\protect\citeauthoryear{Ansher, Barnhardt, Richards, Gurnett  \& Kurth}{Ansher et~al.}{2017}]{KurthPWS}
Ansher J.,  Barnhardt B.,  Richards B.,  Gurnett D.,   Kurth W.,  2017, \mn@doi{10.17189/1519684}

\bibitem[\protect\citeauthoryear{Barth et~al.,}{Barth et~al.}{1997}]{Barth1997}
Barth C.~A.,  et~al., 1997, \mn@doi [Geophysical Research Letters] {10.1029/97GL01927}, 24, 2147

\bibitem[\protect\citeauthoryear{Beth, Galand, Modolo, Jia, Leblanc  \& Huybrighs}{Beth et~al.}{2025}]{Beth2025}
Beth A.,  Galand M.,  Modolo R.,  Jia X.,  Leblanc F.,   Huybrighs H. L.~F.,  2025, \mn@doi [Monthly Notices of the Royal Astronomical Society] {10.1093/mnras/staf313}, 538, 2483

\bibitem[\protect\citeauthoryear{Bockelée-Morvan et~al.,}{Bockelée-Morvan et~al.}{2024}]{Bockelee2024}
Bockelée-Morvan D.,  et~al., 2024, \mn@doi [A\&A] {10.1051/0004-6361/202451599}, 690, L11

\bibitem[\protect\citeauthoryear{Boris}{Boris}{1970}]{Boris1970}
Boris J.~P.,  1970, in Proc. Fourth Conf. Num. Sim. Plasmas. pp 3--67

\bibitem[\protect\citeauthoryear{Buccino et~al.,}{Buccino et~al.}{2022}]{Buccino2022}
Buccino D.~R.,  et~al., 2022, \mn@doi [Geophysical Research Letters] {10.1029/2022GL098420}, 49, e2022GL098420

\bibitem[\protect\citeauthoryear{Carnielli, Galand, Leblanc, Leclercq, Modolo, Beth, Huybrighs  \& Jia}{Carnielli et~al.}{2019}]{Carnielli2019}
Carnielli G.,  Galand M.,  Leblanc F.,  Leclercq L.,  Modolo R.,  Beth A.,  Huybrighs H.,   Jia X.,  2019, \mn@doi [Icarus] {10.1016/j.icarus.2019.04.016}, 330, 42

\bibitem[\protect\citeauthoryear{Carnielli, Galand, Leblanc, Modolo, Beth  \& Jia}{Carnielli et~al.}{2020a}]{Carnielli2020a}
Carnielli G.,  Galand M.,  Leblanc F.,  Modolo R.,  Beth A.,   Jia X.,  2020a, \mn@doi [Icarus] {10.1016/j.icarus.2020.113691}, 343, 113691

\bibitem[\protect\citeauthoryear{Carnielli, Galand, Leblanc, Modolo, Beth  \& Jia}{Carnielli et~al.}{2020b}]{Carnielli2020b}
Carnielli G.,  Galand M.,  Leblanc F.,  Modolo R.,  Beth A.,   Jia X.,  2020b, \mn@doi [Icarus] {10.1016/j.icarus.2020.113918}, 351, 113918

\bibitem[\protect\citeauthoryear{Crameri, Shephard  \& Heron}{Crameri et~al.}{2020}]{Crameri2020}
Crameri F.,  Shephard G.~E.,   Heron P.~J.,  2020, \mn@doi [Nature Communications] {10.1038/s41467-020-19160-7}, 11, 5444

\bibitem[\protect\citeauthoryear{Duling et~al.,}{Duling et~al.}{2022}]{Duling2022}
Duling S.,  et~al., 2022, \mn@doi [Geophysical Research Letters] {10.1029/2022GL101688}, 49, e2022GL101688

\bibitem[\protect\citeauthoryear{Eviatar, {M. Vasyliūnas}  \& {A. Gurnett}}{Eviatar et~al.}{2001}]{Eviatar2001}
Eviatar A.,  {M. Vasyliūnas} V.,   {A. Gurnett} D.,  2001, \mn@doi [Planetary and Space Science] {10.1016/S0032-0633(00)00154-9}, 49, 327

\bibitem[\protect\citeauthoryear{Fahr \& Müller}{Fahr \& Müller}{1967}]{Fahr1967}
Fahr H.,  Müller K.~G.,  1967, \mn@doi [Zeitschrift für Physik] {10.1007/BF01326177}, 200, 343 – 365

\bibitem[\protect\citeauthoryear{Fatemi, Poppe, Vorburger, Lindkvist  \& Hamrin}{Fatemi et~al.}{2022}]{Fatemi2022}
Fatemi S.,  Poppe A.~R.,  Vorburger A.,  Lindkvist J.,   Hamrin M.,  2022, \mn@doi [Journal of Geophysical Research: Space Physics] {10.1029/2021JA029863}, 127, e2021JA029863

\bibitem[\protect\citeauthoryear{Fox}{Fox}{2015}]{Fox2015}
Fox J.~L.,  2015, \mn@doi [Icarus] {10.1016/j.icarus.2015.01.010}, 252, 366

\bibitem[\protect\citeauthoryear{Frank, Ackerson, Lee, English  \& Pickett}{Frank et~al.}{1992}]{Frank1992}
Frank L.~A.,  Ackerson K.~L.,  Lee J.~A.,  English M.~R.,   Pickett G.~L.,  1992, \mn@doi [Space Science Reviews] {10.1007/BF00216858}, 60, 283

\bibitem[\protect\citeauthoryear{Gurnett, Kurth, Shaw, Roux, Gendrin, Kennel, Scarf  \& Shawhan}{Gurnett et~al.}{1992}]{Gurnett1992}
Gurnett D.~A.,  Kurth W.~S.,  Shaw R.~R.,  Roux A.,  Gendrin R.,  Kennel C.~F.,  Scarf F.~L.,   Shawhan S.~D.,  1992, \mn@doi [Space Science Reviews] {10.1007/BF00216861}, 60, 341

\bibitem[\protect\citeauthoryear{Hall, Feldman, McGrath  \& Strobel}{Hall et~al.}{1998}]{Hall1998}
Hall D.~T.,  Feldman P.~D.,  McGrath M.~A.,   Strobel D.~F.,  1998, \mn@doi [The Astrophysical Journal] {10.1086/305604}, 499, 475

\bibitem[\protect\citeauthoryear{Hillenbrand, Ruette, Urbain  \& Savin}{Hillenbrand et~al.}{2022}]{Hillenbrand2022}
Hillenbrand P.-M.,  Ruette N.~d.,  Urbain X.,   Savin D.~W.,  2022, \mn@doi [The Astrophysical Journal] {10.3847/1538-4357/ac41ce}, 927, 47

\bibitem[\protect\citeauthoryear{Huestis}{Huestis}{2008}]{Huestis2008}
Huestis D.~L.,  2008, Planetary and Space Science, 56, 1733

\bibitem[\protect\citeauthoryear{Huntress \& Pinizzotto}{Huntress \& Pinizzotto}{1973}]{Huntress1973}
Huntress W.~T. J.,  Pinizzotto R.~F. J.,  1973, \mn@doi [The Journal of Chemical Physics] {10.1063/1.1680687}, 59, 4742

\bibitem[\protect\citeauthoryear{Irvine \& Latimer}{Irvine \& Latimer}{1997}]{Irvine1997}
Irvine A.,  Latimer C.,  1997, \mn@doi [Int. J. Mass Spectrom. Ion Process.] {10.1016/S0168-1176(97)00025-6}, 164, 115

\bibitem[\protect\citeauthoryear{Jia, Walker, Kivelson, Khurana  \& Linker}{Jia et~al.}{2008}]{Jia2008}
Jia X.,  Walker R.~J.,  Kivelson M.~G.,  Khurana K.~K.,   Linker J.~A.,  2008, \mn@doi [Journal of Geophysical Research: Space Physics] {10.1029/2007JA012748}, 113

\bibitem[\protect\citeauthoryear{Jia, Walker, Kivelson, Khurana  \& Linker}{Jia et~al.}{2009}]{Jia2009}
Jia X.,  Walker R.~J.,  Kivelson M.~G.,  Khurana K.~K.,   Linker J.~A.,  2009, \mn@doi [Journal of Geophysical Research: Space Physics] {10.1029/2009JA014375}, 114

\bibitem[\protect\citeauthoryear{Johnson}{Johnson}{1990}]{Johnson1990}
Johnson R.~E.,  1990, Energetic charged-particle interactions with atmospheres and surfaces.
 Physics and Chemistry in Space Vol. 19, Springer Science \& Business Media

\bibitem[\protect\citeauthoryear{Jones, Birkinshaw  \& Twiddy}{Jones et~al.}{1981}]{Jones1981}
Jones J.,  Birkinshaw K.,   Twiddy N.,  1981, \mn@doi [Chemical Physics Letters] {10.1016/0009-2614(81)85191-3}, 77, 484

\bibitem[\protect\citeauthoryear{Karpas, Anicich  \& Huntress}{Karpas et~al.}{1979}]{Karpas1979}
Karpas Z.,  Anicich V.,   Huntress W.~T. J.,  1979, \mn@doi [The Journal of Chemical Physics] {10.1063/1.437823}, 70, 2877

\bibitem[\protect\citeauthoryear{Kim \& Huntress}{Kim \& Huntress}{1975}]{Kim1975}
Kim J.~K.,  Huntress W.~T. J.,  1975, \mn@doi [The Journal of Chemical Physics] {10.1063/1.430817}, 62, 2820

\bibitem[\protect\citeauthoryear{Kim, Theard  \& Huntress}{Kim et~al.}{1974}]{Kim1974}
Kim J.,  Theard L.,   Huntress W.,  1974, \mn@doi [Int. J. Mass Spectrom. Ion Phys.] {10.1016/0020-7381(74)85001-1}, 15, 223

\bibitem[\protect\citeauthoryear{Kivelson et~al.,}{Kivelson et~al.}{1996}]{Kivelson1996}
Kivelson M.~G.,  et~al., 1996, \mn@doi [Nature] {10.1038/384537a0}, 384, 537

\bibitem[\protect\citeauthoryear{Kivelson, Khurana  \& Volwerk}{Kivelson et~al.}{2002}]{Kivelson2002}
Kivelson M.,  Khurana K.,   Volwerk M.,  2002, \mn@doi [Icarus] {10.1006/icar.2002.6834}, 157, 507

\bibitem[\protect\citeauthoryear{Kivelson, Khurana, Russell, Walker, Joy  \& Mafi}{Kivelson et~al.}{2024}]{KivelsonG}
Kivelson M.,  Khurana K.,  Russell C.,  Walker R.,  Joy S.,   Mafi J.,  2024, \mn@doi{10.17189/gch4-8w75}

\bibitem[\protect\citeauthoryear{Kovalenko, Tran, Rednyk, Rou\'{c}ka, Dohnal, Plašil, Gerlich  \& Glosík}{Kovalenko et~al.}{2018}]{Kovalenko2018}
Kovalenko A.,  Tran T.~D.,  Rednyk S.,  Rou\'{c}ka S.,  Dohnal P.,  Plašil R.,  Gerlich D.,   Glosík J.,  2018, \mn@doi [The Astrophysical Journal] {10.3847/1538-4357/aab106}, 856, 100

\bibitem[\protect\citeauthoryear{Kurth et~al.,}{Kurth et~al.}{2022}]{Kurth2022}
Kurth W.~S.,  et~al., 2022, \mn@doi [Geophysical Research Letters] {10.1029/2022GL098591}, 49, e2022GL098591

\bibitem[\protect\citeauthoryear{Leblanc, Oza, Leclercq, Schmidt, Cassidy, Modolo, Chaufray  \& Johnson}{Leblanc et~al.}{2017}]{Leblanc2017}
Leblanc F.,  Oza A.,  Leclercq L.,  Schmidt C.,  Cassidy T.,  Modolo R.,  Chaufray J.,   Johnson R.,  2017, \mn@doi [Icarus] {10.1016/j.icarus.2017.04.025}, 293, 185

\bibitem[\protect\citeauthoryear{Leblanc et~al.,}{Leblanc et~al.}{2023}]{Leblanc2023}
Leblanc F.,  et~al., 2023, \mn@doi [Icarus] {10.1016/j.icarus.2023.115557}, 399, 115557

\bibitem[\protect\citeauthoryear{Marconi}{Marconi}{2007}]{Marconi2007}
Marconi M.,  2007, \mn@doi [Icarus] {10.1016/j.icarus.2007.02.016}, 190, 155

\bibitem[\protect\citeauthoryear{Millar, Walsh, Van~de Sande  \& Markwick}{Millar et~al.}{2024}]{Millar2022}
Millar T.~J.,  Walsh C.,  Van~de Sande M.,   Markwick A.~J.,  2024, \mn@doi [A\&A] {10.1051/0004-6361/202346908}, 682, A109

\bibitem[\protect\citeauthoryear{Phelps}{Phelps}{1990}]{Phelps1990}
Phelps A.~V.,  1990, \mn@doi [Journal of Physical and Chemical Reference Data] {10.1063/1.555858}, 19, 653

\bibitem[\protect\citeauthoryear{Plainaki et~al.,}{Plainaki et~al.}{2015}]{Plainaki2015}
Plainaki C.,  et~al., 2015, \mn@doi [Icarus] {10.1016/j.icarus.2014.09.018}, 245, 306

\bibitem[\protect\citeauthoryear{Pontoni, Shimoyama, Futaana, Fatemi, Poppe, Wieser  \& Barabash}{Pontoni et~al.}{2022}]{Pontoni2021}
Pontoni A.,  Shimoyama M.,  Futaana Y.,  Fatemi S.,  Poppe A.~R.,  Wieser M.,   Barabash S.,  2022, \mn@doi [Journal of Geophysical Research: Space Physics] {10.1029/2021JA029439}, 127, e2021JA029439

\bibitem[\protect\citeauthoryear{{Prasad} \& {Huntress}}{{Prasad} \& {Huntress}}{1980}]{Prasad1980}
{Prasad} S.~S.,  {Huntress} Jr. W.~T.,  1980, \mn@doi [Astrophysical Journal Supplementary Series] {10.1086/190665}, 43, 1

\bibitem[\protect\citeauthoryear{Rakshit \& Warneck}{Rakshit \& Warneck}{1980}]{Rakshit1980}
Rakshit A.~B.,  Warneck P.,  1980, \mn@doi [J. Chem. Soc.{,} Faraday Trans. 2] {10.1039/F29807601084}, 76, 1084

\bibitem[\protect\citeauthoryear{Roth, Ivchenko, Gladstone, Saur, Grodent, Bonfond, Molyneux  \& Retherford}{Roth et~al.}{2021}]{Roth2021}
Roth L.,  Ivchenko N.,  Gladstone G.~R.,  Saur J.,  Grodent D.,  Bonfond B.,  Molyneux P.~M.,   Retherford K.~D.,  2021, Nature Astronomy, 5, 1043

\bibitem[\protect\citeauthoryear{Roth et~al.,}{Roth et~al.}{2023}]{Roth2023}
Roth L.,  et~al., 2023, \mn@doi [The Planetary Science Journal] {10.3847/PSJ/acaf7f}, 4, 12

\bibitem[\protect\citeauthoryear{Saur et~al.,}{Saur et~al.}{2015}]{Saur2015}
Saur J.,  et~al., 2015, \mn@doi [Journal of Geophysical Research: Space Physics] {10.1002/2014JA020778}, 120, 1715

\bibitem[\protect\citeauthoryear{Shematovich}{Shematovich}{2008}]{Shematovich2008}
Shematovich V.~I.,  2008, \mn@doi [Solar System Research] {10.1134/S0038094608060026}, 42, 473

\bibitem[\protect\citeauthoryear{Shematovich}{Shematovich}{2016}]{Shematovich2016}
Shematovich V.~I.,  2016, \mn@doi [Solar System Research] {10.1134/S0038094616040067}, 50, 262

\bibitem[\protect\citeauthoryear{Smith, Adams  \& Miller}{Smith et~al.}{1978}]{Smith1978}
Smith D.,  Adams N.~G.,   Miller T.~M.,  1978, \mn@doi [The Journal of Chemical Physics] {10.1063/1.436354}, 69, 308

\bibitem[\protect\citeauthoryear{Smith, Spanel  \& Mayhew}{Smith et~al.}{1992}]{Smith1992}
Smith D.,  Spanel P.,   Mayhew C.~A.,  1992, \mn@doi [Int. J. Mass Spectrom. Ion Process.] {10.1016/0168-1176(92)80108-D}, 117, 457

\bibitem[\protect\citeauthoryear{Stahl, Addison, Simon  \& Liuzzo}{Stahl et~al.}{2023}]{Stahl2023}
Stahl A.,  Addison P.,  Simon S.,   Liuzzo L.,  2023, \mn@doi [Journal of Geophysical Research: Space Physics] {10.1029/2023JA032113}, 128, e2023JA032113

\bibitem[\protect\citeauthoryear{Stancil, Schultz, Kimura, Gu, Hirsch  \& Buenker}{Stancil et~al.}{1999}]{Stancil1999}
Stancil P.~C.,  Schultz D.~R.,  Kimura M.,  Gu J.-P.,  Hirsch G.,   Buenker R.~J.,  1999, \mn@doi [Astron. Astrophys. Suppl. Ser.] {10.1051/aas:1999419}, 140, 225

\bibitem[\protect\citeauthoryear{Theard \& Huntress}{Theard \& Huntress}{1974}]{Theard1974}
Theard L.~P.,  Huntress Wesley~T. J.,  1974, \mn@doi [The Journal of Chemical Physics] {10.1063/1.1681453}, 60, 2840

\bibitem[\protect\citeauthoryear{Tran, Rednyk, Kovalenko, Rou\'{c}ka, Dohnal, Plašil, Gerlich  \& Glosík}{Tran et~al.}{2018}]{Tran2018}
Tran T.~D.,  Rednyk S.,  Kovalenko A.,  Rou\'{c}ka S.,  Dohnal P.,  Plašil R.,  Gerlich D.,   Glosík J.,  2018, \mn@doi [The Astrophysical Journal] {10.3847/1538-4357/aaa0d8}, 854, 25

\bibitem[\protect\citeauthoryear{Turner \& Rutherford}{Turner \& Rutherford}{1968}]{Turner1968}
Turner B.~R.,  Rutherford J.~A.,  1968, \mn@doi [Journal of Geophysical Research (1896-1977)] {10.1029/JA073i021p06751}, 73, 6751

\bibitem[\protect\citeauthoryear{Valek et~al.,}{Valek et~al.}{2022}]{Valek2022}
Valek P.~W.,  et~al., 2022, Geophysical Research Letters, 49, e2022GL100281

\bibitem[\protect\citeauthoryear{Vasyliūnas \& Eviatar}{Vasyliūnas \& Eviatar}{2000}]{Vasyliunas2000}
Vasyliūnas V.~M.,  Eviatar A.,  2000, \mn@doi [Geophysical Research Letters] {10.1029/2000GL003739}, 27, 1347

\bibitem[\protect\citeauthoryear{{Viggiano}, {Albritton}, {Fehsenfeld}, {Adams}, {Smith}  \& {Howorka}}{{Viggiano} et~al.}{1980}]{Viggiano1980}
{Viggiano} A.~A.,  {Albritton} D.~L.,  {Fehsenfeld} F.~C.,  {Adams} N.~G.,  {Smith} D.,   {Howorka} F.,  1980, \mn@doi [The Astrophysical Journal] {10.1086/157766}, 236, 492

\bibitem[\protect\citeauthoryear{Vorburger, Fatemi, Galli, Liuzzo, Poppe  \& Wurz}{Vorburger et~al.}{2022}]{Vorburger2022}
Vorburger A.,  Fatemi S.,  Galli A.,  Liuzzo L.,  Poppe A.~R.,   Wurz P.,  2022, \mn@doi [Icarus] {10.1016/j.icarus.2021.114810}, 375, 114810

\bibitem[\protect\citeauthoryear{Vorburger, Fatemi, {Carberry Mogan}, Galli, Liuzzo, Poppe, Roth  \& Wurz}{Vorburger et~al.}{2024}]{Vorburger2024}
Vorburger A.,  Fatemi S.,  {Carberry Mogan} S.~R.,  Galli A.,  Liuzzo L.,  Poppe A.~R.,  Roth L.,   Wurz P.,  2024, \mn@doi [Icarus] {10.1016/j.icarus.2023.115847}, 409, 115847

\bibitem[\protect\citeauthoryear{Waite~Jr. et~al.,}{Waite~Jr. et~al.}{2024}]{Waite2024}
Waite~Jr. J.~H.,  et~al., 2024, \mn@doi [Journal of Geophysical Research: Planets] {10.1029/2023JE007859}, 129, e2023JE007859

\bibitem[\protect\citeauthoryear{Wakelam et~al.,}{Wakelam et~al.}{2012}]{Wakelam2012}
Wakelam V.,  et~al., 2012, \mn@doi [The Astrophysical Journal Supplement Series] {10.1088/0067-0049/199/1/21}, 199, 21

\bibitem[\protect\citeauthoryear{Weber, Dalleska, Tjelta, Fisher  \& Armentrout}{Weber et~al.}{1993}]{Weber1993}
Weber M.~E.,  Dalleska N.~F.,  Tjelta B.~L.,  Fisher E.~R.,   Armentrout P.~B.,  1993, \mn@doi [The Journal of Chemical Physics] {10.1063/1.464593}, 98, 7855

\makeatother
\end{thebibliography}
\bsp	

\appendix 

\section{Collision scheme}\label{Algo}
We summarise here the different steps in the test-particle including collisions. 
\begin{algorithm}
\caption{Ion motion from $t$ to $t+\Delta t$}
\label{<your label for references later in your document>}
\begin{algorithmic}[1]
\State Ion with position $\vec{x}(t)$ and velocity  $\vec{\varv}(t)$
\State Locate the ion in the MHD simulation 
\State Interpolate $\vec{E}$ and $\vec{B}$ at position $\vec{x}$ from values $\vec{E}_i$ and $\vec{B}_i$ at the vertices of the MHD cell, associating them a weight $w_m$
\newline $\vec{E}(\vec{x}) \gets \sum_m^8 w_m\vec{E}_m$
\newline $\vec{B}(\vec{x}) \gets \sum_m^8 w_m\vec{B}_m$
\State Apply Boris' algorithm to update $\vec{x}$ and $\vec{\varv}$ from $t$ to $t+\Delta t$
\newline $\vec{\varv}(t+\Delta t) \overset{\vec{E},\vec{B}}{\gets} \vec{\varv}(t)$ and $\vec{x}(t+\Delta t) \overset{\vec{E},\vec{B}}{\gets} \vec{x}(t)$
\State Locate the ion in the exospheric simulation 
\State Derive the ion-neutral collision frequency $\nu_{i,n}$ between the ion species $i$ and the neutral species $n$ based on the neutral density in the exospheric cell
\newline $\nu_{i,n}=\sigma_{i,n} ({\varv}_\text{rel}){\varv}_\text{rel}n_n$ or $k_{i,n}(T)n_n$
\State Calculate the collision probability $p_\text{collision}$ of the ion $i$ with one of the $N$ neutral species $n$ during $\Delta t$
\newline $p_\text{collision}=1-\exp(-\sum_n^{N} \nu_{i,n}\Delta t)$
\State Draw a random number $p\in\mathcal{U}_{[0,1]}$
\If{$p < p_\text{collision}$}  \Comment{\vphantom{/}\hfill// Ion collides with one neutral species}
    \State Draw a random number $r\in\mathcal{U}_{[0,1]}$
    \State Initialisation: $p_0 \gets 0$ and $k \gets 0$
    \While{$p_0<r$} 
        \State $k \gets k+1$\hfill\Comment{\vphantom{/}// $k$: index of the examined neutral species}
        \State $p_k\gets \nu_{i,k}/\sum_n^N \nu_{i,n}$\hfill \Comment{\vphantom{/}\hspace{\algorithmicindent}// Probability to collide with $k$}
        \State $p_0 \gets p_0+p_k$ 
    \EndWhile 
    \newline \Comment{\vphantom{/}\hspace{\algorithmicindent}// Ion species $i$ collides with neutral species $k$}
    \newline \Comment{\vphantom{/}\hspace{\algorithmicindent}// $A+B^+\longrightarrow$ \text{products}?}
    \If{number of products $>1$}
        \State Draw a random number $s\in\mathcal{U}_{[0,1]}$
        \State Initialisation: $b_0 \gets 0$ and $l \gets 0$
        \While{$b_0<s$}
            \State $l \gets l+1$ \hfill \Comment{\vphantom{/}\hspace{0.5em}// Examining the reaction $l$}
            \State $b_0 \gets b_0+b_l$ \hfill \Comment{\vphantom{/}\hspace{0.5em}// $b_l$: branching ratio for reaction $l$}
        \EndWhile 
        \newline\Comment{\vphantom{/}\hspace{\algorithmicindent}// Ion product $j$ from reaction $l$}
        \Else
        \newline\Comment{\vphantom{/}\hspace{\algorithmicindent}// Ion product $j$ from the unique reaction}
    \EndIf
    \newline\Comment{\vphantom{/}\hspace{\algorithmicindent}// The ion $i$ is turned into a new ion $j$}
    \State Reinitialise the ion properties, mass $m$ and charge $q$
    \newline \vphantom{/}\hspace{\algorithmicindent}$m \gets m_j$ and $q \gets q_j$
    \State Update $\vec{\varv}$ following the collision, position $\vec{x}$ is conserved
\Else\newline\Comment{\vphantom{/}\hfill// Ion does not collide}
\EndIf 
\State Back to 1
\end{algorithmic}
\end{algorithm}
We have implemented ion-neutral collision in our test-particle model. As ion-ion interaction is neglected and ions do not have any feedback on the neutral background, which is imposed and steady-state, a collision between an ion and a neutral species can be easily implemented.

\section{Kinetic rate coefficient}\label{Appendixreaction}
We provide here the list of ion-neutral reactions from \url{https://umistdatabase.net/} \citep{Millar2022}. $\Delta H$ has been calculated separately. For `almost' athermic `reversible' reactions (i.e. $\Delta H \sim \pm 0.01$~eV), namely H$^+$ + O $\longleftrightarrow$ H + O$^+$ and O$_2$H$^+$ + H$_2$ $\longleftrightarrow$ O$_2$ + H$_3^+$, we set $\Delta H$ to 0. The rate coefficient is given by:
\begin{equation}
    k_{in}(T)[10^{-10} \text{cm}^{3}\,\text{s}^{-1}]=\alpha \left(\dfrac{T}{300}\right)^\beta\exp(-\gamma / T )  
\end{equation}
with $\alpha$, $\beta$, and $\gamma$ are tabulated in Table \ref{tablereaction}. The absence of values corresponds to 0.
\begin{table*}
\caption{List of ion-neutral reactions for our test-particle model. Values for $\alpha$ [$10^{-16}$\,m$^{-3}$\,s$^{-1}$], $\beta$, and $\gamma$ are taken from \url{https://umistdatabase.net/}. $\Delta H$ is the excess energy associated with each reaction, n$^\circ$ refers to the reaction's number on the website. References to the relevant publications are given.}\label{tablereaction}
\centering
\begin{tabular}{p{2em} p{0.1em} p{1em} p{1em} p{1em} p{0.1em} p{2em} | p{2em} p{2em} p{2em} | l l l}
    \multicolumn{7}{l|}{}&\multicolumn{1}{c}{$\alpha$}&\multicolumn{1}{c}{$\beta$}&\multicolumn{1}{c|}{$\gamma$}&$\Delta H$&n$^{\circ}$&References\\
\hline
\multirow{6}{*}{\phantom{O$_2$}H$^+$$\left\lbrace\begin{array}{l}
                \\
                \\
                \\
                \\
                \\
                \\
                \end{array}\right. $} & + & H & $\longrightarrow$ &$h\nu$& + & H$_2^+$ & &\multicolumn{1}{c}{N/A}&&\\
 & + & H$_2$ & $\longrightarrow$ & none&  & & & &&\\
 & + & O & $\longrightarrow$ & H & + & O$^+$ & $\phantom{0}6.86$&$\phantom{-}0.26$&$224.3$&0&\href{https://umistdatabase.uk/react/417}{417}&\citet{Stancil1999}\\
 & + & HO & $\longrightarrow$ & H& + & HO$^{+}$ & $21\phantom{.00}$&$-0.50$&&0.58&\href{https://umistdatabase.uk/react/419}{419}&\citet{Prasad1980}\\
 & + & H$_2$O & $\longrightarrow$ & H& + & H$_2$O$^+$ & $69\phantom{.00}$&$-0.50$&&0.98&\href{https://umistdatabase.uk/react/388}{388}&\citet{Smith1992}\\
 & + & O$_2$ & $\longrightarrow$ & H& + & O$_2^+$ & $20\phantom{.00}$&&&1.53&\href{https://umistdatabase.uk/react/416}{416}&\citet{Smith1992}\\
 \hline
\multirow{9}{*}{\phantom{O$_2$}H$_2^+$$\left\lbrace\begin{array}{l}
                \\
                \\
                \\
                \\
                \\
                \\
                \\
                \\
                \\
                \end{array}\right. $}  & + & H & $\longrightarrow$ & H$_2$& + & H$^+$ & $\phantom{0}6.4\phantom{0}$ &  & &1.83&\href{https://umistdatabase.uk/react/500}{500}&\citet{Karpas1979}\\
 & + & H$_2$ & $\longrightarrow$ & H& + & H$_3^+$ & $20.8\phantom{0}$&&&1.68&\href{https://umistdatabase.uk/react/3189}{3189}&\citet{Theard1974}\\
 & + & O & $\longrightarrow$ & H & + & HO$^+$ & $15\phantom{.00}$&&&2.34&\href{https://umistdatabase.uk/react/3201}{3201}&\citet{Prasad1980}\\
 & + & HO & $\longrightarrow$ & H$_2$& + & HO$^{+}$ & $\phantom{0}7.6\phantom{0}$&$-0.50$&&2.41&\href{https://umistdatabase.uk/react/417}{467}&\citet{Prasad1980}\\
 &   &    & $\longrightarrow$ & H& + & H$_2$O$^{+}$ & $\phantom{0}7.6\phantom{0}$&$-0.50$&&3.43&\href{https://umistdatabase.uk/react/3202}{3202}&\citet{Prasad1980}\\
 & + & H$_2$O & $\longrightarrow$ & H$_2$& + & H$_2$O$^+$ & $39\phantom{.00}$&$-0.50$&&2.80&\href{https://umistdatabase.uk/react/458}{458}&\citet{Kim1975}\\
 &   &  & $\longrightarrow$ & H& + & H$_3$O$^+$ & $34\phantom{.00}$&$-0.50$&&4.43&\href{https://umistdatabase.uk/react/3191}{3191}&\citet{Kim1975}\\
 & + & O$_2$ & $\longrightarrow$ & H$_2$& + &O$_2^+$  & $\phantom{0}8\phantom{.00}$&&&3.36&\href{https://umistdatabase.uk/react/466}{466}&\citet{Kim1975}\\
  &  &  & $\longrightarrow$ & H& + & O$_2$H$^+$ &$19\phantom{.00}$ &&&1.67&\href{https://umistdatabase.uk/react/3200}{3200}&\citet{Kim1975}\\
\hline
\multirow{7}{*}{\phantom{O$_2$}H$_3^+$$\left\lbrace\begin{array}{l}
                \\
                \\
                \\
                \\
                \\
                \\
                \\
                \end{array}\right. $}  & + & H & $\longrightarrow$ & none&  & &  && &  &\\
 & + & H$_2$ & $\longrightarrow$ & none&  & &  & & &  &\\
  & + & O  & $\longrightarrow$ & H$_2$ & + & HO$^+$& $\phantom{0}4.65$ &$-0.14$&$0.67$&1.69&\href{https://umistdatabase.uk/react/3548}{3548}&\citet{Hillenbrand2022}\\
   &  &  & $\longrightarrow$ & H & + & H$_2$O$^+$& $\phantom{0}2.08$ &$-0.40$&$4.86$&0.66&\href{https://umistdatabase.uk/react/3547}{3547}&\citet{Hillenbrand2022}\\
 & + & HO & $\longrightarrow$ & H$_2$& + & H$_2$O$^+$&$13\phantom{.00}$ &$-0.50$&&1.75&\href{https://umistdatabase.uk/react/3550}{3550}&\citet{Prasad1980}\\
 & + & H$_2$O & $\longrightarrow$ & H$_2$ & + & H$_3$O$^+$& $59\phantom{.00}$ &$-0.50$&&2.75&\href{https://umistdatabase.uk/react/3499}{3499}&\citet{Kim1974}\\
 & + & O$_2$ & $\longrightarrow$ & H$_2$& + & O$_2$H$^{+}$ & $\phantom{0}9.3\phantom{0}$ &&$100$&0&\href{https://umistdatabase.uk/react/3546}{3546}&\citet{Adams1984}\\
\hline
\multirow{7}{*}{\phantom{O$_2$}O$^+$$\left\lbrace\begin{array}{l}
                \\
                \\
                \\
                \\
                \\
                \\
                \\
                \end{array}\right. $} & + & H & $\longrightarrow$ & O& + & H$^+$ & $\phantom{0}5.66$ & $\phantom{-}0.36$&$-8.6$&0&\href{https://umistdatabase.uk/react/503}{503}&\citet{Karpas1979}\\
 & + & H$_2$ & $\longrightarrow$ & H& + & HO$^+$ & $13.5\phantom{0}$&&&0.53&\href{https://umistdatabase.uk/react/3274}{3274}&\citet{Kovalenko2018}\\
  & + & O & $\longrightarrow$ & sym.&  & & &&\\
 & + & HO & $\longrightarrow$ & O& + & HO$^{+}$ & $\phantom{0}3.6\phantom{0}$&$-0.50$&&0.60&\href{https://umistdatabase.uk/react/675}{675}&\citet{Prasad1980}\\
 &   &    & $\longrightarrow$ & H& + & O$_2^{+}$ & $\phantom{0}3.6\phantom{0}$&$-0.50$&&2.25&\href{https://umistdatabase.uk/react/4536}{4536}&\citet{Prasad1980}\\
 & + & H$_2$O & $\longrightarrow$ & O& + & H$_2$O$^+$ & $32\phantom{.00}$&$-0.50$&&1.00&\href{https://umistdatabase.uk/react/667}{667}&\citet{Smith1978}\\
 & + & O$_2$ & $\longrightarrow$ & O& + & O$_2^+$ & $\phantom{0}0.19$&&&1.55&\href{https://umistdatabase.uk/react/673}{673}&\citet{Smith1978}\\
\hline
\multirow{7}{*}{\phantom{$_2$}HO$^+$$\left\lbrace\begin{array}{l}
                \\
                \\
                \\
                \\
                \\
                \\
                \\
                \end{array}\right. $} & + & H & $\longrightarrow$ & none&  & &  && &  &\\
 & + & H$_2$ & $\longrightarrow$ & H& + & H$_2$O$^+$ & $12.7\phantom{0}$&$\phantom{-}0.18$&&1.02&\href{https://umistdatabase.uk/react/3277}{3277}&\citet{Tran2018}\\
 & + & O & $\longrightarrow$ & H & + & O$_2^+$& $\phantom{0}7.1\phantom{0}$ &&&1.65&\href{https://umistdatabase.uk/react/4638}{4638}&\citet{Prasad1980}\\
 & + & HO & $\longrightarrow$ & O& + & H$_2$O$^{+}$ & $\phantom{0}7\phantom{.00}$&$-0.50$&&1.09&\href{https://umistdatabase.uk/react/4676}{4676}&\citet{Prasad1980}\\
 & + & H$_2$O & $\longrightarrow$ & HO& + & H$_2$O$^+$ & $15.9\phantom{0}$&$-0.50$&&0.40&\href{https://umistdatabase.uk/react/700}{700}&\citet{Huntress1973}\\
 &  &  & $\longrightarrow$ & O& + & H$_3$O$^+$ & $13\phantom{.00}$&$-0.50$&&2.10&\href{https://umistdatabase.uk/react/4667}{4667}&\citet{Huntress1973}\\
 & + & O$_2$ & $\longrightarrow$ & HO& + & O$_2^+$ & $\phantom{0}5.9\phantom{0}$&&&0.95&\href{https://umistdatabase.uk/react/705}{705}&\citet{Jones1981}\\
\hline
\multirow{6}{*}{H$_2$O$^+$$\left\lbrace\begin{array}{l}
                \\
                \\
                \\
                \\
                \\
                \\
                \end{array}\right. $} & + & H & $\longrightarrow$ & none&  & &  & & & &\\
 & + & H$_2$ & $\longrightarrow$ & H& + & H$_3$O$^+$ & $\phantom{0}9.7\phantom{0}$&&&1.62&\href{https://umistdatabase.uk/react/3250}{3250}&\citet{Tran2018}\\
 & + & O & $\longrightarrow$ & H$_2$ & + & O$_2^+$& $\phantom{0}0.4\phantom{0}$ &&&0.63&\href{https://umistdatabase.uk/react/4621}{4621}&\citet{Viggiano1980}\\
 & + & HO & $\longrightarrow$ & O& + & H$_3$O$^{+}$ & $\phantom{0}6.9\phantom{0}$&$-0.50$&&1.69&\href{https://umistdatabase.uk/react/4687}{4687}&\citet{Prasad1980}\\
 & + & H$_2$O & $\longrightarrow$ & HO& + & H$_3$O$^+$ & $21\phantom{.00}$&$-0.50$&&1.00&\href{https://umistdatabase.uk/react/3317}{3317}&\citet{Huntress1973}\\
 & + & O$_2$ & $\longrightarrow$ & H$_2$O& + & O$_2^+$ & $\phantom{0}4.6\phantom{0}$&&&0.55&\href{https://umistdatabase.uk/react/485}{485}&\citet{Rakshit1980}\\
\hline
\multirow{6}{*}{H$_3$O$^+$$\left\lbrace\begin{array}{l}
                \\
                \\
                \\
                \\
                \\
                \\
                \end{array}\right. $} & + & H & $\longrightarrow$ & none&  & &  && &  &\\
 & + & H$_2$ & $\longrightarrow$ & none&  & &  && &  &\\
 & + & O & $\longrightarrow$ & none&  & &  && &  &\\
 & + & HO & $\longrightarrow$ & none&  & &  && &  &\\
 & + & H$_2$O & $\longrightarrow$ & none&  & &  && &  &\\
 & + & O$_2$ & $\longrightarrow$ & none&  & &  && &  &\\
\hline
\multirow{6}{*}{\phantom{H$_3$}O$_2^+$$\left\lbrace\begin{array}{l}
                \\
                \\
                \\
                \\
                \\
                \\
                \end{array}\right. $} & + & H & $\longrightarrow$ & none&  & &  && &  &\\
 & + & H$_2$ & $\longrightarrow$ & none & & &  && &  &\\
 & + & O & $\longrightarrow$ & none&  & &  && &  &\\
 & + & HO & $\longrightarrow$ & none&  & &  && &  &\\
 & + & H$_2$O & $\longrightarrow$& none&  & &  && &  &\\
 & + & O$_2$ & $\longrightarrow$ & sym. &  & &  && &  &\\
\hline
\multirow{6}{*}{O$_2$H$^+$$\left\lbrace\begin{array}{l}
                \\
                \\
                \\
                \\
                \\
                \\
                \end{array}\right. $} & + & H & $\longrightarrow$ & none&  & &  && &  &\\
 & + & H$_2$ & $\longrightarrow$ & O$_2$& + & H$_3^+$& $\phantom{0}6.4\phantom{0}$ &&&0&\href{https://umistdatabase.uk/react/3276}{3276}&\citet{Adams1984}\\
 & + & O & $\longrightarrow$ & O$_2$& + & HO$^+$& $\phantom{0}6.2\phantom{0}$ &&&0.67&\href{https://umistdatabase.uk/react/4637}{4637}&\citet{Prasad1980}\\
 & + & HO & $\longrightarrow$ & O$_2$& + & H$_2$O$^+$& $\phantom{0}6.1\phantom{0}$ &$-0.50$&&1.76&\href{https://umistdatabase.uk/react/4694}{4694}&\citet{Prasad1980}\\
 & + & H$_2$O & $\longrightarrow$ & O$_2$& + & H$_3$O$^+$& $\phantom{0}8.2\phantom{0}$ &$-0.50$&&2.76&\href{https://umistdatabase.uk/react/3369}{3369}&\citet{Prasad1980}\\
 & + & O$_2$ & $\longrightarrow$ & none&  & &  &\\
\end{tabular}
\end{table*}

\section{Updating ion velocity post-collision}\label{AppendixPostcoll}
After an ion-neutral collision, the ion product from the reaction has a new velocity that should be calculated based on the initial properties of the reactants. In addition, as discussed in Appendix \ref{Appendixreaction}, the ion-neutral collisions considered here are a priori exothermic (spontaneous in the gas phase). In reality, some reactions are only triggered above a certain energy threshold (hence endothermic) when collisions are treated from a pure kinetic aspect involving the energy-dependent cross-sections that are ignored here. As they are exothermic, the excess energy of the reaction should be redistributed: It is mainly dissipated through excitation and change in the kinetic energy of the products. We assume no excitation of the products as they are no data to support our modelling here. To determine the post-collision ion velocity, we assume that the excess energy $\Delta H$ is redistributed into kinetic energy only. Our approach is considered quasi-elastic. Assuming energy conservation before and after the collisions, one gets
\begin{equation}
\label{conservation}
    \dfrac{1}{2}m_1\varv_1^2+ \dfrac{1}{2}m_2\varv_2^2= \dfrac{1}{2}m_3\varv_3^2+ \dfrac{1}{2}m_4\varv_4^2-\Delta H 
\end{equation}
where $\Delta H$ is the excess energy ($\Delta H>0$ for exothermic reactions), $\vec{\varv}_s$ is the velocity of the species $s$ (products or reactants), and $m_s$ is their mass. We assume that the centre of mass velocity is conserved $\vec{\varv}_\text{CM}$, as no external force is applied to the system during the collision, given by
\begin{equation}
\vec{\varv}_\text{CM}=\dfrac{m_1\vec{\varv}_1+m_2\vec{\varv}_2}{m_1+m_2}=\dfrac{m_3\vec{\varv}_3+m_4\vec{\varv}_4}{m_3+m_4}
\end{equation}
We define $\vec{u}_s$ as the velocity in the centre of mass frame such that $\vec{\varv}_s=\vec{\varv}_\text{CM}+\vec{u}_s$. In the centre of mass frame  $m_1\vec{u}_1=-m_2\vec{u}_2$ and $m_3\vec{u}_3=-m_4\vec{u}_4$. In addition, there is no change in mass, $m_1+m_2=m_3+m_4=M$. From Eq.~\ref{conservation}:
\begin{align*}
    m_1\varv_1^2 +m_2\varv_2^2&= m_3\Vert\vec{\varv}_\text{CM}+\vec{u}_3\Vert^2+ m_4\Vert\vec{\varv}_\text{CM}+\vec{u}_4\Vert^2-2\Delta H\\
    m_1\varv_1^2 +m_2\varv_2^2&= M\varv^2_\text{CM}+m_3u_3^2+m_4u_4^2-2\Delta H\\
    \dfrac{(m_1+m_2)}{M}(m_1\varv_1^2+m_2\varv_2^2) &= \dfrac{\Vert m_1\vec{\varv}_1+m_2\vec{\varv}_2\Vert^2}{M}+m_3u_3^2+\underbrace{m_4u_4^2}_{m_3^2u_3^2/m_4}-2\Delta H\\
&=\dfrac{m^2_1\varv_1^2+2\vec{\varv}_1\cdot\vec{\varv}_2+m^2_2\varv_2^2}{M}\\
&\hspace{5em}\phantom{=~}+\left(m_3+\dfrac{m^2_3}{m_4}\right)u_3^2-2\Delta H\\
    \dfrac{m_1m_2}{M}(\varv^2_1-2\vec{\varv}_1\cdot\vec{\varv}_2+\varv^2_2)&= \dfrac{M}{m_4}m_3u_3^2-2\Delta H\\
    \overbrace{\dfrac{m_1m_2}{M}\Vert\underbrace{\vec{\varv}_1-\vec{\varv}_2}_{\vec{\varv}_\text{rel.}}\Vert^2}^{2E_\text{rel.}}&= \dfrac{M}{m_4}m_3u_3^2-2\Delta H\\
    u_3^2&=\dfrac{m_4}{m_3M}\left(2E_\text{rel.}+2\Delta H\right)
    \end{align*}
such that
\begin{align*}
    u_3^2&=\left[\dfrac{2m_4}{m_3M}\right]\left(E_\text{rel.}+\Delta H\right)\\
    u_4^2&=\left[\dfrac{2m_3}{m_4M}\right]\left(E_\text{rel.}+\Delta H\right)
\end{align*}
where $E_\text{rel.}=\mu\Vert\vec{\varv}_1-\vec{\varv}_2\Vert^2/2$ is by definition the total kinetic energy in the centre of mass frame and $\mu=m_1m_2/M$ is the reduced mass before the collision and is conserved during charge exchange reactions, unlike proton transfer. The result is consistent with endothermic reactions only triggered above a certain relative speed (with the condition $E_\text{rel.}>\Delta H$ or $\Vert\vec{\varv}_1-\vec{\varv}_2\Vert>\sqrt{2\Delta H/\mu}$). As $\Delta H$ is usually provided in eV and mass in u (or Da), a useful conversion is 1~J\,kg\,=\,$9.65\times 10^7$~eV\,u. 

The new velocity is then $\vec{\varv}_3=\vec{\varv}_\text{CM}+{u}_3\vec{U}$  where $\vec{U}$ is a random vector of norm 1 equivalent at drawing a random point on a sphere centred on 0 of radius 1. Note that in the low relative speed limit ($\Vert\vec{\varv}_1-\vec{\varv}_2\Vert\approx 0$), $u_3=\sqrt{m_4/m_3}\sqrt{2\Delta H/M}$.  This shows that an ion-neutral reaction such as H$_2$O + H$_2$O$^+$ $\longrightarrow$ HO + H$_3$O$^+$ + 1\,eV (cf. Appendix \ref{AppendixPostcoll}) in slow relative speed limit releases H$_3$O$^+$ above 2~km\,s$^{-1}$.

This also showed that a priori ions should be a mix of several populations with different temperatures. For instance, H$_3$O$^+$ is formed through different reactions with different $\Delta H$ from 1.00~eV to 4.43~eV. However, H$_3$O$^+$ is often the most massive of the products; hence, it recoups less than half of this energy. For example, in the low relative velocity limit, for H$_2$O + H$_2$O$^+$, 0.47~eV would be given to H$_3$O$^+$. However, for H$_2^+$ + H$_2$O, although the reaction is the most energetic, H$_3$O$^+$ gets back 0.22~eV only.

\section{Molecular properties}\label{AppendixPAEI}
\begin{table}
\centering
\begin{tabular}{l r r}
Species&PA (eV)&IE (eV)\\
\hline
$\mathrm{H}$&2.67&13.598\\
$\mathrm{H_2}$&4.38&	15.425\\
$\mathrm{O}$&5.03&13.618\\
$\mathrm{HO}$&6.15&13.017\\
$\mathrm{H_2O}$&7.16&12.621\\
$\mathrm{O_2}$&4.37&12.070\\
\end{tabular}
\caption{Proton affinity (PA) and ionisation energy (IE) of the different neutral species detected at Ganymede.} 
\end{table}
As $\mathrm{H}$ has the lowest proton affinity (PA), $\mathrm{H_2^+}$ reacts with all neutrals to transfer one proton. In contrast, having the highest PA, $\mathrm{H_2O}$ steals a proton from any protonated molecule. As $\mathrm{H_2}$ and $\mathrm{O_2}$ have extremely close PA, the proton transfer reaction is almost athermic and therefore reversible. Similar analysis can be done by looking at the ionisation energy. On the one hand, $\mathrm{O_2}$ has the lowest ionisation energy (IE) in the ground state. Therefore, $\mathrm{O_2^+}$ is unlikely to be neutralised through charge exchange (except with O$_2$). Nevertheless, O$_2^+$ can react with neutral species if it is in an excited state. On the other hand, $\mathrm{H_2}$ has the largest ionisation energy, leading to the neutralisation of $\mathrm{H_2^+}$. Having both a large IE and a low PA is why $\mathrm{H_2^+}$ may often lead to two different ion products by reacting with one given neutral species as it will easily either capture an electron or transfer its proton. $\mathrm{O^+}$ with the second highest IE (though close to that of H$^+$) makes it very reactive through charge exchange, leading to its loss. By a simple analysis, it is already possible to anticipate the main ion species:  $\mathrm{H_3O^+}$ based on PA and $\mathrm{O_2^+}$ based on IE. 

The priority order for both processes are:
\begin{itemize}
    \item PA: H$_2^+$, O$_2$H$^+$, H$_3^+$, HO$^+$, H$_2$O$^+$, H$_3$O$^+$ (you tend to form the latter following successive proton transfer)
    \item IE: H$_2^+$, O$^+$, H$^+$, HO$^+$, H$_2$O$^+$, O$_2^+$ (you tend to form the latter following successive charge-exchange)
\end{itemize}
O$^+$ and H$^+$ are not involved in proton transfer reactions, while H$_3^+$, H$_3$O$^+$, and O$_2$H$^+$ are not in charge-exchange ones.

\section{Energy spectra}\label{Appendixspectra}
\begin{figure}
\includegraphics[width=\columnwidth]{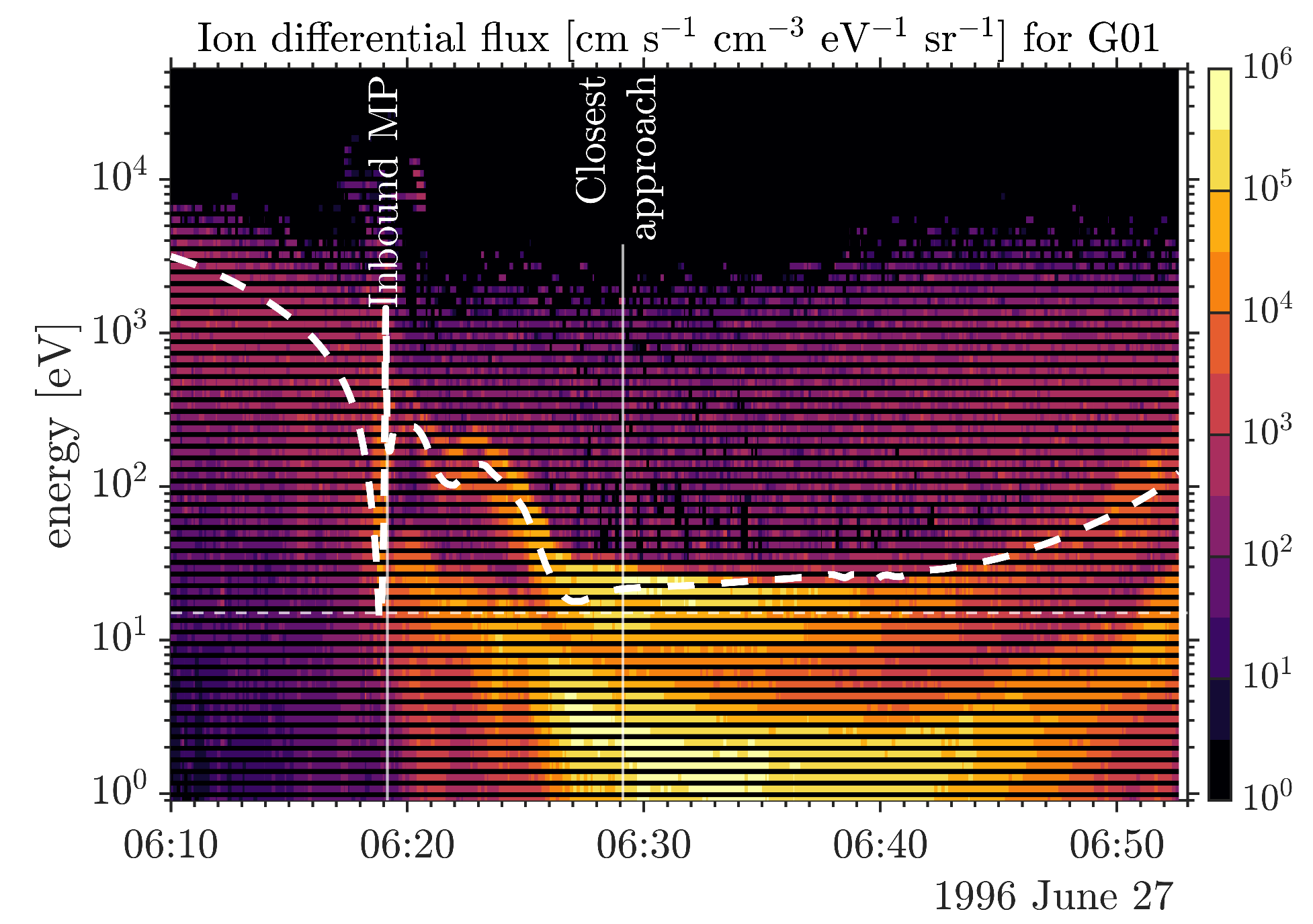}\\
\includegraphics[width=\columnwidth]{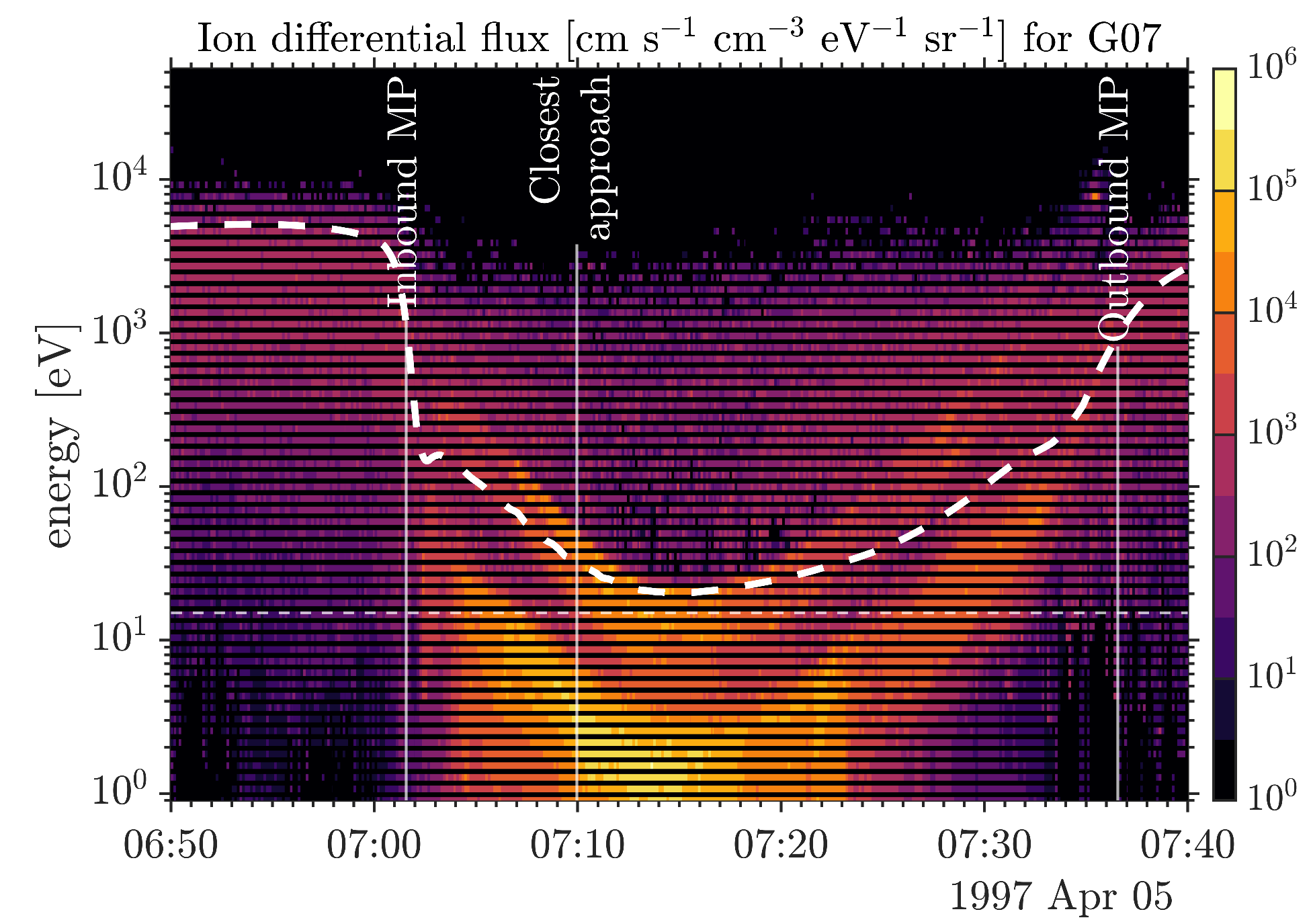}\\
\includegraphics[width=\columnwidth]{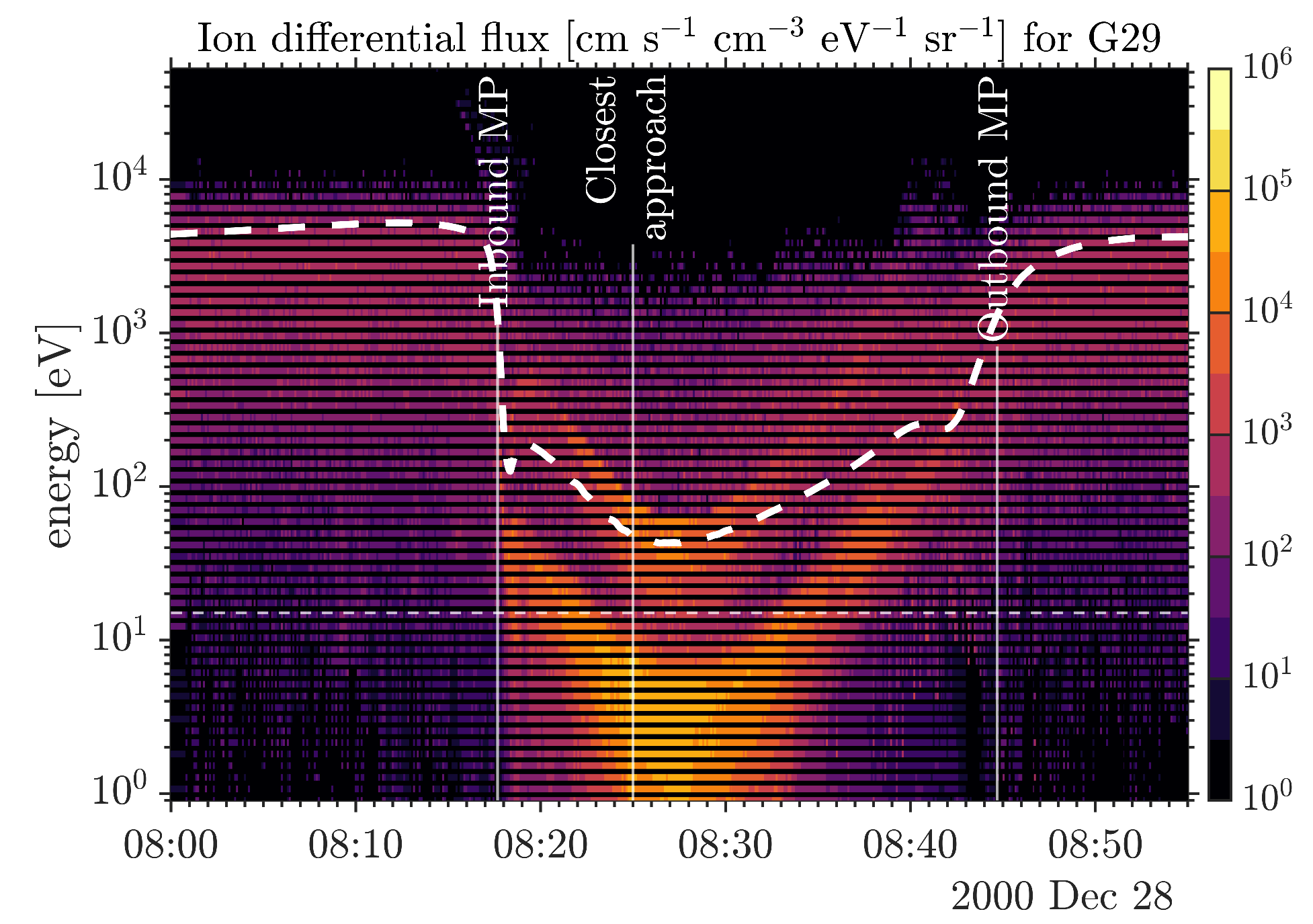}
\caption{Simulated ion energy spectra as a function of time for G01 (left), G07 (middle), and G29 (right) flybys with ion-neutral chemistry. The white dashed line is the kinetic energy of an O$_2^+$ drifting at the local $\Vert\vec{E}\times\vec{B}\Vert/B^2$ speed in the spacecraft frame. To be compared with Fig. 7 from \citet{Beth2025}. \label{Figapp5}}
\end{figure}

For completeness, we have simulated similar energy spectra than those in \citet{Beth2025} for G01, G07, and G29 flybys, including ion-neutral collisions. Fig. \ref{Figapp5} should be compared with Fig. 7 in \citet{Beth2025}: Except being slightly noisier owing to fewer macroparticles, the spectra do not exhibit noticeable differences. Although we have simulated new ions namely H$_3^+$, H$_3$O$^+$, and O$_2$H$^+$, their mass is not drastically different from those already simulated in the collisionless case: $m_\text{H$_3^+$}/m_\text{H$_2^+$}=1.5$, $m_\text{H$_3$O$^+$}/m_\text{H$_2$O$^+$}=1.06$, and $m_\text{O$_2$H$^+$}/m_\text{O$_2^+$}=1.03$. In addition, as ions already spread over several energy bins, H$_3^+$ ion energy blends with that of H$_2^+$, H$_3$O$^+$ with H$_2$O$^+$/HO$^+$/H$^+$, and O$_2$H$^+$ with O$_2^+$.

\label{lastpage}
\end{document}